\documentclass[twoside,11pt]{article}

\usepackage{jmlr2e}

\usepackage{amssymb}
\let\proof\relax

\usepackage{amsmath}
\usepackage{epsfig}
\usepackage{graphicx}
\usepackage{float}
\usepackage{multirow}
\usepackage{color}
\usepackage{lineno}
\usepackage[normalem]{ulem} 
\usepackage{makeidx}
\usepackage{xspace}
\usepackage{wrapfig}
\usepackage{cleveref}
\usepackage{subcaption}
\usepackage{booktabs}
\usepackage{adjustbox}
\usepackage{mathtools}
\usepackage{fixmath}
\usepackage[ruled,vlined]{algorithm2e}
\usepackage{amsthm}
\usepackage{enumitem}
\usepackage{lastpage}

\usepackage[table]{xcolor}
\definecolor{colorA}{rgb}{0.24705882352941178, 0.5647058823529412, 0.8549019607843137    }
\definecolor{colorB}{rgb}{1.0, 0.6627450980392157, 0.054901960784313725  }
\definecolor{colorC}{rgb}{0.7411764705882353, 0.12156862745098039, 0.00392156862745098 }
\definecolor{colorD}{rgb}{0.5803921568627451, 0.6431372549019608, 0.6352941176470588  }
\definecolor{colorE}{rgb}{0.5137254901960784, 0.17647058823529413, 0.7137254901960784}
\definecolor{colorF}{rgb}{0.6627450980392157, 0.4196078431372549, 0.34901960784313724 }
\definecolor{colorG}{rgb}{0.9058823529411765, 0.38823529411764707, 0.0}
\definecolor{colorH}{rgb}{0.7254901960784313, 0.6745098039215687, 0.4392156862745098}
\definecolor{colorI}{rgb}{0.44313725490196076, 0.4588235294117647, 0.5058823529411764}
\definecolor{colorJ}{rgb}{0.5725490196078431, 0.8549019607843137, 0.8666666666666667}
\definecolor{colorK}{rgb}{0.25098039215686274, 0.6901960784313725, 0.6509803921568628}

\usepackage{mathtools,tikz,caption}
\DeclareRobustCommand\sampleline[1]{%
  \tikz\draw[#1] (0,0) (0,\the\dimexpr\fontdimen22\textfont2\relax)
  -- (2em,\the\dimexpr\fontdimen22\textfont2\relax);%
}
\newtheorem{Marginal Independence Testing}{Marginal Independence Testing}

\newtheorem*{bdtest*}{do-null Independence Testing}
\newtheorem{Definition}{Definition}
\newtheorem{Theorem}{Theorem}
\newtheorem{Remark}{Remark}

\newtheorem{Proposition}{Proposition}
\newtheorem{Lemma}{Lemma}

\newtheorem{Corollary}{Corollary}

\newenvironment{Proof}{\textit{\bf Proof}}{\hfill\BlackBox\\[6pt]}

\newcommand{\tcat}{\text{cat}}
\newcommand{\tcont}{\text{cont}}
\newcommand{\tbin}{\text{bin}}

\newcommand{\cmid}{\,|\,}

\newcommand{\densratio}{{\frac{p^*(x)}{p(x\cmid z)}}}
\newcommand{\densratioprime}{{\frac{p^*(x')}{p(x'\cmid z')}}}
\newcommand{\densratiodiscretei}{{\frac{p^*(x_i)}{p(x_i\cmid z_i)}}}
\newcommand{\densratiodiscretej}{{\frac{p^*(x_j)}{p(x_j\cmid z_j)}}}
\newcommand{\estdens}{{\hat{h}_n(x,z)}}
\newcommand{\estdensprime}{{\hat{h}_n(x',z')}}
\newcommand{\estdensi}{{\hat{h}_n(x_i,z_i)}}
\newcommand{\estdensj}{{\hat{h}_n(x_j,z_j)}}

\newcommand{\estasympslow}{{\mathcal{O}\left(\frac{1}{n^{\alpha}}\right)}}
\newcommand{\estasymp}{{\mathcal{O}\left(\frac{1}{n^{2\alpha}}\right)}}
\SetKwInput{KwInput}{Input}
\SetKwInput{KwOutput}{Output}
\SetKwInput{KwReturn}{Return}

\begin{document}

\jmlrheading{25}{2024}{1-\pageref{LastPage}}{11/21; Revised
2/24}{5/24}{21-1409}{Robert Hu, Dino Sejdinovic, Robin Evans}
\ShortHeadings{A Kernel Test for Causal Association via Noise Contrastive Backdoor Adjustment}{Hu, Sejdinovic and Evans}
\title{A Kernel Test for Causal Association via \\Noise Contrastive Backdoor Adjustment}

\author{\name Robert Hu \email robyhu@amazon.co.uk \\
       \addr  Amazon
       \AND
        \name Dino Sejdinovic \email dino.sejdinovic@adelaide.edu.au \\
       \addr School of Computer and Mathematical Sciences\\
       University of Adelaide\\
       Adelaide 5005, Australia
              \AND
        \name Robin J.~Evans \email evans@stats.ox.ac.uk \\
       \addr Department of Statistics\\
       University of Oxford\\
       Oxford OX1 3LB, UK}

\editor{Silvia Chiappa}

\maketitle

\begin{abstract}%
Causal inference grows increasingly complex as the dimension of confounders increases. Given treatments $X$, outcomes $Y$, and measured confounders $Z$, we develop a non-parametric method to test the \textit{do-null} hypothesis that, after an intervention on $X$, there is no marginal dependence of $Y$ on $X$, against the general alternative. Building on the Hilbert-Schmidt Independence Criterion (HSIC) for marginal independence testing, we propose backdoor-HSIC (bd-HSIC), an \emph{importance weighted} HSIC which combines \emph{density ratio estimation} with kernel methods. Experiments on simulated data verify the correct size and that the estimator has power for both binary and continuous treatments under a large number of confounding variables. Additionally, we establish convergence properties of the estimators of covariance operators used in bd-HSIC. We investigate the advantages and disadvantages of bd-HSIC against parametric tests as well as the importance of using the do-null testing in contrast to marginal or conditional independence testing. A complete implementation can be found at \hyperlink{https://github.com/MrHuff/kgformula}{\texttt{https://github.com/MrHuff/kgformula}}.
\end{abstract}

\begin{keywords}
  causal inference, noise contrastive estimation, kernel methods, backdoor adjustment, hsic, observational data
\end{keywords}

\section{Introduction and Related Work}
Modern causal inference often considers very large data sets with many observed confounding variables that may have vastly different properties. These settings are considered in a wide range of applications where randomized controlled trials are not always readily available: these include epidemiology \citep{doi:10.2105/AJPH.2004.059204}, brain imaging \citep{2020_xray}, retail \citep{retail_causal}, and entertainment platforms \citep{Liang2016CausalIF}. The inference setting is often complex, with high dimensional confounding variables needing to be accounted for. In such complex settings, non-parametric inference schemes that answer a simple query of causal association from observational data are needed as an initial step before more sophisticated causal relationships can be established. \newline \\ G-computation \citep{robins1986new} is a classical method for estimating a causal effect from observational studies involving variables that are both mediators and confounders. Its popularity persists to this day \citep{daniel2013methods, keil2020quantile}, because it allows one to test for a non-null causal effect using a variety of postulated models. The causal effect can also be identified for models fulfilling the so-called \emph{backdoor criterion} \citep{pearl_2009} and \emph{ignorability assumptions} \citep{10.2307/2335942}.

In this paper, we propose a non-parametric approach to testing for the presence of a causal effect. In the Reproducing Kernel Hilbert Space (RKHS) and machine learning literature, the Hilbert-Schmidt Independence Criterion (HSIC) introduced by \cite{10.1007/11564089_7} is a widely used approach to non-parametric testing of independence. As the HSIC has good power properties and it is applicable to multivariate settings as well as to random variables taking values in generic domains, we use it as a foundation in this paper. Using g-computation principles, we introduce an extension of HSIC that can be applied to causal association testing. 

Variations of HSIC have been proposed to test for conditional independence \citep{doran2014permutation,10.5555/3020548.3020641,uai_kernel,StroblZhangVisweswaran+2019} and while testing for conditional independence and causal association is different, some of the techniques used for conditional independence testing can be carried over to causal association. Kernel methods have been applied to tests for causal association in the binary treatment case in \cite{10.5555/3546258.3546420}, and for average treatment effect estimation by \cite{10.1093/biomet/asad042}. In this paper, we will extend the approach of \citeauthor{10.5555/3546258.3546420} to a general treatment setting. This extension is challenging as instead of modeling propensity scores, we work with general conditional densities and as such need to consider a suite of density ratio estimation techniques. We hope to bring further insights on how MMD-based approaches proposed in \cite{10.1093/biomet/asad042} can be used for hypothesis testing.

Non-parametric tests for causal association generally come with the additional difficulty of correctly simulating the null distribution through permutations, as direct permutations often break confounding relationships as we show in Section \ref{permutation_break}. In this paper, we also aim to give a formal treatment on the permutation aspects of non-parametric kernel tests and expand upon ideas in \cite{rosenbaum1984conditional} to consider cases beyond binary treatment. \newline \\ We summarize our contributions as follows: 
\begin{enumerate}
    \item We introduce bd-HSIC, which is derived analogously to HSIC but instead uses importance weighted covariance operators, and further establish the convergence properties of the corresponding estimators in the causal setting, coupled with a novel optimization strategy to improve effective sample size.
    \item We provide a novel permutation strategy for bd-HSIC that yields a permutation test with theoretically correct size for arbitrary treatment types.
    \item We demonstrate that bd-HSIC has correct size and good power for different types of treatments and a large number of confounders when testing the \textit{do-null}.
    \item We analyze bd-HSIC by providing ablation studies and characterize under which circumstances it becomes invalid.
    \end{enumerate}
    
The rest of the paper is organized as follows: \Cref{background} describes the problem setting and provides a background on HSIC, \Cref{method} presents bd-HSIC and establishes convergence properties of the associated estimators, \Cref{method2} presents additional details on the estimation procedure of bd-HSIC, \Cref{results} provides the experimental results and we conclude the paper in \Cref{conclusion}.
 
\section{Background}
\subsection{Terminology} We start by reviewing some terminology commonly used in causal inference:

\setlist[description]{font=\normalfont\itshape\space}

\begin{description}[style=unboxed]
\item[Outcome:] Intuitively represents the outcome of interest possibly caused by the treatment. It is denoted by $Y$ and could be continuous such as blood pressure, or binary if it represented as recovery or not from disease. 
    \item[Treatment:] Treatments intuitively refer to the variables whose effect on the outcome $Y$ we are trying to infer. We denote \emph{treatments} by $X$ and in the case where treatment is placebo-controlled we can consider $X \in \{0,1\}$ representing control and active treatment respectively. In general however, treatment could be a random variable taking values in continuous or multivariate domains.
        \item[Observed confounding variables:] Intuitively represents patient characteristics that may affect both treatment and outcome. We denote these variables with $Z$ and these could be continuous variables such as height and weight, or binary variables such as sex.  
    \item[Assignment:] In many confounded settings, certain patient characteristics ($Z$) affect \emph{treatment assignment}. As an example, one can imagine socioeconomic factors having an effect on the availability of medical treatment. This creates a bias in treatment assignment which needs to be adjusted for. 
\end{description}

\begin{wrapfigure}[12]{r}{0.25\textwidth}
\vspace{-2.5em}
  \begin{center}
    \includegraphics[width=0.2\textwidth]{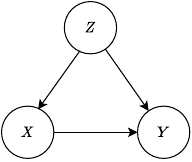}
  \end{center}
  \caption{Graph showing the relationship between treatments $X$, outcomes $Y$, and confounders $Z$}
      \label{first_fig_intuition}
\end{wrapfigure}
\subsection{Setup}
\label{background}

Consider a situation where we observe treatments $X$, outcomes $Y$ and potential confounders $Z$ defined on measurable spaces $\mathcal X$, $\mathcal Y$, and $\mathcal Z$, respectively.  We assume these quantities are observed as $\left\{ (z_i,x_{i},y_{i})\right\}_{i=1}^{n}\sim p$, where $p$ is some joint probability density on the product space $\mathcal{Z}\times\mathcal{X}\times\mathcal{Y}$. We are interested in establishing a causal relationship between treatments $X$ and outcomes $Y$, meaning that we manage to isolate whether there precisely exists a dependency between $X$ and $Y$ whilst holding all pre-treatment variables constant. In ideal circumstances, we would not have any confounders and such a relationship can be established straightforwardly, using e.g.~regression methods. However, such circumstances are unusual outside randomized trials and would require that the treatment is assigned to units by some exogenous process. In the more common case of observational studies, dependencies among $X,Y,Z$ are often depicted as in \Cref{first_fig_intuition}. The existence of confounders $Z$ complicates establishing the causal relationship between $X$ and $Y$, as they introduce dependence between $X$ and $Y$ which is difficult to disentangle from the postulated causal effect.  \newline \\ \emph{A motivating real-world problem} \quad In development economics, it is of great importance to establish causes of infant mortality \citep{Ensor2010TheIO}. The causes may often have a non-linear association with the mortality while being confounded by circumstantial factors such as the socio-economical background of the parents and the medical history of the mother. \newline \\ In this setting, it is also important to test for \emph{distributional differences} different policies might induce, as tests based on average treatment effect may fail to capture such differences as discussed in \citep{10.5555/3546258.3546420,dr_cfme}. As the mean embedding framework offers a natural way to non-parametrically quantify distributional differences, kernel-based test statistics will be our preferred strategy. We will illustrate throughout the paper the importance of a non-parametric test that is able to capture non-linear dependencies for different treatments under a large number of confounders.

\begin{Definition} (do-null hypothesis) Let
\begin{equation*}
    p(y \cmid \text{do}(x)):=\int p(y\cmid x,z) \, p(z) \, dz
\end{equation*}
where in general $p(y|x)\neq p(y\cmid \text{do}(x))$ since $X$ and $Z$ may be dependent. We are interested in testing if the interventional distribution $Y \mid do(X=x)$ does not depend on the value of the treatment variable $X$. We refer to this hypothesis as a \emph{do-null hypothesis}, which can be stated as $Y \perp X \mid do(X=x)$. In terms of distributions, we can consider
\begin{equation}
\label{eq:donullhypothesis}
H_{0}: p(y\cmid \text{\it do}(x))=p^*(y)
\end{equation}
as our null hypothesis versus the general alternative, where $p^*$ is an arbitrary distribution that does not depend upon the value of $X$. We consider $H_0$ for all values of $do(x)$ and the observational regime. Note that our null hypothesis \emph{does not} imply that $p(y\cmid do(x)) = p(y)$, which is why we introduce $p^*$. 
\end{Definition}

For a remark on why we need $p^*$, see Appendix \ref{pstar}. \newline \\ In this 
paper, we will introduce a test statistic for the hypothesis \eqref{eq:donullhypothesis}, but first, we review HSIC \citep{10.1007/11564089_7}, which serves as the foundation for our contribution.

\subsection{Hilbert-Schmidt Independence Criterion}
The Hilbert-Schmidt Independence Criterion (HSIC) is a powerful non-parametric test of independence for high-dimensional data. It considers the problem of empirically establishing whether there is any form of departure from independence between two random variables taking values on generic domains.

\begin{Definition}[Marginal Independence Testing]
Let $P_{x y}$ be a Borel probability measure defined on a domain $\mathcal{X} \times \mathcal{Y}$ and let $P_{x}$ and $P_{y}$ be the respective marginal distributions on $\mathcal{X}$ and $\mathcal{Y}$. Given an i.i.d.~sample $(X, Y)=$ $\left\{\left(x_{1}, y_{1}\right), \ldots,\left(x_{m}, y_{m}\right)\right\}$ of size $m$ drawn 
according to $P_{x y},$ does $P_{x y}$ factorize as $P_{x} P_{y}$? We usually consider the null hypothesis to be \[H_0:P_{xy}=P_xP_y \] against the general alternative \[H_1:P_{xy}\not=P_xP_y. \]
\end{Definition}
Since we do not have access to $P_{xy}$, $P_x$, or $P_y$, we need to estimate or represent these distributions through either parametric or non-parametric means. A convenient way for representing distributions is to use the RKHS formalism \citep{10.5555/559923}. \newline \\ HSIC can intuitively be understood as a covariance between RKHS representations of random variables. 
\begin{Definition}[Reproducing Kernel Hilbert Spaces] Let $\mathcal{X}$ be a non-empty set and $\mathcal{H}$ a Hilbert space of functions $f: \mathcal{X} \rightarrow \mathbb{R}$. Then $\mathcal{H}$ is called a \emph{reproducing kernel Hilbert space} endowed with dot product $\langle\cdot, \cdot\rangle$ if there exists a function $k: \mathcal{X} \times \mathcal{X} \rightarrow \mathbb{R}$ with the following properties:
\begin{enumerate}
    \item $k$ has the reproducing property 
    \[
    \langle f, k(x, \cdot)\rangle=f(x),\;\; \forall f \in \mathcal{H}, x \in \mathcal{X};
\]
    \item $k$ spans $\mathcal{H},$ that is, $\mathcal{H}=\overline{\operatorname{span}\{k(x, \cdot) | x \in \mathcal{X}\}}$ where the bar denotes the completion of the space.
\end{enumerate}
\end{Definition}
\noindent We generally refer to the function $k$ as a \textit{kernel}. For certain choices of $k$, the corresponding RKHS is \emph{characteristic}, i.e. the mapping of a probability measure $\mu$,  $\mu \mapsto \int_{X}k(x,\cdot)d\mu(x)$ is injective, i.e. $\mu$ is mapped to a unique element in the RKHS. We refer to \citet{JMLR:v12:sriperumbudur11a} for more details. This property is very practical, as it allows us to embed probability distributions into the RKHS and calculate their expectations based on observations. These embeddings are called kernel mean embeddings, see \citet{Muandet_2017} for a thorough exposition.

\begin{Definition}
Let $\mathcal{X}$ be a measurable space and let $\mathcal H_{\mathcal{X}}$ be an RKHS on $\mathcal{X}$ with kernel $k$. Let $P$ be a Borel probability measure on $\mathcal{X}$.  An element $\mu_x\in \mathcal H_\mathcal{X}$ such that $\mathbb{E}_{x\sim P}[f(x)]=\langle f,\mu_x\rangle,\;\forall f \in \mathcal H _\mathcal{X}$ is called the \textbf{kernel mean embedding} of $P$ in $\mathcal H_\mathcal{X}$, where $\mu_x: P \mapsto \int_{\mathcal{X}} k(x,\cdot )dP(x)$.
\end{Definition}

\noindent A sufficient condition for the existence of a kernel mean embedding is that $\mathbb{E}_{x\sim P}[\sqrt{k(x,x)}]<\infty$, which is satisfied for, e.g. bounded kernel functions.

\noindent Given some observations $\{x_i\}_{i=1}^n \sim P $, the empirical mean embedding of $P$ is estimated as:
\begin{equation*}
    \hat{\mu}_x = \frac{1}{n}\sum_{i=1}^nk(x_i,\cdot).
\end{equation*}
Kernel mean embeddings intuitively allow us to estimate expectations under each of $p_{xy}$, $p_x$, and $p_y$. In order to test for marginal dependence, we consider a test statistic based on $\operatorname{Cov}[f(X),g(Y)]$, where $f,g$ are arbitrary continuous functions evaluating random variables $X,Y$. Instead of picking individual functions $f$,$g$, we consider the representation of their covariance using the RKHS.
\begin{Definition}
Let $(X,Y)$ be a pair of random variables defined on $\mathcal X \times \mathcal Y$ and let  $\mathcal H_{\mathcal{X}}$ and  $\mathcal H_{\mathcal{Y}}$ be RKHSs on $\mathcal X$ and $\mathcal Y$, respectively. An operator $C_{X,Y}: \mathcal{H}_{\mathcal{X}} \rightarrow \mathcal{H}_{\mathcal{Y}}$ such that

$$
\langle g,C_{X,Y}f\rangle = \operatorname{Cov}[f(X),g(Y)], \;\forall f\in\mathcal H_{\mathcal X}, g \in \mathcal {H}_\mathcal{Y}
$$
is called a \textbf{cross-covariance operator} of $X$ and $Y$. We note that $C_{X,Y}$ is the property of the joint distribution of pair $(X,Y)$. We denote it simply by $C$ when there is no ambiguity.
The Hilbert-Schmidt Independence Criterion (HSIC) is then defined as the squared Hilbert-Schmidt (HS) norm of $C$, i.e. 
\begin{equation*}
    \text{HSIC}(X,Y):= \left\|C\right\|_{\mathrm{HS}}^{2}= \sum_i\sum_j\langle Cu_i,v_j \rangle^2
\end{equation*}
with $u_i$ and $v_j$ being orthogonal basis of $\mathcal{H}_{\mathcal{X}}$ and $\mathcal{H}_{\mathcal{Y}}$ respectively. 
\end{Definition}
\noindent It is readily shown via reproducing property that $C$ can be written as
$$
C:=\mathbb{E}\left[\left(k(X,\cdot)\right) \otimes\left(l(Y,\cdot)\right)\right]-\mathbb{E}\left[k(X,\cdot)\right] \otimes\mathbb E\left[l(Y,\cdot)\right],
$$
where $k,l$ are kernels of $\mathcal H_\mathcal{X}$ and $\mathcal H_\mathcal{Y}$ respectively, and $\otimes$ denotes the outer product. An alternative view of HSIC is that it measures the squared RKHS distance between the kernel mean embedding of $P_{xy}$ and $P_xP_y$. For sufficiently expressive kernels \citep{JMLR:v12:sriperumbudur11a}, this distance is zero if and only if $X$ and $Y$ are independent. To see that $C$ indeed is a Hilbert-Schmidt operator, we refer to \cite{Muandet_2017}.

\noindent Given a sample $\{(x_i,y_i)\}_{i=1}^n$ from the joint distribution $P_{xy}$, an estimator\footnote{This is the most commonly used, biased estimator of HSIC. An unbiased estimator also exists, cf. \cite{10.1145/1273496.1273600}} of HSIC is given by:
\begin{equation*}\begin{aligned}
\widehat{\mathrm{HSIC}}(X, Y) &=\frac{1}{n^{2}} \sum_{i, j=1}^{n} k(x_{i}, x_{j}) l(y_{i}, y_{j})+\frac{1}{n^{2}} \sum_{i, j=1} k(x_{i}, x_{j}) \frac{1}{n^{2}} \sum_{k, l=1}^{n} l(y_{k}, y_{l}) \\
&-\frac{2}{n^{3}} \sum_{i, j, k=1}^{n} k(x_{i}, x_{j}) l(y_{i}, y_{l}).
\end{aligned}\end{equation*}
The estimator above serves as the test statistic, for complete derivation, we refer to \cite{10.1007/11564089_7}. To estimate its distribution under the null hypothesis, we resort to repeatedly permuting $y_i$'s to obtain $\left\{\left(x_{i}, y_{\pi(i)}\right)\right\}_{i=1}^{n}$ where $\pi$ is a random permutation, and recomputing HSIC on this permuted data set. We note that the asymptotic null distribution of HSIC has a complicated form \citep{Zhang_2017}, and it is hence standard practice to use a permutation approach to approximate it. For more details on HSIC, we refer to \citep{10.1007/11564089_7}.

\subsection{Difference between the do-null, marginal independence and conditional independence}
While the do-null intuitively bears many similarities to marginal independence and conditional independence, we briefly illustrate the difference between these independencies and provide an experimental demonstration that conditional independence tests (RCIT) and marginal independence tests (HSIC) cannot be used to test for the do-null. 

\subsubsection{The do-null is not marginal independence $X\perp Y$.}

We contrast the do-null with marginal independence $X\perp Y$, by constructing a data set where the do-null is true, but there is marginal dependence between $X$ and $Y$. We illustrate the dependency in \Cref{marg_break}. \newline \\ We further plot size ($\alpha=0.05$) against sample size when applying HSIC and bd-HSIC (true weights) in \Cref{plot_14_hsic}. HSIC is not calibrated under the do-null. \newline \\ This can further be interpreted as equality in distribution of $Y$ for all values $do(x)$ excluding the observational regime, meaning our hypothesis does not test for marginal independence between $X$ and $Y$. 

\subsubsection{The do-null is not conditional independence $X \perp Y \mid Z$.}
\begin{figure}[htb!]
    \centering
        \begin{subfigure}{0.24\textwidth}
        \includegraphics[width=\linewidth,,height=2.5cm]{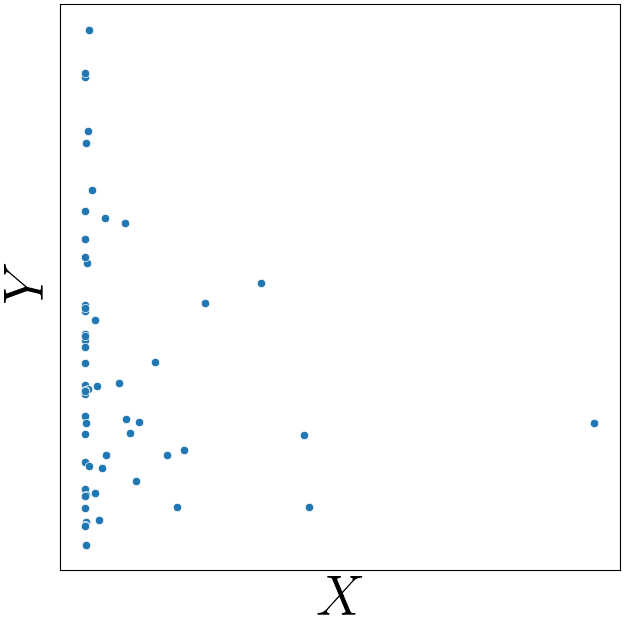}
        \caption{Marginal dependency but true do-null \\ \quad}
        \label{marg_break}
    \end{subfigure}\hfill%
    \begin{subfigure}{0.24\textwidth}
        \includegraphics[width=\linewidth,height=2.5cm]{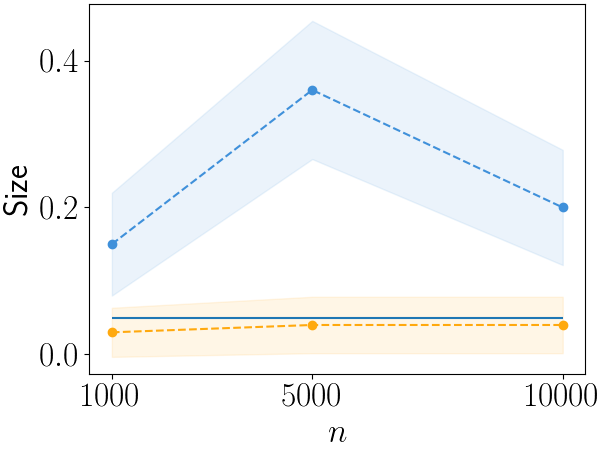}
        \caption{bd-HSIC(true weights):{\color{colorB} \sampleline{line width=2pt}} vs HSIC: {\color{colorA} \sampleline{line width=2pt}}}
    \label{plot_14_hsic}
    \end{subfigure}\hfill%
        \begin{subfigure}{0.24\linewidth}
\centering

            \includegraphics[width=\linewidth,height=2.5cm]{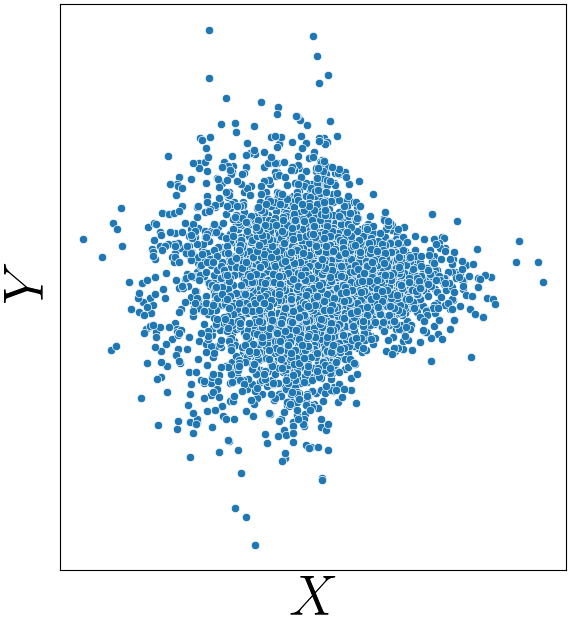}
        \caption{Conditional dependency but true do-null}
                \label{cond_break}
\end{subfigure}\hfill%
\begin{subfigure}{0.24\linewidth}
\centering
    \includegraphics[width=\linewidth,height=2.5cm]{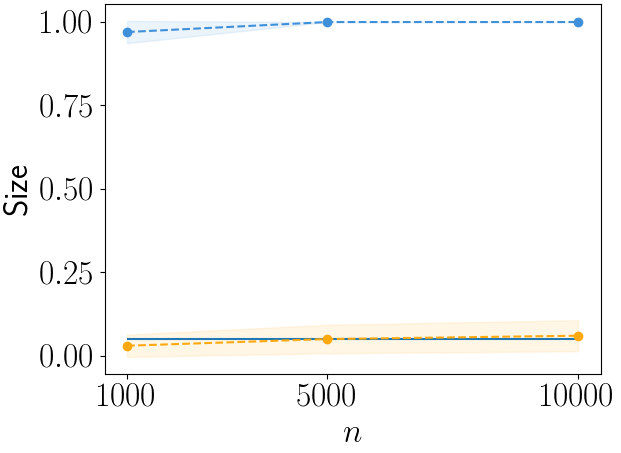}
        \caption{bd-HSIC(true weights):{\color{colorB} \sampleline{line width=2pt}} vs RCIT: {\color{colorA} \sampleline{line width=2pt}}}
    \label{plot_15_linear}
\end{subfigure}
\caption{Difference between the do-null, marginal independence and conditional independence. See Appendix \ref{cont_generation} on how the data in \Cref{marg_break} and \Cref{cond_break} is generated.}
\end{figure}

We similarly contrast the do-null with conditional independence  $X \perp Y \mid Z$. We simulate a data set such that there is a conditional dependence $X \not\perp Y \mid Z$ while the do-null is true. See \Cref{cond_break} for an illustration of the dependency between $X$ and $Y$.  We apply the ``RCIT" method, a kernel-based conditional independence test proposed in \cite{StroblZhangVisweswaran+2019} and demonstrate in \Cref{plot_15_linear} that it rejects almost all the time when applied to the data under the do-null. 

\section{Backdoor-HSIC}
\label{method}
Our proposed method has two parts, a weighted HSIC test statistic and a density ratio estimation procedure. In this section, we introduce the test statistic, which we term \textit{backdoor-HSIC} (bd-HSIC).

\subsection{Overview}

\subsubsection{The do-operator and identifiability}

We start with reviewing the meaning of the do-operator and how it lays the foundation for bd-HSIC. In the remainder of the paper, we will assume that all relevant probability distributions admit densities.

\begin{figure}[!htb]
    \centering
    \includegraphics{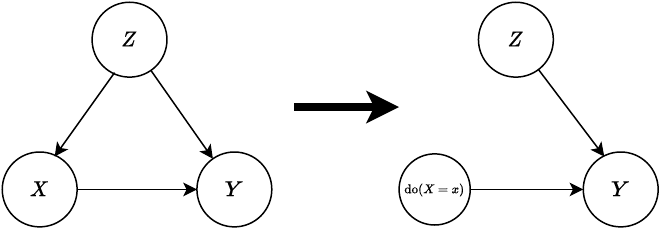}
    \caption{Illustration of the interventional distribution $p(y\mid do(x))$}
    \label{fig:do_null}
\end{figure}
In \Cref{fig:do_null}, we first consider the graphical representation of the problem of establishing the relationship between $X$ and $Y$ under observed confounders $Z$. We apply the do-operation \citep{pearl_2009} on our treatment $X$, in order to remove any dependency between $X$ and $Z$. This adjusts for the effect of confounding on our treatment. The resulting distribution for the above graph, which we denote $p(z, y\cmid \text{\it do}(x))$, can be seen as the conditional distribution of $Y,Z$ given $X$ where we have made $Z$ and $X$ independent, but not affected the conditional distribution of $Y$ given $Z,X$. This distribution can then be used to understand the \emph{causal} relationship between $X$ and $Y$. \newline \\ We want to ensure the causal effect adjusted for confounders $Z$ is identifiable for the graph in \Cref{fig:do_null}. We revisit the \textit{backdoor criterion} presented by \citet{pearl_2009}.

\begin{Definition}
%
A set of variables ${Z}$ satisfies the back-door criterion relative to an ordered pair of variables $(A,B)$ 
in a DAG $G$ if:
\begin{itemize}
    \item no node in $Z$ is a descendant of 
    $A$; and
    \item $Z$ blocks every path between $A$ and $B$
    that contains an arrow into $A$.
\end{itemize}
Similarly, if $X$ and $Y$ are two disjoint subsets of nodes in $G$, then $Z$ is said to satisfy the back-door criterion relative to $(X, Y)$ if it satisfies the criterion relative to any pair $(A,B)$ such that $A \in X$ and $B \in Y$.
\end{Definition}
We see that our confounders $Z$ satisfy the backdoor criterion. By Theorem 3.3.2 in \cite{pearl_2009} the causal effect is identifiable as
\begin{eqnarray*}
p^{*}(y|x)=p(y | do(x))  &=& \int p(y|\text{\it do}(x),z)p(z\cmid \text{\it do}(x)) dz\\&=&\int p(y|\underbrace{x,z}_{\mathclap{\text{Only causal association between $x$ and $y$}}})p(z\cmid \text{\it do}(x))dz \\ &=&\int p(y|x,z)p(\underbrace{z}_{\mathclap{\text{Dependency between $x$ and $z$ gone}}})dz.
\end{eqnarray*}
Given some arbitrary density $p^*(x)$, this defines a joint density $p^{*}(z,x,y)=p(z)p^*(x)p(y|x,z)$. \newline \\ \emph{The core idea} \quad is to calculate the HSIC between $X$ and $Y$ under the interventional regime, a distribution we denote by $p^*$.
 and conduct a permutation test to establish whether $p^*(y) = p^*(y\cmid x)$. Since we consider the interventional distribution $p^*(y\cmid x)$, care must be taken to obtain the correct mean embedding. It turns out that one can use importance weights to express the mean embedding of $p^*(y\cmid x)$, where the importance weights are defined as
\begin{equation}
    w(x,z) = \frac{p^*(x)}{p(x\cmid z)},
\end{equation}
which defines a \emph{density ratio}. Note that the dependence upon $z$ complicates the necessary permutation procedure. Our method then comprises of two steps:
\begin{enumerate}
    \item Estimate the importance weights $w(x,z)$.
    \item Run a permutation test and calculate a p-value to establish whether $p^*(y) = p^*(y\cmid x)$.
\end{enumerate}
We present our proposed method in detail for step 2 in the remainder of this section, and step 1 together with the overall procedure in \Cref{method2}.

\subsection{Estimation}
We are now ready to present the first link between HSIC and $p(y\cmid \text{\it do}(x))$, as we are interested in expectations under $p^{*}$ using samples from $p$. 
We define the expectation operator $\mathbb{E}_p$ in the usual way:
\begin{align*}
    \mathbb{E}_p \left[f(X_1, \ldots, X_k)\right] &= \int\cdots\int f(x_1, \ldots, x_k) p(x_1, \ldots, x_k) \, dx_1 \ldots dx_k.
\end{align*}

We would also like to calculate the expectation under the interventional distribution $p^*$, using samples from $p$. This can be done by using a weight function $w(z,x) = p^*(x)/p(x \cmid z)$ so that 
\begin{align*}
    \mathbb{E}_{p^*} [f(Z, X, Y)] &= \mathbb{E}_{p} [w(X,Z) \cdot f(Z, X, Y)].
\end{align*}

\begin{Proposition}
\label{prop_1}
Consider continuous and bounded real-valued functions $f,g$. The covariance between $f(X)$ and $g(Y)$ under $p^*$ can be calculated as
\begin{eqnarray*}
\operatorname{Cov}_{p^*}\left[f(X),g(Y)\right] & = & \mathbb{E}_{p^*}[f(X)g(Y)]-\mathbb{E}_{p^*}[f(X)] \, \mathbb{E}_{p^*}[g(Y)]\\
 & = & \mathbb{E}_{p}[Wf(X)g(Y)]-\mathbb{E}_{p^*}[f(X)] \, \mathbb{E}_{p}[Wg(Y)],
\end{eqnarray*}
where $W = w(X,Z)$, provided that $p(X|Z)> 0, \forall X,Z$ s.t.~$p^*(X) > 0$ and the integrals exist. Using these weights we can now calculate any expectation term under $p^*$ in the covariance estimator.
\end{Proposition}

\begin{Proof}
We show that $\mathbb{E}_{p^*}[g(Y)]$ and $\mathbb{E}_{p^*}[f(X)g(Y)]$ indeed can be calculated as $\mathbb{E}_{p}[Wg(Y)]$ and $\mathbb{E}_{p}[Wf(X)g(Y)]$ respectively. To see this, we have that
\begin{eqnarray*}
\mathbb{E}_{p}[Wf(X)g(Y)] & = & \iiint f(x)g(y)\frac{p^*(x)}{p(x|z)}p(x,y,z) \, dx \, dy\,dz\\
& = & \iiint f(x)g(y) p(z) p^*(x) p(y \cmid z,x) \,dz \, dx \, dy\\
& = & \iiint f(x)g(y) p(z) p^*(z,x,y) \,dz \, dx \, dy\\
 & = & \mathbb{E}_{p^*}[f(X)g(Y)]. 
\end{eqnarray*}
The case for $f(x)=1$ then also follows.
\end{Proof}

In the above example, we will need to estimate importance weights $w_{i}$ from observations $x_{i}\in \mathcal X$ and $z_{i}\in \mathcal Z$ such that $w_{i}=p^*(x_{i})/p(x_{i}|z_{i})$. Under the do-null, we have that $\mathbb{E}_p[W f(X) g(Y)]=$ $\mathbb{E}_{p^*}[f(X)] \mathbb{E}_p[W g(Y)]$.


\subsubsection{Choosing the marginal distribution of $X$}

Calculating the covariance between two arbitrary functions can generally be challenging and may require parametric assumptions. This is obviously undesirable, so similarly to HSIC, we will consider cross-covariance operators, which represent the covariances between two functions of the variables. The key object of interest is the cross-covariance operator of treatments $X$ and outcomes $Y$ under $p^*$ in the interventional regime, which we denote by $C_{p^*}$. This operator plays an analogous role to that of the cross-covariance under an observational distribution $p$ in standard independence testing. In particular, the squared HS norm of $C_{p^*}$ is the population HSIC under $p^*$ and hence the size of this operator measures departure from the do-null hypothesis. Similarly to HSIC, whenever we use characteristic kernels, we have that $\Vert C_{p^*} \Vert^2 =0 \iff p^*(y)= p(y\cmid \text{\it do}(x))$. Of course, we are unable to estimate this quantity directly since we do not have access to samples from $p^*(y|x)$. The following immediate corollary to Proposition \ref{prop_1} relates $C_{p^*}$ to expectations under $p$.
\begin{Corollary}
The cross-covariance operator $C_{p^*}$ of $X$ and $Y$ under $p^*(y|x)$ satisfies 
\begin{equation*}
   \langle f, C_{p^*} g\rangle = \mathbb{E}_{p}[W_{p^*}f(X)g(Y)]-\mathbb{E}[f(X)]\mathbb{E}_{p}[W_{p^*}g(Y)],\;\forall f\in\mathcal H_\mathcal X, g\in\mathcal H_\mathcal Y.
\end{equation*}
and  $C_{p^*}$ can be expressed as
\begin{equation*}
    C_{p^*}:= \mathbb{E}_{p}[W_{p^*}k(X,\cdot)\otimes l(Y,\cdot)]-\mathbb{E}[k(X_{p^*},\cdot)]\otimes \mathbb{E}_{p}[W_{p^*}l(Y,\cdot)]
\end{equation*}
with $W_{p^*} = \frac{p^*(X)}{p(X \cmid Z)}$.
\end{Corollary}

\noindent Following this corollary, we can empirically estimate $C_{p^*}$ using the following expression:
\begin{equation}
\label{estimator}
    \widehat{C}_{p^*} = \frac{1}{n}\sum_{i=1}^{n}\tilde{w}_{i}k(\cdot,x_{i})\otimes l(\cdot,y_{i})-\left(\frac{1}{n}\sum_{j=1}^{n}k\left(\cdot,x_{j}^{p^*}\right)\right)\otimes\left(\frac{1}{n}\sum_{i=1}^{n}\tilde{w}_{i}l\left(\cdot,y_{i}\right)\right),
\end{equation}
where $\left\{ \left(z_i,x_{i},y_{i}\right)\right\}_{i=1}^n \sim p(z,x,y)$
and $\left\{ x_{j}^{p^*}\right\} \sim p^*$. In the case where both $p^*(x_i)$ and $p(x_i|z_i)$ are known, one would simply use the ``true weights'' $w_i = \frac{p^*(x_i)}{p(x_i|z_i)}$. We can show that the resulting estimator (\ref{estimator}) is consistent.
\begin{Theorem}
\label{bigger_lemma}
Assuming $\text{Var}\left(\frac{p^*(X)}{p(X\mid Z)}\right)<\infty$, and that  $\mathbb{E}_{X,X'}[k(X,X')]<\infty$, $\mathbb{E}_{Y,Y'}[l(Y,Y')]<\infty$, then $\widehat{C}_{p^*}$ using true weights is a consistent estimator of $C_{p^*}$ and satisfies
\begin{equation*}
    \mathbb{E}\left[\Vert C_{p^*} - \widehat{C}_{p^*}\Vert^2_{\mathrm{HS}} \right]= \mathcal{O}\left(\frac{1}{n}\right).
\end{equation*}
\begin{Proof}
See \Cref{lemma_proof}.
\end{Proof}
\end{Theorem}
In practice, however, the true weights are rarely available and they would need to be estimated using density ratio estimation techniques, which we shall discuss in detail in \Cref{method2}. The weights estimation corresponds to estimating a function $h$, s.t.\ $\tilde{w}_{i}=h(x_i,z_i)$. We note that estimating $h$ will need to be performed on \emph{a different data set} than
the one used to estimate the covariance in \eqref{estimator}, to ensure independence between $h$  
and the samples used for testing independence. 
We now give a result regarding the convergence rate when density ratios are estimated, which shows how the convergence rate of the density ratio estimator affects that of the corresponding estimator of the cross-covariance operator.
\begin{Theorem}
\label{big_theorem_consistency}
Take the conditions of Theorem \ref{bigger_lemma}, and assume also that $\hat{h}_n(x,z)$ is a consistent estimator of the density ratio $\frac{p^*(x)}{p(x\cmid z)}$ with uniform convergence rate $\mathcal{O}(\frac{1}{n^\alpha})$ for $\alpha >0$, i.e.~ $$ \lim_{n\to \infty} \sup_{x,z} \left| \hat{h}_n(x,z) - \densratio\right| \propto \mathcal{O}\left(\frac{1}{n^\alpha}\right).$$
Then 
\begin{equation*}
    \mathbb{E}\left[\left\Vert C_{p^*} - \widehat{C}_{p^*}\right\Vert^2_{\mathrm{HS}} \right] = \mathcal{O}\left(\frac{1}{n^{\min(1,\alpha)}}\right).
\end{equation*}
\begin{Proof}
See \Cref{theorem_proof}.
\end{Proof}
\end{Theorem}

\noindent In summary, the convergence rate for the bd-HSIC estimator using estimated weights is at worst the slower rate between the estimator and the uniform convergence rate of the weight estimates. Now that we have established how to estimate the cross-covariance operator $C_{p^*}$, analogously to HSIC, we will use the squared HS norm of $\hat C_{p^*}$ as our test statistic.
\begin{Remark} To the best of our knowledge, no density ratio estimator we consider in this work provides any uniform convergence guarantees. Theorem \ref{big_theorem_consistency} only gives an indicative convergence rate if such density ratio estimation guarantees could be established. In the absence of such bounds, consistency will be hard to guarantee without other strong assumptions. {\color{black}
The result can be further refined to settings where uniform convergence is not assumed for the density ratio estimator, however, we leave this to future work.}
\end{Remark}

\begin{Proposition}
\label{prop:estimator}
Let $\circ$ denote the element-wise matrix product and $\mathbf{K}_{++}$ denote the sum of all elements in the matrix $\mathbf{K}$. The squared HS norm of the estimator in \eqref{estimator} is given by
\begin{equation}
\label{big_equation}
    \Vert\widehat{C}_{p^*}\Vert_{\text{HS}}^2 =\frac{1}{n^2} \tilde{\mathbf{w}}^{\top}\left(\mathbf{K}\circ \mathbf{L}\right)\tilde{\mathbf{w}}+ \frac{1}{n^4} (\mathbf{K}^{X_{p^*},X_{p^*}})_{++}(\mathbf{L}\circ \tilde{\mathbf{W}})_{++} -\frac{2}{n^3}\cdot \tilde{\mathbf{w}}^{\top}\left(\mathbf{K}^{X,X_{p^*}} \mathbf{1}_{n}\circ \mathbf{L}\tilde{\mathbf{w}}\right)
\end{equation}
where $\tilde{\mathbf{w}} = [h(x_1,z_1), \hdots, h(x_n,z_n)]$, $\tilde{\mathbf{W}}=\tilde{\mathbf{w}}^{\top}\tilde{\mathbf{w}}$, $\mathbf{K}= [k(x_i,x_j)]_{i,j = 1}^n$, $\mathbf{L}= [l(y_i,y_j)]_{i,j = 1}^n$, $\mathbf{K}^{X_{p^*},X_{p^*}} = [k(x^{p^*}_i,x^{p^*}_j)]_{i,j=1}^{n}$, $\mathbf{K}^{X,X_{p^*}} = [k(x_i,x^{p^*}_j)]_{i,j=1}^{n}$ and $\mathbf{1}_{n}$ is a vector of ones with length $n$. 
\begin{Proof}
See \Cref{proof_thm1}. 
\end{Proof}
\end{Proposition}

Equation \ref{big_equation} can be viewed as a weighted version of HSIC. We henceforth will refer to this as \emph{bd-HSIC} and use it as our test statistic. We note that a weighted form of HSIC has previously been considered in a different context -- when testing for independence on right-censored data \citep{rindt2020kernel}.

\subsubsection{The choice of the $p^*$-marginal}
\label{why_marginal}

\begin{figure}
    \centering
    \includegraphics[width=\linewidth]{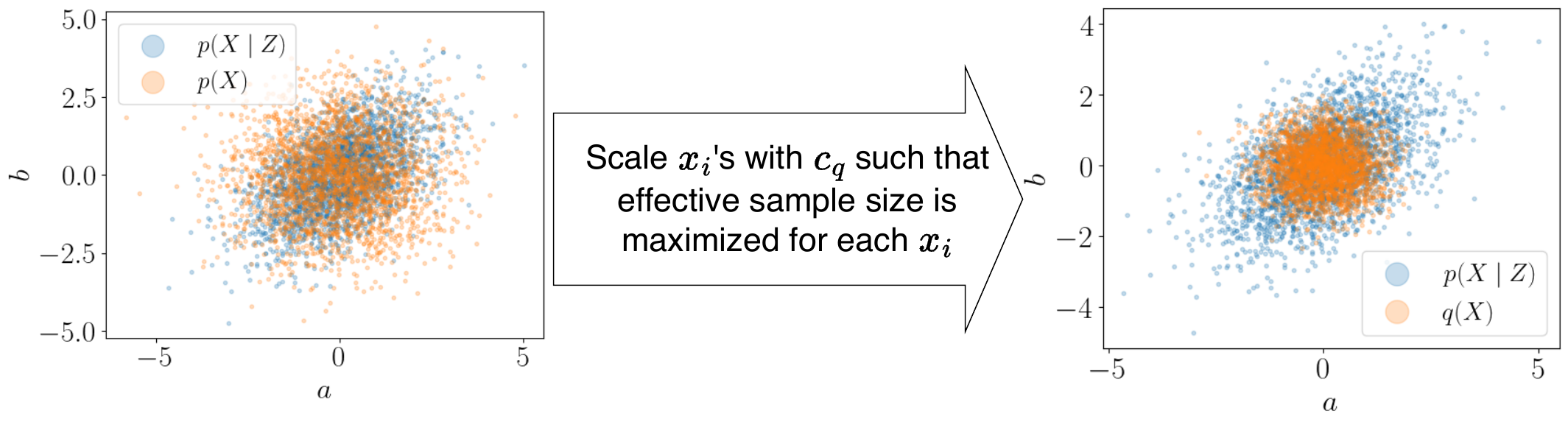}
    \caption{In the left example plot, $p(X)$ has samples where $p(X\cmid Z)$ has little support. By constructing $p^*(x)$, samples from $p^*(x)$ have support where $p(X\cmid Z)$ has support, which yields a higher effective sample size. In this example, we consider $p(X)$ and $p(X\cmid Z)$ a 2d distribution with observations on the form $x = (a,b)$. The axes represent the values for $a$ and $b$.}
    \label{q_dist_mot}
\end{figure}

\textit{Using $p^*(x)$ vs $p(x)$} In general, we would expect $p^*(x)$ to provide a higher effective sample size of $w$'s if chosen appropriately. To maximize effective sample size we can choose $p^*(x)$ such that samples $x_i^{p^*} \sim p^*(x)$ are given by $x_i^{p^*} = c_{p^*} \cdot x_i$. For continuous $x_i$ with mean 0, this can be viewed as scaling the variance of samples $x_i$. We illustrate this in \Cref{q_dist_mot}.
We describe how to choose an optimal $c_{p^*}$ for continuous densities in the paragraphs below. 
\textit{Choice of $c_{p^*}$}
 To find an optimal $c_{p^*}$ for univariate and multivariate $X,Z$, we optimize the effective sample size of $w_i = \frac{p^*(x_i)}{p(x_i\cmid z_i)}$. We choose the effective sample size to be as large as possible to try to maximize the power of the test. Since the marginals of $X,Z$ are generally unknown, we derive a heuristic based on Gaussian distributions.

\begin{Proposition}
Assuming standard normal marginals for univariate $X,Z$ with correlation $\rho$. Then $c_{p^*}$ maximizes effective sample size (ESS)
$$
\text{ESS}:=\frac{\left(\sum_{i} w_{i}\right)^{2}}{\sum_{i} w_{i}^{2}}
$$ 
where $w_{i}$ is the weight used to re-sample observations. The optimal $c_{p^*}$ is then given by $c_{p^*}=\sqrt{1-2\rho^2}$, where $\rho$ is the correlation between $X$ and $Z$.
\end{Proposition}

\begin{Proof}
See \Cref{c_{p^*}_1_proof}.
\end{Proof}
Here, under the assumption of joint normality, the optimal $c_{p^*}$ is found analytically.

\begin{Proposition}
Assuming $\left(X,Z\right) \sim \mathcal{N}_{m+n}(0,\Sigma)$, where we take $\Sigma = 
\begin{bmatrix}
\Sigma_{xx} & \Sigma_{xz}\\
\Sigma_{zx} & \Sigma_{zz}
\end{bmatrix}
$ and assume that $\Sigma_{xx}=I_{m}$ and $\Sigma_{zz}=I_{n}$, the optimal new covariance matrix is found by maximizing the quantity 
\begin{equation*}
    \operatorname{det}(D) \operatorname{det}\left(2 T^{-1}-\left(I_{m}-\Sigma_{x z} \Sigma_{z x}\right)^{-1}-B D^{-1} B^\top\right)
\end{equation*}
with respect to the positive definite matrix $T$, where  $B:=\left(I_{m}-\Sigma_{x z} \Sigma_{z x}\right)^{-1} \Sigma_{x z}$ and $D:=I_{n}-\Sigma_{z x}\left(I_{m}-\Sigma_{x z} \Sigma_{z x}\right)^{-1} \Sigma_{x z}$.
\end{Proposition}

\begin{Proof}
See \Cref{c_{p^*}_2_proof}.
\end{Proof}
In practice in the multivariate case, we choose $T = I_m \cdot c_{p^*}$, and perform the optimization using gradient descent with $c_{p^*} \in \mathbb{R}_{>0}$.

\subsection{The importance of using the correct permutation}
\label{permutation_break}

For HSIC, \cite{https://doi.org/10.1002/sta4.364} shows that permutation tests indeed provide the correct size under the null. However, when permuting under confounders their Theorem 2 does not hold anymore due to the extra dependency on $Z$ between $X$ and $Y$. \newline \\ To see this, consider i.i.d.~samples $(z_i,x_i,y_i)\sim p$ and that $H_0$ is true, i.e.~$p(y\cmid do(x))=p^*(y)$. If you would permute on $x_i$'s or $y_i$'s directly it would violate the exchangeability and hence the guarantee that the test will have the correct size $\alpha$ in expectation, due to the additional dependencies with $z_i$'s. Permuting $y_i$'s directly breaks the required dependency with $z_i$. Further permuting $x_i$'s directly breaks the interventional distribution $do(X=x)$ as it is obtained by re-weighting $x_i$'s according to $\frac{p^*(x_i)}{p(x_i\mid z_i)}$. Thus we have that, for a random permutation $D$, 
$$(\underbrace{D\mathbf{x}}_{{\text{permuting } x_i\text{'s}}},\mathbf{y})\overset{d}{\neq}(\mathbf{x},\mathbf{y})$$ 
and 
$$
 (\mathbf{x},\underbrace{D\mathbf{y}}_{{\text{permuting } y_i \text{'s}}}) \overset{d}{\neq} (\mathbf{x},\mathbf{y}).
$$
Hence the size of the test is not guaranteed to be at most $\alpha$. \newline\\ To circumvent this, we consider a similar approach to \cite{doran2014permutation} who permute only where the covariates $Z$ are similar. As bd-HSIC uses the density ratio as an importance weight, we show that permuting $Y,Z$ against $X$ where $p(x_i\cmid z_i)$ is similar yields an exchangeable sample.

\begin{Theorem}
(Correct type 1 error rate for bd-HSIC using finite permutations) Assume we have i.i.d.\ samples $(z_i,x_i,y_i) \sim p$ and that  $H_0$ is true, i.e.~$p(y\cmid do(x))=p^*(y)$. If a permutation test with finite samples is applied to bd-HSIC to test for causal association at level $\alpha$, this test will reject with probability at most $\alpha$ if only $y_i$'s with the same conditional density $p(x_i|z_i)$ are being permuted.
\end{Theorem}
\begin{Proof}
Let $D'$ be a permutation such that only $y_i$'s with the same conditional density $p(\cdot|z_i)$ are being permuted. As the samples are permuted within $p(\cdot|z_i)$, the following holds
$$(\mathbf{x},\mathbf{y}) \overset{d}{=} (\mathbf{x},\underbrace{D\mathbf{y}}_{{\text{permuting } y_i \text{'s}}})$$ 
for any permutation $D$. The computed test statistics (unpermuted and permuted data) are now exchangeable under the null, and thus the rankings will be uniformly distributed.  
\end{Proof}

It might seem counter-intuitive that the weights $w_i$ from the density ratio estimation cannot be used for the permutation test. We provide a corollary below why using such $w_i$'s would fail. 

\begin{Corollary}
Assuming a non-constant true density ratio weights, permuting ``within'' the density ratio weights $w_i$ yields incorrect permutations.
\end{Corollary}
\begin{Proof}
We provide an example that breaks exchangeability when permuting within density ratio weights. Consider binary $Z$ and $X$ with $p=0.5$ marginally, with $Y$ being a clone of $Z$. Let $p(X = x\mid Z=z)$ be defined as in Table \ref{tab:probs} with $p\neq 0.5$. 
As $Y$ is a clone of $Z$, $p(X=x\mid Y=y)$ has the same probabilities as above. Under these assumptions, it is easy to see that $(x,y)$ samples $(0,0)$ and $(1,1)$ (as well as $(1,0)$ and $(0,1)$) will have to permute since $p(X=0\mid Y=0) = p(X=1\mid Y=1) = p$.  
\begin{table}
\centering
\begin{tabular}{l|cc}
      & $z=0$ & $z=1$  \\ \hline
$x=0$ & $p$   & $1-p$   \\
$x=1$ & $1-p$ & $p$
\end{tabular}
\caption{Probabilities for $p(X = x\mid Z=z)$}
\label{tab:probs}
\end{table}
If one considers a situation in which the same number of pairs ($0,0$ and $1,1$) and ($1,0$ and $0,1$) appear in a sample, the resulting conditional distribution under permutation will be $p(Y=y \mid X=x) = 0.5$ for all $x,y \in \{0,1\}$. Since we took $p\neq 0.5$, the permuted samples are no longer exchangeable with the original sample, even though the do-null clearly holds. 
\end{Proof}

\subsubsection{MMD clustering}
In practice, we are unlikely to get a group of samples with exactly the same density ratio. To approximately permute $y_i$'s within $p(x_i\cmid z_i)$, we propose a maximum mean discrepancy \citep[MMD,][]{10.1007/11564089_7} based k-means clustering method \citep{MacQueen1967}, which uses non-parametric kernel conditional density estimation. The procedure can be summarized in Algorithm \ref{MMD-clustering}.

\begin{algorithm}[!htb]
\SetAlgoLined
\KwInput{Training data $\{x_i\}_{i=1}^{n_{\text{tr}}}$, labels $\{z_i\}_{i=1}^{n_{\text{tr}}}$, test data $\{z_j\}_{j=1}^{n_{\text{test}}}$, regularization parameter $\lambda$, number of clusters $k$}
\KwOutput{Cluster assignments $c_j$ for test data points}

\textbf{Training Phase:}

Estimate the conditional mean embedding: $\hat\mu_{X\mid Z=z}(\cdot) = \sum_{i=1}^{n_{\text{tr}}} w_i(z) l(x_i,\cdot)$\\
Compute $\mathbf{w}(z) = (\mathbf{L}_{Z}+\lambda I)^{-1}\mathbf{l}_{z}$, where $\mathbf{l}_{z}=(l(z_1,z),\hdots,l(z_{n_{\text{tr}}},z))^{\top}$. \newline \\ Run k-means to get cluster centers $C_k$, by optimizing the MMD metric: 
$\sum_i \Vert  \hat\mu_{X\mid Z=z_i}(\cdot) - \hat\mu_{X\mid Z=C_k}(\cdot) \Vert^2 = \sum_{i=1}^{n_{\text{tr}}} (\mathbf{w}(z_i)-\mathbf{w}(C_k))^{\top}\mathbf{L}_X(\mathbf{w}(z_i)-\mathbf{w}(C_k))$.

\textbf{Testing Phase:}
\For{$j = 1$ to $n_{\text{test}}$}{
    $c_j = \min_{k} (\mathbf{w}(z_j)-\mathbf{w}(C_k))^{\top}\mathbf{L}_X(\mathbf{w}(z_j)-\mathbf{w}(C_k))$.
}

\KwReturn{Cluster assignments $c_j$ for test data points}
\caption{Clustering using Conditional Mean Embedding}
\label{MMD-clustering}
\end{algorithm}
One permutes $y_j$'s within assigned clusters to get a $D'$ permutation. To select the optimal number of clusters, we maximize the silhouette score \citep{ROUSSEEUW198753} over a fixed number of clusters.

\begin{Remark}

We note that MMD-clustering procedure uses Conditional Mean Embedding (CME) estimators (since the true $p(x\cmid z)$ is unknown). \cite{li2023optimal} show these estimators to be consistent with rate $\mathcal{O}(\frac{\log n}{n})$, under realistic smoothness assumptions. It is an interesting avenue for further research to investigate if the clustering obtained using estimated CMEs can be related to the clustering obtained using the true CMEs in the large sample limit.

\end{Remark}

\subsection{Consistency of bd-HSIC} \quad So far we have provided results on correct type 1 errors for permutation tests. To prove the asymptotic consistency of bd-HSIC, it suffices to show that Lemma 1 in \cite{https://doi.org/10.1002/sta4.364} holds for bd-HSIC, assuming correctly estimated weights and access to $p(x_i |z_i)$. Consistency then follows from Theorem 3 in \cite{https://doi.org/10.1002/sta4.364}.

\begin{Proposition}
\label{prop_consistency}
 Let $\psi$ be a random permutation of $y_i$'s such that they only are permuted within groups that have the same $p(x_i \cmid z_i)$. We write $\Vert\widehat{C}_{p^*}\Vert_{\text{HS}}^2(\psi)$ to denote the HS-norm of $\widehat{C}_{p^*}$, under the permutation $\psi$ to all $y_i$. Then
$$\Vert\widehat{C}_{p^*}\Vert_{\text{HS}}^2(\psi) \rightarrow 0
$$
in probability.
\end{Proposition}

\begin{Proof}
See Appendix \ref{proof_consistency}.
\end{Proof}

\begin{Remark}

{\color{black}
It should be noted that we have not used samples $x^{p^*} \sim p^*$ in the test statistic for the proposition above. If one were to use $x^{p^*}$ in the test statistic, the proof strategy can be repeated up to the final sum of $A_n + B_n -2C_n$, which instead will be $$\underbrace{\mathbb{E}\left[l(Y,Y')\right] \mathbb{E}\left[k(X,X')\right]}_{A_n} + \underbrace{\mathbb{E}\left[l(Y,Y')\right] \mathbb{E}\left[k(X^{p^*},{X'}^{p^{*}})\right]}_{B_n} - 2\underbrace{\mathbb{E}\left[l(Y,Y')\right] \mathbb{E}\left[k(X,X^{p^*})\right]}_{C_n} $$
which generally will not sum to 0 as $n\to \infty$, implying that we lose consistency. Thus, we use samples $x^{p^*}$ to empirically improve power at the cost of being biased. }

\end{Remark}

It is worth noting that the assumptions here are a couple of steps removed from the practical algorithm -- in particular, they ~require access to true weights and a perfect permutation strategy. However, the results do indicate that we can expect the power of the test to improve with sample size when weights are estimated in a consistent manner and when the clustering approach to permutations consists of clusters with approximately equal conditional densities.

\section{Estimation of Weights}
\label{method2}
Since we will generally never have access to the true weights needed for backdoor adjustment, we need to estimate them from observed data. We denote estimated weights as $\tilde{w}_i$. In this section, we describe how to estimate these $\tilde{w}_i$ for both categorical and continuous $X$.

\subsection{Categorical treatment variable $X$}
For categorical $x_i$ we estimate $w_i$ using the ratio $\frac{p^*(x_i)}{p(x_i\mid z_i)}$, where we take $p^*(x)=p(x)$. We first estimate $p(x)$ by simply taking the empirical probabilities for each category using the training data. For $p(x\cmid z)$, we fit a probabilistic classifier mapping from $z$ to each class of $x$. When $X$ consists of multiple categorical dimensions, i.e. $d_X>1$, we consider the $p^*(x)$ and $p(x\cmid z)$ over the joint space of $\mathcal{X}_{1}\times \hdots \times \mathcal{X}_{d_X}$. In the cases where $d_X$ is large ($\geq8$), we assume each $x^d$ to be independent of each other and take $p(x)=\prod_{d=1}^{d_X}p(x_d)$ and $p(x\cmid z) = \prod_{d=1}^{d_X} p(x_d \cmid z)$.

\subsection{Continuous treatment variable $X$}

In the continuous case, we can no longer estimate $p(x\cmid z)$ or $p^*(x)$ using a classifier straightforwardly. This complicates the estimation of the density ratio $w = \frac{p^*(x)}{p(x \cmid z)}$, as we could either estimate $p^*(x)$ and $p(x\cmid z)$ separately using density estimation or estimate the density ratio directly. We review some existing methods below. \newline \\ \emph{Direct density estimation} of $p^*(x)$ and $p(x\cmid z)$ allows for a broad range of methods such as normalizing flows \citep{pmlr-v37-rezende15}, generative adversarial networks \citep{goodfellow2014generative} and kernel density estimation \citep{10.1214/10-AOS799} among many. While these methods provide accurate density estimation, they tend to be computationally expensive and hard to train \citep{mescheder2018training}. In this paper, we do not explore them further since densities themselves are not of direct interest. \newline \\\emph{RuLSIF} \citep{yamada2011relative} propose using kernel ridge regression to directly estimate the density ratio $r(x)= \frac{p_1(x)}{p_2(x)}$ between distributions $p_1(x)$ and $p_2(x)$. While this method offers an analytical approach to estimating the density ratio, a regression may often not be flexible enough to learn our density ratio of interest in a high dimensional setting. We compare against RuLSIF in the experiment section. \newline \\\emph{Noise contrastive density estimation (NCE)} \citep{JMLR:v13:gutmann12a} considers the problem of estimating an unknown density $p_{\text{true}}(\mathbf{x};\theta)$ with parameters $\theta$ from samples $\mathbf{x}\in \mathbb{R}^d$. The key idea of NCE is to convert a density estimation problem to a classification problem by selecting an auxiliary noise contrastive distribution $p_{\text{noise}}(\mathbf{x})$ to compare with samples from $p_{\text{true}}$. This noise contrastive distribution is used to train a density ratio $\hat{p}(\mathbf{x};\theta')$ with parametrization $\theta'$ of $p_{\text{true}}$ to distinguish between fake samples $\mathbf{z}\sim p_{\text{noise}}$ and observations $\mathbf{x}_i \sim p_{\text{true}}$ through binary classification. We can then derive $p_{\text{true}}$ by using this estimated density ratio. NCE has numerous desirable properties, including consistency under mild assumptions. We will propose and use a slightly modified NCE method to estimate our desired density ratio. We detail this method in the next section. \newline\\ \emph{Telescoping density ratio (TRE)} \citep{rhodes2020telescoping} considers the problem of estimating the density ratio between distributions $p_0$ and $p_m$ using samples $\mathbf{x}_0 \sim p_0$ and $\mathbf{x}_m \sim p_m$. However, these density ratio problems tend to become pathological when $p_0$ and $p_m$ are too far apart, exhibiting a phenomenon coined \emph{density chasm}. The main idea of TRE is then to decompose this density ratio into several sub-tasks through a telescoping product

\begin{equation*}
\frac{p_{0}(\mathbf{x})}{p_{m}(\mathbf{x})}=\frac{p_{0}(\mathbf{x})}{p_{1}(\mathbf{x})} \frac{p_{1}(\mathbf{x})}{p_{2}(\mathbf{x})} \cdots \frac{p_{m-2}(\mathbf{x})}{p_{m-1}(\mathbf{x})} \frac{p_{m-1}(\mathbf{x})}{p_{m}(\mathbf{x})},
\end{equation*}
and estimate each individual ratio with separate estimators $r_{k}\left(\mathbf{x} ; \theta_{k}\right) \approx p_{k}(\mathbf{x}) / p_{k+1}(\mathbf{x})$ for $k=0, \ldots, m-1$; we can then compose the original density ratio as 
\begin{equation*}
r(\mathbf{x} ; \theta)=\prod_{k=0}^{m-1} r_{k}\left(\mathbf{x} ; \theta_{k}\right) \approx \prod_{k=0}^{m-1} \frac{p_{k}(\mathbf{x})}{p_{k+1}(\mathbf{x})}=\frac{p_{0}(\mathbf{x})}{p_{m}(\mathbf{x})}.
\end{equation*}
To train each estimator $r_{k}\left(\mathbf{x} ; \theta_{k}\right)$ we require a gradual transformation of samples between $\mathbf{x}_0$ and $\mathbf{x}_m$, resulting in intermediate samples $\textbf{x}_k \sim p_k,  \quad k=0,\hdots,m-1$. We define these samples as a linear combination of $\textbf{x}_0$ and $\textbf{x}_m$ 

\begin{equation*}
\mathbf{x}_{k}=\sqrt{1-\alpha_{k}^{2}} \mathbf{x}_{0}+\alpha_{k} \mathbf{x}_{m}, \quad k=0, \ldots, m
\end{equation*}
where the $\alpha_{k}$'s form an increasing sequence from 0 to 1. The training objective 
\begin{equation*}
\begin{aligned}
\mathcal{L}_{\mathrm{TRE}}(\theta) &=\frac{1}{m} \sum_{k=0}^{m-1} \mathcal{L}_{k}\left(\theta_{k}\right) \\
\mathcal{L}_{k}\left(\theta_{k}\right) &=-\mathbb{E}_{\mathbf{x}_{k} \sim p_{k}} \log \left(\frac{r_{k}\left(\mathbf{x}_{k} ; \theta_{k}\right)}{1+r_{k}\left(\mathbf{x}_{k} ; \theta_{k}\right)}\right)-\mathbb{E}_{\mathbf{x}_{k+1} \sim p_{k+1}} \log \left(\frac{1}{1+r_{k}\left(\mathbf{x}_{k+1} ; \theta_{k}\right)}\right)
\end{aligned}
\end{equation*}
is the average of all $m$ losses of the subtasks.

\subsection{NCE for bd-HSIC}
\label{density_ratios}
In the do-null context, we have access to samples $\left\{(x_{i},y_{i},z_{i})\right\} _{i=1}^{n}\sim p$. To calculate $\Vert\widehat{C}_{p^*}\Vert^2$ we need to estimate the density ratio $\tilde{w}_i = \frac{p^*(x_i)}{p(x_i|z_i)}$ from our observations. Here $p^*(x)$ is a chosen marginal distribution of $X$. We can express the density ratio as 
$$
w_i=\frac{p^*(x_i)}{p(x_i|z_i)}=\frac{p^*(x_i)p(z_i)}{p(x_i,z_i)}.
$$ 
If we take $p^*(x)=p(x)$, the problem translates into finding the density ratios between the product of the marginals and the joint density. 

We can make use of the NCE framework by taking joint samples $D_1=\{(x_i,z_i)\}_{i=1}^{n_1}$ and approximate samples from the product of the marginals $D_2=\left\{\left(x_{i}, z_{\pi(i)}\right)\right\}_{i=1}^{n_2}\sim p(x)p(z)$, for a randomly drawn permutation $\pi$. We obtain $D_1$ and $D_2$ by splitting the data set to ensure independence between positive samples and negative samples in NCE. \newline \\ By setting $p_{\text{true}}({x}, z)=p(x,z)$ and $p_{\text{noise}}(x,z)=p(x)p(z)$, we can parametrize the noise contrastive classifier as $h(x,z;\theta)= \sigma(\ln{(\frac{1}{\nu}\frac{p(x)p(z)}{p(x,z;\theta)})})=\sigma(\ln{(\frac{1}{\nu}r(x,z;\theta))})$, where $r(x,z;\theta)$ is a classifier parametrized by $\theta$, $\sigma(x)=\frac{1}{1+e^{-x}}$ and $\nu=\frac{n_2}{n_1}$, the ratio between samples from the product of the marginal to the joint distribution. By using NCE to directly estimate the density ratio between the product of the marginals and the joint density we gain the advantage that we do not need to specify an explicit noise contrastive distribution (and potentially introduce bias), as we can already obtain samples approximately through permutation. The problem is then reduced to a classification problem where we have to discriminate between samples from the product of marginals and the joint distribution. We introduce two modifications which take advantage of using the marginal $p^*(x)$. 

\subsubsection{NCE-$p^*$} 
Consider any $p^*(x)\neq p(x)$. We then build a classifier to discriminate data sets $D_{1}=\left\{ \left(x_{i},z_{i}\right)\right\} $
from $D_{2}^{p^*}=\{ (x_{j}^{p^*},z_{j})\} $ where
$\{ x_{j}^{p^*}\} \sim p^*$ independently of $z_{j}$, so
$D_{2}^{p^*}$ contains samples from $p^*(x)\,p(z)$. When we pass any new pair $(x,z)$ (i.e.~regardless where it comes from, and in particular it can come from $p(x,z)$)
to the classifier, it gives us the density ratio $$\frac{p^*(x)p(z)}{p(x,z)}=\frac{p^*(x)}{p(x \cmid z)},$$ which is then parametrized as 
$$h(x,z;\theta)= \sigma\left(\ln{\left\{\frac{1}{\nu}\frac{p^*(x)p(z)}{p(x,z;\theta)}\right\} }\right) = \sigma\left(\ln{\left\{ \frac{1}{\nu}r(x,z;\theta) \right\}}\right).$$ 

\subsubsection{TRE-$p^*$}

We can apply TRE to density ratio estimations between joint samples $\textbf{x}_0 = \left(x , z \right)$ and product of marginals $\textbf{x}_m = \left(X^{p^*} , z \right)$. For our particular context involving a chosen $p^*(x)$, we generate intermediate samples $\mathbf{x}_k = \left(\sqrt{1-\alpha_{k}^{2}} x+\alpha_{k} X^{p^*},z \right)$ by fixing $z$ for $k=0,\hdots, m$.

It should be noted that neither TRE-$p^*$ or NCE-$p^*$ provides any guarantees in regard to uniform convergence rates of consistency.

\subsection{Mixed treatment $X$}

When $X$ contains both continuous and categorical treatments, modifications to the continuous method are needed. We observe that if we take $X = x_{\tcat}\cup x_{\tcont}$ we have

\begin{align}   
    w(x_{\tcat},x_{\tcont})=\frac{p(x_{\tcat},x_{\tcont})}{p(x_{\tcat},x_{\tcont}\cmid z)}&=& \frac{p(x_{\tcat},x_{\tcont})}{p(x_{\tcat}\cmid z,x_{\tcont})p(x_{\tcont}\cmid z)}\\ &= & \underbrace{\frac{p(x_{\tcat} \cmid x_{\tcont})}{p(x_{\tcat}\cmid z,x_{\tcont})}}_{\text{Classifiers}} \cdot \underbrace{\frac{p(x_{\tcont})}{p(x_{\tcont}\cmid z)}}_{\text{NCE}}.
\end{align}
We note that we can decompose the density ratio into a \emph{product} of density ratios estimated using classifiers and density ratio estimation methods. We will use this as the main method for mixed treatment data since this composition allows us to simplify the problem by avoiding estimating density ratios over joint categorical and continuous treatment data which could induce density chasms \citep{rhodes2020telescoping}. We compare our proposed method to dimension-wise \emph{mixing}, proposed in \citet{rhodes2020telescoping}. The same techniques can also be applied to NCE-$p^*$.

\subsection{Algorithmic procedure}
We describe the procedures for estimating weights $\tilde{w}$ and our proposed testing procedure.

\subsubsection{Training the density ratio estimator}

We train our density ratio estimators $r(x,z; \theta)$ through gradient descent. We parameterize all our estimators as Neural Networks (NN), due to their ability to fit almost any function. We summarize the training procedure for categorical data in Algorithm \ref{cat-train} and continuous data in Algorithm \ref{NCE-train}. We take our validation criteria in Algorithm \ref{NCE-train} to be out-of-sample loss.

\begin{algorithm}[!htb]
\SetAlgoLined
\KwInput{Data $\mathcal{D}=\{x_i,z_i,x^{p^*}_i\}_{i=1}^{n//2}$}
\KwOutput{Trained estimator $r(x,z;\theta')$}
Partition data into training and validation $\mathcal{D}=\mathcal{D}_{tr}\cup\mathcal{D}_{val}$\\
Initialise estimator $r(x,z,\theta')$\\ 

\If{$D>8$}{
\For{each categorical $x^d$}{
    Estimate $p(x^d)$ using empirical probabilities\\
    Estimate $p(x^d \cmid z)$ using a classifier with parameters $\theta_d$\\
    Set $r^d(x^d,z;\theta_d) = \frac{p(x^d)}{p(x^d \cmid z; \theta_d)}$
}
Set $r(x,z) = \prod_{d=1}^{n}r^d(x^d,z;\theta_d)$
}
\Else{
    Estimate $p(x)$ over joint space\\
    Estimate $p(x \cmid z)$ using a classifier $\theta$\\
    Set $r(x,z) = \frac{p(x)}{p(x \cmid z; \theta)}$
}

\KwReturn{Trained estimator $r(x,z)$}
\caption{Training a density ratio estimator for categorical $X$}
\label{cat-train}
\end{algorithm}

\begin{algorithm}[!htb]
\SetAlgoLined
\KwInput{Data $\mathcal{D}=\{x_i,z_i,x^{p^*}_i\}_{i=1}^{n//2}$}
\KwOutput{Trained estimator $r(x,z;\theta)$}
Partition data into training and validation $\mathcal{D}=\mathcal{D}_{tr}\cup\mathcal{D}_{val}$\\
Initialize estimator $r(x,z,\theta)$\\ 
Partition data into joint and product of the margin samples $\mathcal{D}_{tr} = \mathcal{D}_{tr}^{\text{pom}} \cup \mathcal{D}_{tr}^{\text{joint}}$, $\mathcal{D}_{val} = \mathcal{D}_{val}^{\text{pom}} \cup \mathcal{D}_{val}^{\text{joint}}$\\
\While{validation criteria $\nu$ not converged}{
Sample positive and negative samples $\delta_{+} \subset \mathcal{D}_{tr}^{\text{pom}},\delta_{-} \subset \mathcal{D}_{tr}^{\text{joint}} $\;
Calculate loss $l = \mathcal{L}(r(\delta_+),r(\delta_-))$\;
Gradient Descent $\theta = \theta + \frac{\partial l}{\partial \theta}$\;
Calculate validation loss $\mathcal{L}(r(\mathcal{D}_{val}^{\text{pom}}),r(\mathcal{D}_{val}^{\text{joint}}))$\;
}
\KwReturn{Trained estimator $r(x,z)$}
\caption{Training an NCE-based density ratio estimator}
\label{NCE-train}
\end{algorithm}

\subsubsection{Testing procedure}

The entire procedure of the test can be summarized in Algorithm \ref{algo}. 

\begin{algorithm}[!htb]
\SetAlgoLined
\KwInput{Data $\left\{ (x_{i},y_{i},z_{i})\right\} _{i=1}^{n}$, Distribution $p^*(x)$, density ratio estimator $r(\cdot)$, number of permutations $n_q$}
\KwOutput{p-value for $H_0$}
Find optimal $c_{p^*}$ for continuous $X$\\
Sample $\{X^{p^*}_i\}^n_{i=1} \sim p^*$\\
Partition data into $\mathcal{D}_1 = \left\{ (x_{i},y_{i},z_{i},X^{p^*}_i)\right\} _{i=1}^{\lfloor{n/2}\rfloor}$ and $\mathcal{D}_2 = \left\{ (x_{i},y_{i},z_{i},X^{p^*}_i)\right\} _{i=\lfloor{n/2}\rfloor+1}^{n}$\\
Train $r$ on $\mathcal{D}_1$\\ 
Estimate $p(x\cmid z)$ on $\mathcal{D}_1$ using modified k-means\\
Get weights $\{\tilde{w}_i\}_{i=\lfloor{n/2}\rfloor+1}^{n} = r(\left\{ (x_{i},z_{i},X^{p^*}_i)\right\} _{i=\lfloor{n/2}\rfloor+1}^{n})$\\
Calculate $\Vert\widehat{C}_{p^*}\Vert^2$ using $\mathcal{D}_2$ \\ 
Calculate permuted test statistics $\{\Vert\widehat{C}_{p^*}\Vert^2_1,\hdots, \Vert\widehat{C}_{p^*}\Vert^2_{n_q}\}$\\
Calculate the p-value as $p = 2\min\left(1-\frac{ 1+  \sum_{i=1}^{n_q} 1_{\Vert\widehat{C}_{p^*}\Vert^2 <  \Vert\widehat{C}_{p^*}\Vert^2_i} }{1+n_q},\frac{ 1+  \sum_{i=1}^{n_q} 1_{\Vert\widehat{C}_{p^*}\Vert^2 <  \Vert\widehat{C}_{p^*}\Vert^2_i}}{1+ n_q}\right)$\\
\KwReturn{$p$}
\caption{Testing $H_0: p(y\cmid \text{\it do}(x)) = p^*(y)$}
\label{algo}
\end{algorithm}
It should be noted that we partition the data such that the data used for the estimation of weights are independent of the data used for the permutation test.

\section{Simulations}
\label{results}

We run experiments using bd-HSIC in the following contexts of do-null testing:
\begin{enumerate}
    \item Linear $X,Y$ dependencies under multiple treatments and treatment types, multiple confounders and multiple outcomes
    \item Non-linear $X,Y$ dependencies under multiple treatments, multiple confounders and multiple outcomes
\end{enumerate}
To extend the exposition of bd-HSIC, we contrast against post-double selection (PDS), which serves as a representative benchmark with pathologies that bd-HSIC attempts to amend. 

\subsection{Comparison against semi-parametric methods}
We compare against the popular post-double selection (PDS) method \citep{10.1093/restud/rdt044}, a semi-parametric lasso-based model that is widely used for causal estimation. PDS considers the following setup

\begin{equation}
\begin{array}{ll}
y_{i}=d_{i} \alpha_{0}+g\left(z_{i}\right)+\zeta_{i}, & \quad\mathbb{E}\left[\zeta_{i} \mid z_{i}, d_{i}\right]=0 \\
d_{i}=m\left(z_{i}\right)+v_{i}, & \quad \mathbb{E}\left[v_{i} \mid z_{i}\right]=0
\end{array}
\end{equation}
where $y_{i}$ is the outcome variable, $d_{i}$ is the policy/treatment variable whose impact $\alpha_{0}$ is the quantity of interest, $z_{i}$ represents confounding factors , and $\zeta_{i}$ and $v_{i}$ are disturbances. The problem is then recast into a linear form

\begin{equation}
\begin{array}{l}
y_{i}=d_{i} \alpha_{0}+\underbrace{x_{i}^{\prime} \beta_{g 0}+r_{g i}}_{g\left(z_{i}\right)}+\zeta_{i}, \\
d_{i}=\underbrace{x_{i}^{\prime} \beta_{m 0}+r_{m i}}_{m\left(z_{i}\right)}+v_{i},
\end{array}
\end{equation}
where $x_{i}^{\prime} \beta_{g 0}$ and $x_{i}^{\prime} \beta_{m 0}$ are approximations to $g(z_{i})$ and $m(z_{i})$, and $r_{g i}$ and $r_{m i}$ are the corresponding approximation errors. PDS then uses lasso to estimate $m(z_i)$ and $g(z_i)$ and for selecting non-zero control variables, and then regresses $y_i$ on $d_i$ together with the union of selected non-zero control variables. Note that PDS is a lasso-based model, which makes it susceptible to non-linear confounding effects and causal effects.  Given the above contexts, the coverage of PDS includes linear $X,Y$ dependencies with multiple treatments under multiple confounders. \newline \\ To make comparisons straightforward, we only compare to PDS in univariate treatment, confounder and outcome cases. 

\subsection{Comparisons against CfME and \cite{10.1093/biomet/asad042}}
We additionally compare against CfME proposed in \cite{10.5555/3546258.3546420} for the binary treatment case and \cite{10.1093/biomet/asad042} for general treatment. While it is not immediately clear how to adapt \cite{10.1093/biomet/asad042} for a hypothesis test, we provide a derivation in Appendix \ref{bd-cme}. We will refer to the newly derived test as \emph{backdoor-CME} (bd-CME). For both these cases, we consider direct permutation on $Y$ following the implementation of CfME. \newline \\ An inherent limitation of CfME is that it only can be used for binary treatment cases. For bd-CME, which is entirely kernel dependent, problems could arise when the data consists of both categorical and continuous variables, as it becomes unclear how to adequately select kernels for both of these data types. One could of course consider a product kernel, but such a kernel may break certain dependencies that render the test invalid.

\subsubsection{Ablation study}
We further compare to $\tilde{w}_i \sim \text{Uniform}(0,1)$. While this choice of weights is not a very principled approach, it serves as a reference to see whether bd-HSIC actually needs correctly estimated weights to have power. 

\subsubsection{Studying the size under $H_0$}
When the null hypothesis is true, we would expect the p-value distribution of the test run on several data sets to have the correct size. Here we use level $\alpha=0.05$.

\subsubsection{Studying the power under $H_1$}
Under the alternative, a desired property is high power across all parameters of the data generation. We will conduct experiments to demonstrate when the test has high power and when it does not. We calculate power for level $\alpha=0.05$. \newline \\ We present our results in bivariate plots, where we plot $\beta_{XY}$ (x-axis) against the power at level $\alpha=0.05$ (y-axis). An ideal plot would be a discontinuous function with height 0.05 at $\beta_{XY}=0$ and height 1.0 for $\beta_{XY}>0$. We would then expect that as the sample size increases, the plot would approach this ideal.

\subsection{Results}
We divide our experiments into two steps: we first take the \emph{true weights} and use them in the subsequent permutation test for bd-HSIC. In this initial step, we chose $p^*=p$. This is intended as a unit test to validate that the parameter selection used for data generation is working. It should be noted that generally the choice of $p^*$ and estimation of $c_{p^*}$ must be done cautiously to avoid double use of data. \newline \\ In the second step we estimate weights from the data and investigate the effectiveness of our proposed density ratio estimation procedure. We consider the following methods for weight estimation: RuLSIF, random uniform, NCE-$p^*$ and TRE-$p^*$ as described in \Cref{density_ratios} and \Cref{method2}.

\subsubsection{Binary treatment $X$}
\begin{figure}[htp]
    \centering
\begin{subfigure}[t]{\linewidth}
\begin{subfigure}[H]{0.04\linewidth}
\hfill
\end{subfigure}
\begin{subfigure}[H]{0.33\linewidth}
\centering
\rotatebox{0}{\scalebox{0.75}{$n=1000$}}
\end{subfigure}%
\begin{subfigure}[H]{0.33\linewidth}
\centering
\rotatebox{0}{\scalebox{0.75}{$n=5000$}}
\end{subfigure}%
\begin{subfigure}[H]{0.33\linewidth}
\centering
\rotatebox{0}{\scalebox{0.75}{$n=10000$}}
\end{subfigure}
\end{subfigure}
\begin{subfigure}{0.33\linewidth}
\includegraphics[width=\linewidth]{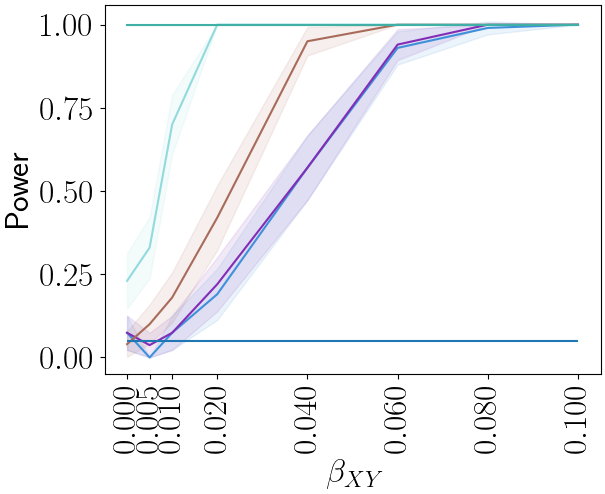}
\centering
\end{subfigure}%
\begin{subfigure}{0.33\linewidth}
\includegraphics[width=\linewidth]{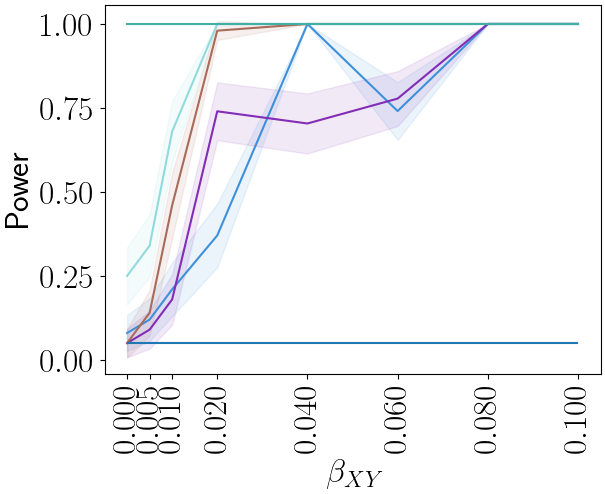}
\centering
\end{subfigure}%
\begin{subfigure}{0.33\linewidth}
\includegraphics[width=\linewidth]{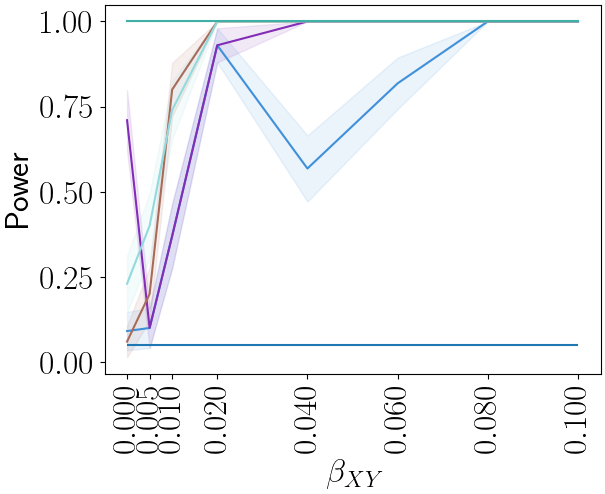}
\centering
\end{subfigure}%

    \caption{bd-HSIC(NCE-$p^*$): {\color{colorA} \sampleline{line width=2pt}} $\quad$ bd-HSIC(true weights): {\color{colorE} \sampleline{line width=2pt}} $\quad$
    PDS: {\color{colorF} \sampleline{line width=2pt}} $\quad$
    CfME: {\color{colorJ} \sampleline{line width=2pt}} $\quad$
    bd-CME: {\color{colorK} \sampleline{line width=2pt}} \newline Binary treatment for $n=1000,5000,10000$ using an RBF kernel in bd-HSIC. Both CfME and bd-CME has incorrect size. The horizontal line denotes when power is 0.05 as a visual reference. }
    \label{plot_1_bin}
\end{figure}

We simulate univariate binary data according to \Cref{bin_generation}. We vary the dependence $\beta_{XY}$ between $X$ and $Y$ to be $\{0.0, 0.005, 0.01,  0.02,0.04,0.06,0.08,0.1\}$. We plot the power for the level $\alpha=0.05$ against 
$\beta_{XY}$ in \Cref{plot_1_bin}. 

\subsubsection{Continuous treatment $X$}
\begin{figure}[htp!]
\begin{subfigure}[t]{\linewidth}
\begin{subfigure}[H]{0.04\linewidth}
\hfill
\end{subfigure}
\begin{subfigure}[H]{0.24\linewidth}
\centering
\rotatebox{0}{\scalebox{0.75}{$d_Z=1$}}
\end{subfigure}%
\begin{subfigure}[H]{0.24\linewidth}
\centering
\rotatebox{0}{\scalebox{0.75}{$d_Z=3$}}
\end{subfigure}%
\begin{subfigure}[H]{0.24\linewidth}
\centering
\rotatebox{0}{\scalebox{0.75}{$d_Z=15$}}
\end{subfigure}%
\begin{subfigure}[H]{0.24\linewidth}
\centering
\rotatebox{0}{\scalebox{0.75}{$d_Z=50$}}
\end{subfigure}
\end{subfigure}
\begin{subfigure}[H]{\linewidth}
\raisebox{1.5cm}{\rotatebox[origin=c]{90}{\scalebox{0.75}{$n=1000$}}}%
\includegraphics[width=0.24\linewidth]{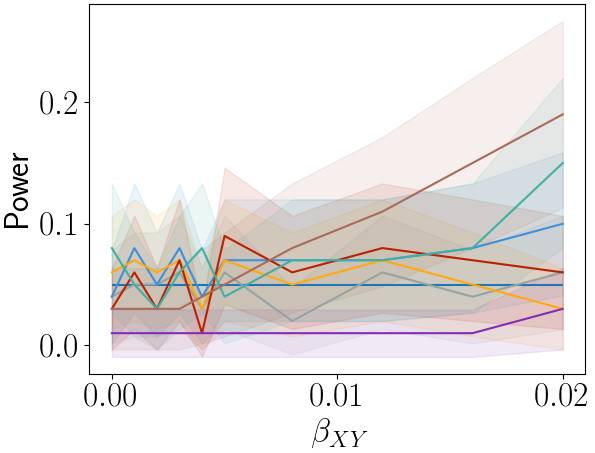}%
\includegraphics[width=0.24\linewidth]{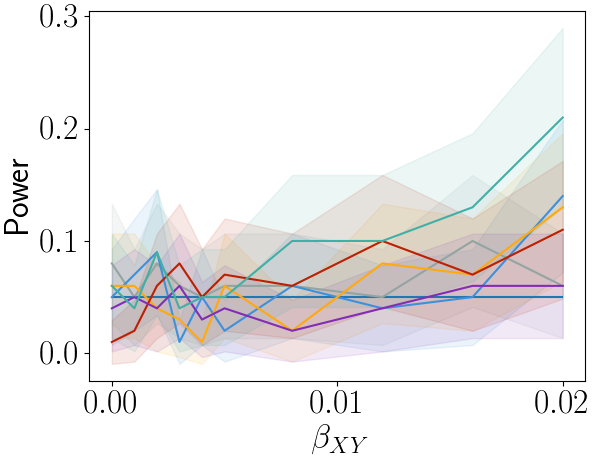}%
\includegraphics[width=0.24\linewidth]{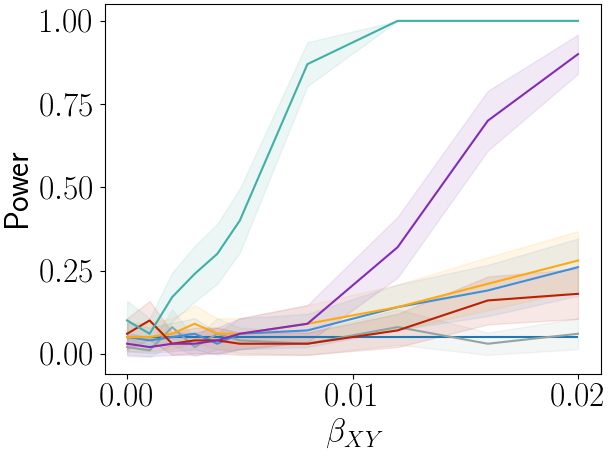}%
\includegraphics[width=0.24\linewidth]{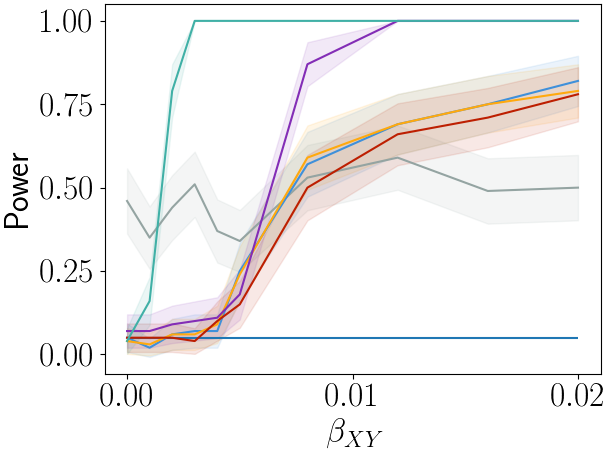}%

\end{subfigure}
\begin{subfigure}[H]{\linewidth}
\raisebox{1.5cm}{\rotatebox[origin=c]{90}{\scalebox{0.75}{$n=5000$}}}%
\includegraphics[width=0.24\linewidth]{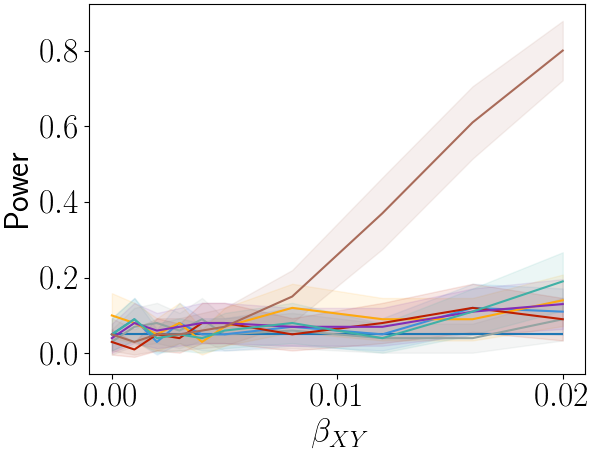}%
\includegraphics[width=0.24\linewidth]{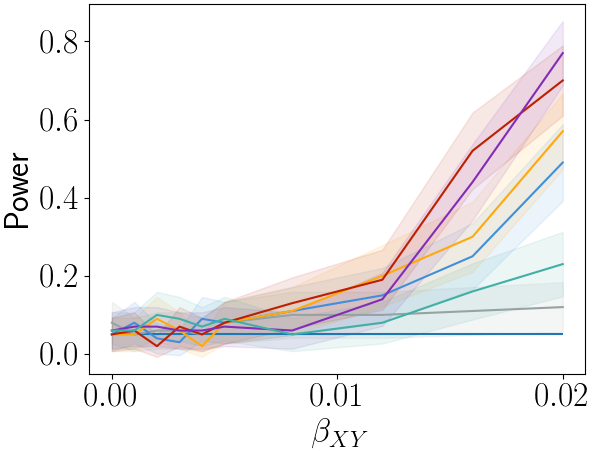}%
\includegraphics[width=0.24\linewidth]{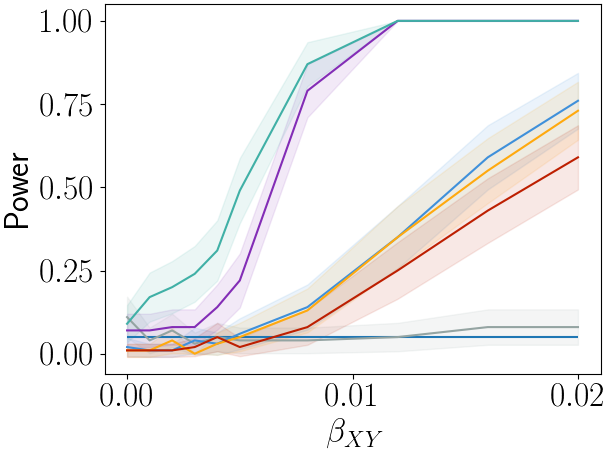}%
\includegraphics[width=0.24\linewidth]{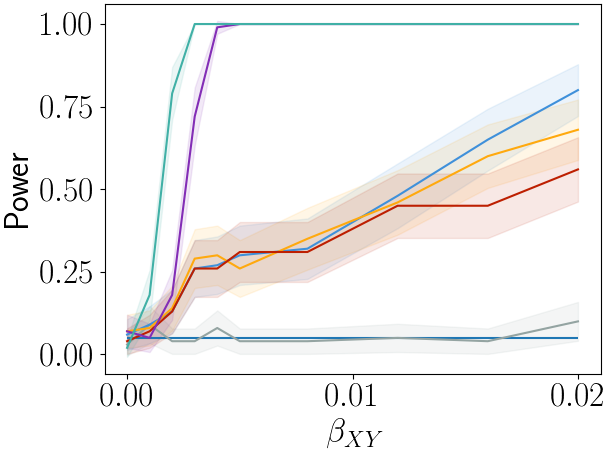}%

\end{subfigure}
\begin{subfigure}[H]{\linewidth}
\raisebox{1.5cm}{\rotatebox[origin=c]{90}{\scalebox{0.75}{$n=10000$}}}%
\includegraphics[width=0.24\linewidth]{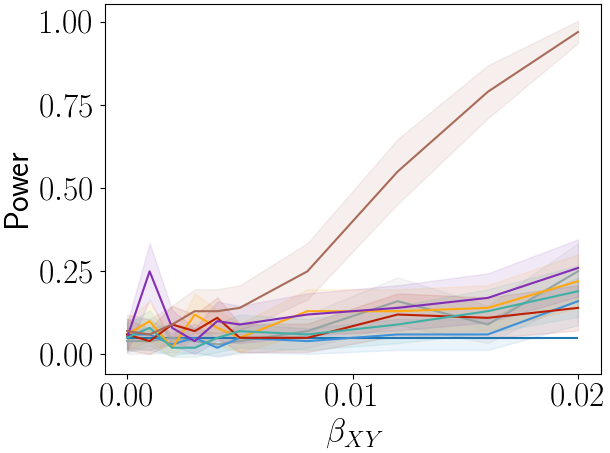}%
\includegraphics[width=0.24\linewidth]{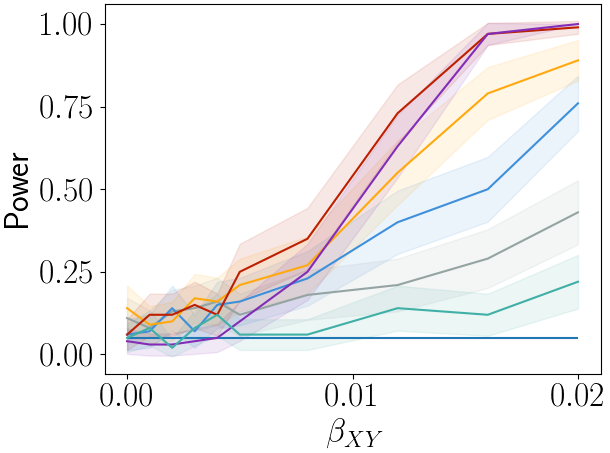}%
\includegraphics[width=0.24\linewidth]{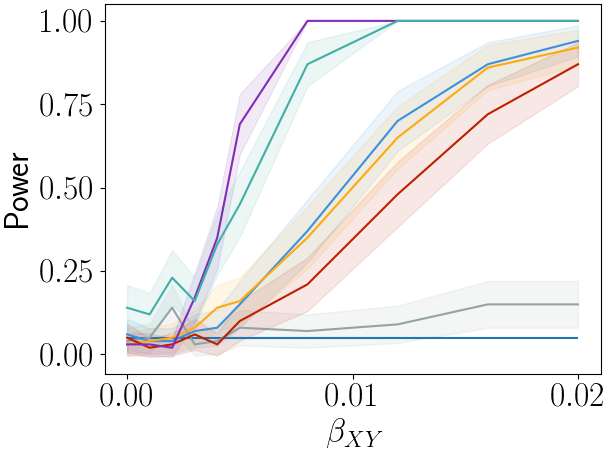}%
\includegraphics[width=0.24\linewidth]{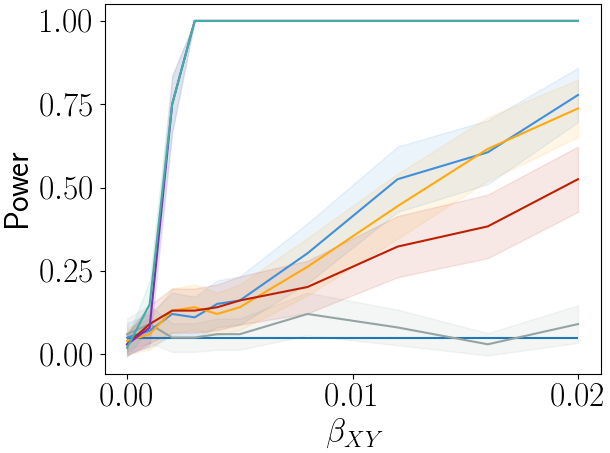}%

\end{subfigure}
\caption{bd-HSIC(NCE-$p^*$): {\color{colorA} \sampleline{line width=2pt}} $\quad$  bd-HSIC(TRE-$p^*$): {\color{colorB} \sampleline{line width=2pt}} $\quad$ bd-HSIC(random uniform): {\color{colorC} \sampleline{line width=2pt}} $\quad$ bd-HSIC(RuLSIF): {\color{colorD} \sampleline{line width=2pt}}$\quad$ bd-HSIC(true weights): {\color{colorE} \sampleline{line width=2pt}} $\quad$ PDS: {\color{colorF} \sampleline{line width=2pt}} $\quad$ bd-CME: {\color{colorK} \sampleline{line width=2pt}}\newline  Continuous treatment results. We find that RuLSIF has incorrect size under the null and that uniform weights has less power. NCE-$p^*$ seems to have the best power while being calibrated under the null. We generally note that random uniform weights have less or no power when compared to ``true weights'' and estimated weights.}
\label{cont_res}
\end{figure}

\emph{Linear dependency between $X$ and $Y$}  \newline The data are simulated for $(d_z,d_x,d_y) \in \{(1,1,1),(3,3,3),(15,3,3),(50,3,3)\}$. In our experiments, we found that a strong confounding effect (i.e.~large $\beta_{XZ}$) led to a smaller effective sample size, making it harder to obtain a consistent test under $H_0$. For $H_1$, the difficulty was mostly controlled by the magnitude of $\beta_{XY}$, where small magnitudes often led to a test with little or no power. We have chosen rejection sampling parameters $\theta,\phi,\beta_{XZ},\beta_{YZ}$ such that the tests are non-trivial but not a failure mode, where $\theta,\phi$ control variance of the marginal distribution of the treatment and the variance of proposal distribution respectively. For exact simulation details, we refer to the appendix. We consider $\beta_{XY}\in [0.0,0.05]$, with $\beta_{XY}=0.0$ corresponding to $H_0$. We present results for continuous treatment in \Cref{cont_res}. We note that bd-CME has inflated type 1 errors for $d_Z=15$.

\begin{figure}
    \centering
    \begin{subfigure}{.33\textwidth}
        \includegraphics[width=\linewidth]{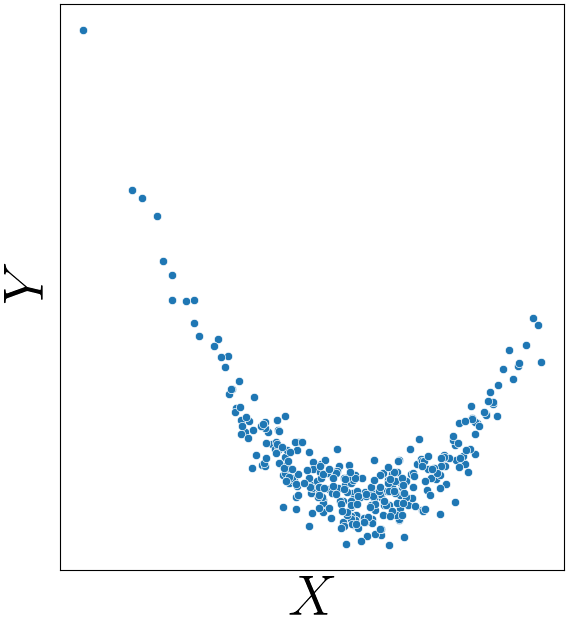}
        \caption{U-shaped dependency, $Y = X^2\beta_{XY}$ \\$\quad$ }
        \label{banana}
    \end{subfigure}%
    \hspace{1in}
    \begin{subfigure}{.33\textwidth}
        \includegraphics[width=\linewidth]{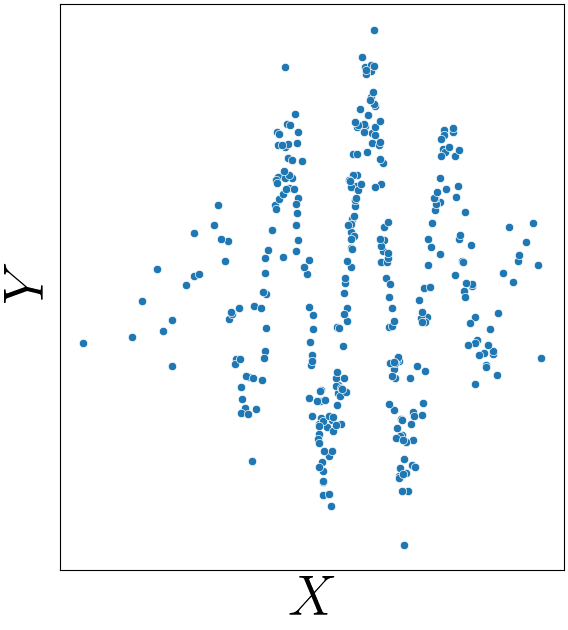}
        \caption{General symmetric non-linear dependency,  $Y = \exp(-0.1X^2)\cos(X\pi)\beta_{XY}$}
                \label{cosine}
    \end{subfigure}
    \caption{Samples from $p^*$ under $H_1$ where the dependency is non-linear. }
    \label{nonlinear_illustration_simulation}
\end{figure}

\begin{figure}[htp!]
\begin{subfigure}[t]{\linewidth}
\begin{subfigure}[H]{0.04\linewidth}
\hfill
\end{subfigure}
\begin{subfigure}[H]{0.33\linewidth}
\centering
\rotatebox{0}{\scalebox{0.75}{$n=1000$}}
\end{subfigure}%
\begin{subfigure}[H]{0.33\linewidth}
\centering
\rotatebox{0}{\scalebox{0.75}{$n=5000$}}
\end{subfigure}%
\begin{subfigure}[H]{0.33\linewidth}
\centering
\rotatebox{0}{\scalebox{0.75}{$n=10000$}}
\end{subfigure}%
\end{subfigure}
\begin{subfigure}[H]{\linewidth}
\raisebox{2.0cm}{\rotatebox[origin=c]{90}{\scalebox{0.75}{$d_Z=1$}}}%
\includegraphics[width=0.33\linewidth]{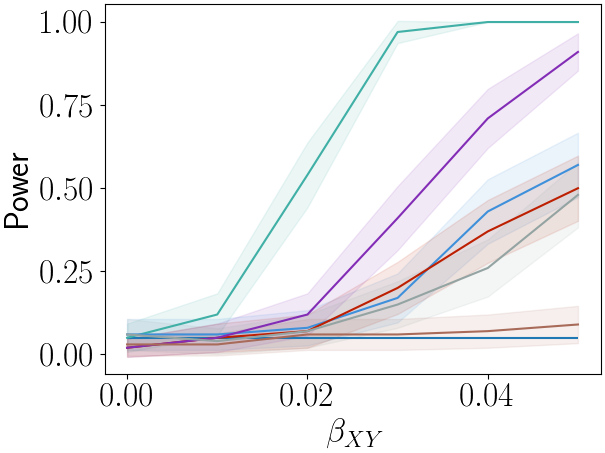}%
\includegraphics[width=0.33\linewidth]{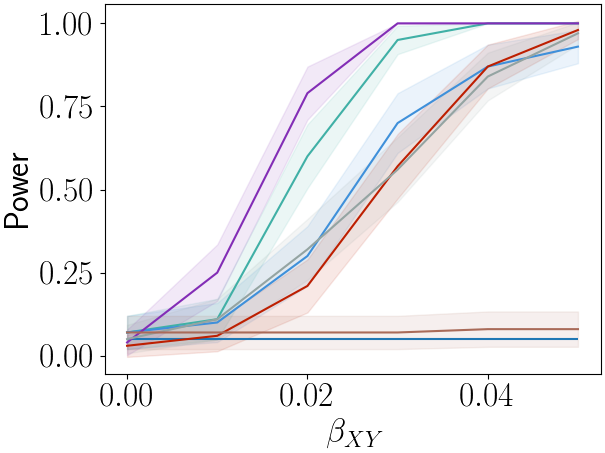}%
\includegraphics[width=0.33\linewidth]{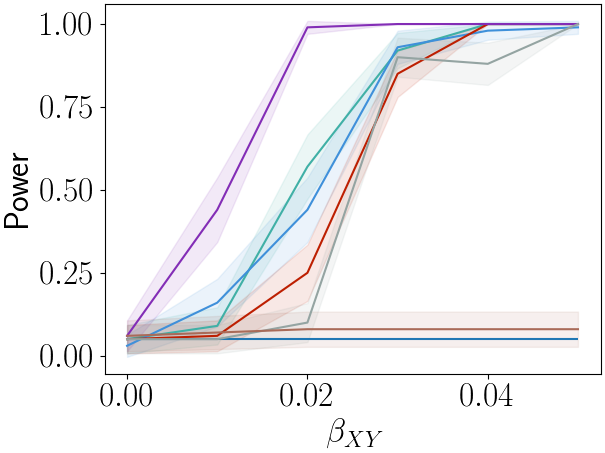}%
\end{subfigure}
\begin{subfigure}[H]{\linewidth}
\raisebox{2.0cm}{\rotatebox[origin=c]{90}{\scalebox{0.75}{$d_Z=50$}}}%
\includegraphics[width=0.33\linewidth]{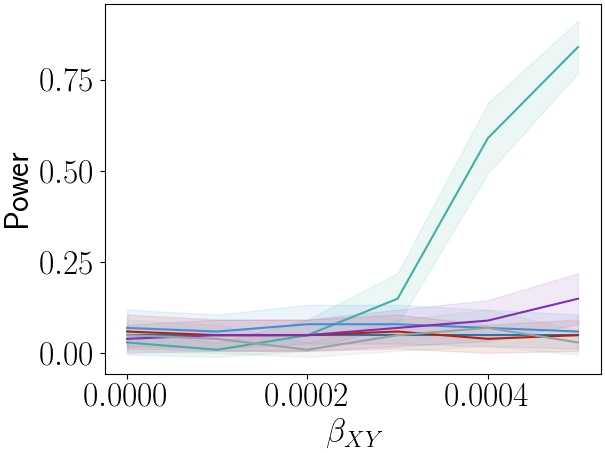}%
\includegraphics[width=0.33\linewidth]{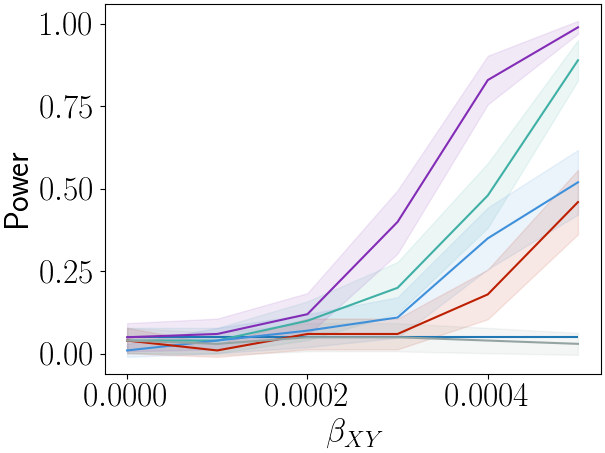}%
\includegraphics[width=0.33\linewidth]{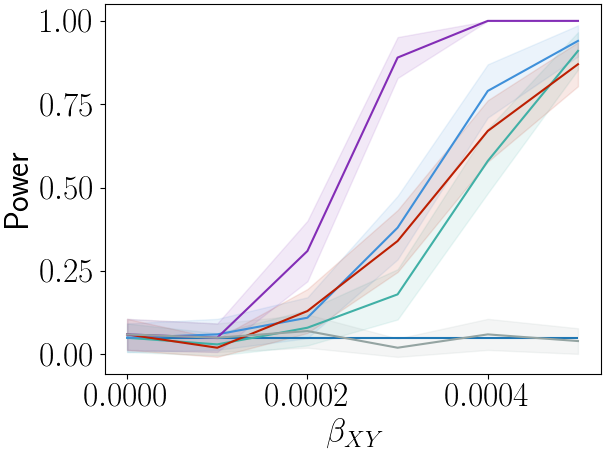}%
\end{subfigure}
\caption{bd-HSIC(NCE-$p^*$): {\color{colorA} \sampleline{line width=2pt}} $\quad$ bd-HSIC(random uniform): {\color{colorC} \sampleline{line width=2pt}} $\quad$bd-HSIC(RuLSIF): {\color{colorD} \sampleline{line width=2pt}} $\quad$bd-HSIC(true weights): {\color{colorE} \sampleline{line width=2pt}} $\quad$PDS: {\color{colorF} \sampleline{line width=2pt}} $\quad$
    bd-CME: {\color{colorK} \sampleline{line width=2pt}} \newline \\ Experiments for U-shaped dependency between $X$ and $Y$. }
\label{non_linear_1}
\end{figure}

\begin{figure}[htp!]
\begin{subfigure}[t]{\linewidth}
\begin{subfigure}[H]{0.04\linewidth}
\hfill
\end{subfigure}
\begin{subfigure}[H]{0.33\linewidth}
\centering
\rotatebox{0}{\scalebox{0.75}{$n=1000$}}
\end{subfigure}%
\begin{subfigure}[H]{0.33\linewidth}
\centering
\rotatebox{0}{\scalebox{0.75}{$n=5000$}}
\end{subfigure}%
\begin{subfigure}[H]{0.33\linewidth}
\centering
\rotatebox{0}{\scalebox{0.75}{$n=10000$}}
\end{subfigure}%
\end{subfigure}
\begin{subfigure}[H]{\linewidth}
\raisebox{2.0cm}{\rotatebox[origin=c]{90}{\scalebox{0.75}{$d_Z=1$}}}%
\includegraphics[width=0.33\linewidth]{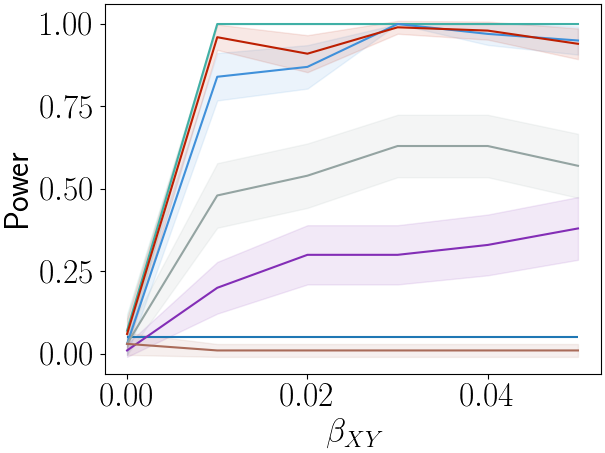}%
\includegraphics[width=0.33\linewidth]{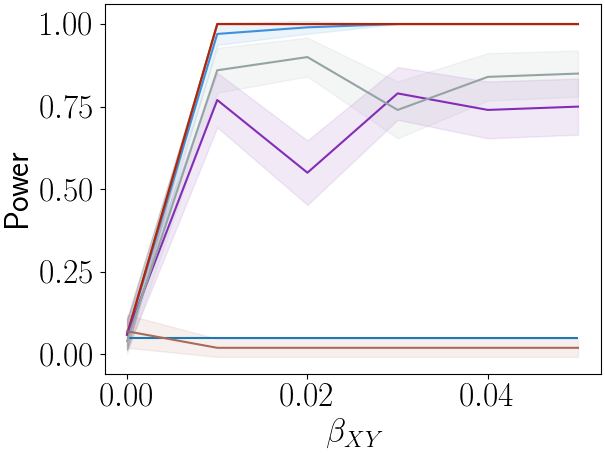}%
\includegraphics[width=0.33\linewidth]{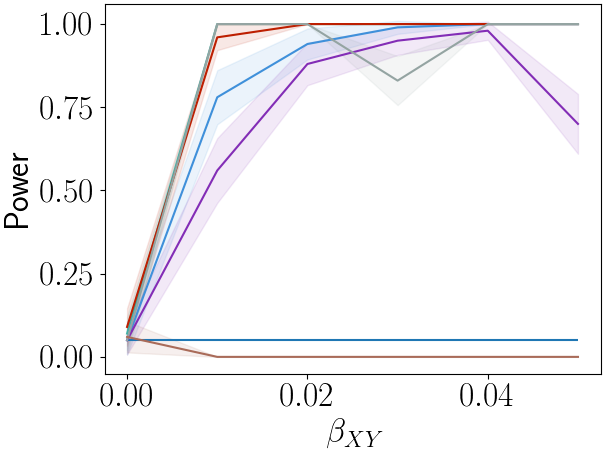}%
\end{subfigure}
\begin{subfigure}[H]{\linewidth}
\raisebox{2.0cm}{\rotatebox[origin=c]{90}{\scalebox{0.75}{$d_Z=50$}}}%
\includegraphics[width=0.33\linewidth]{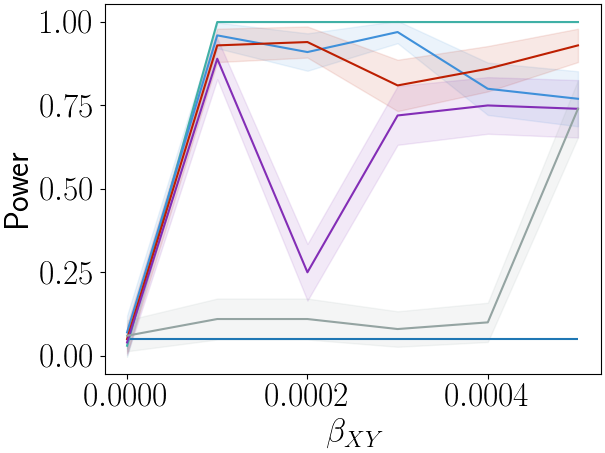}%
\includegraphics[width=0.33\linewidth]{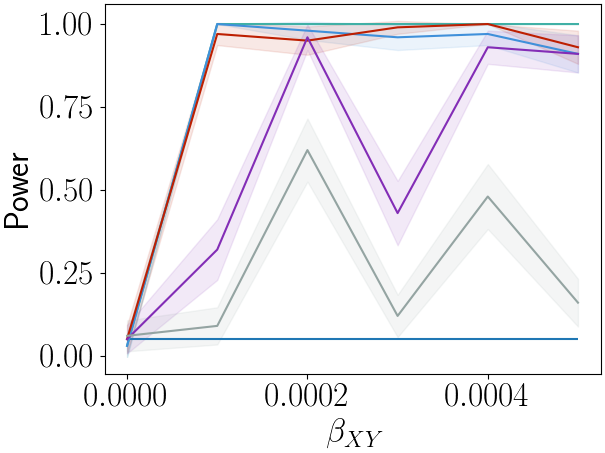}%
\includegraphics[width=0.33\linewidth]{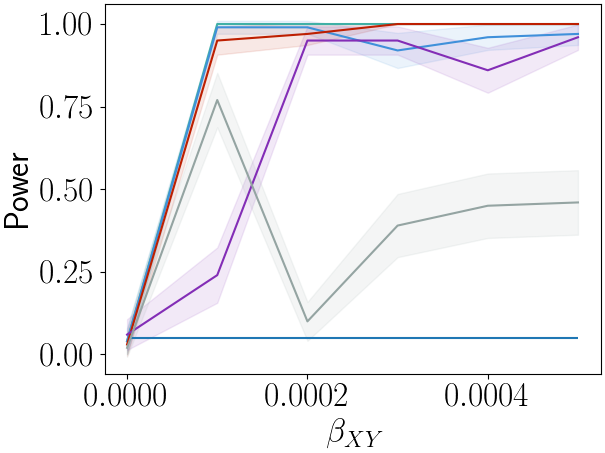}%
\end{subfigure}
\caption{bd-HSIC(NCE-$p^*$): {\color{colorA} \sampleline{line width=2pt}} $\quad$ bd-HSIC(random uniform): {\color{colorC} \sampleline{line width=2pt}} $\quad$ bd-HSIC(RuLSIF): {\color{colorD} \sampleline{line width=2pt}} $\quad$ bd-HSIC(true weights): {\color{colorE} \sampleline{line width=2pt}} $\quad$ PDS: {\color{colorF} \sampleline{line width=2pt}} $\quad$
    bd-CME: {\color{colorK} \sampleline{line width=2pt}} \newline \\ Experiments for general non-linear symmetric dependency between $X$ and $Y$. }
\label{non_linear_2}
\end{figure}

\noindent\emph{Non-linear dependency between $X$ and $Y$} \newline Here we simulate data for $(d_z,d_x,d_y) \in \{(1,1,1),(50,3,3)\}$ using the same parameters as in the linear case. The only difference now is that we consider a non-linear dependency between $X$ and $Y$ illustrated in \Cref{nonlinear_illustration_simulation}. \Cref{banana} illustrates a U-shaped dependency between $X$ and $Y$, which can be found in relationships between happiness vs. age \citep{Kostyshak2017NonParametricTO}, and BMI vs. fragility \citep{Watanabe2020AUR} to name a few. The U-shaped dependency can be generalized to symmetric non-linear relationships between $X$ and $Y$ illustrated in \Cref{cosine}. We show the results in \Cref{non_linear_1} and \Cref{non_linear_2}. We note that PDS has no power against the alternative when the dependency between $X$ and $Y$ is symmetric and non-linear, which is expected due to the linear nature of PDS.

 \subsubsection{Mixed treatment $X$}
The data are simulated for $(d_z,d_x,d_y) \in \{(2,2,2),(4,4,3),(15,6,6),(50,8,8)\}$. We simulate the mixed data according to Algorithm \ref{mixed_generation}. Here we fix half of the $X$'s to be continuous and the other half binary. We consider $\beta_{XY}\in [0.0,0.1]$. We follow the same principles as in the continuous case when selecting $\theta,\phi,\beta_{XZ},\beta_{YZ}$. 

\begin{figure}[htp!]
\begin{subfigure}[t]{\linewidth}
\begin{subfigure}[H]{0.04\linewidth}
\hfill
\end{subfigure}%
\begin{subfigure}[H]{0.32\linewidth}
\centering
\rotatebox{0}{\scalebox{0.75}{$d_Z=2$}}
\end{subfigure}%
\begin{subfigure}[H]{0.32\linewidth}
\centering
\rotatebox{0}{\scalebox{0.75}{$d_Z=15$}}
\end{subfigure}%
\begin{subfigure}[H]{0.32\linewidth}
\centering
\rotatebox{0}{\scalebox{0.75}{$d_Z=50$}}
\end{subfigure}%
\end{subfigure}
\begin{subfigure}[H]{\linewidth}
\raisebox{1.5cm}{\rotatebox[origin=c]{90}{\scalebox{0.75}{$n=1000$}}}%
\includegraphics[width=0.32\linewidth]{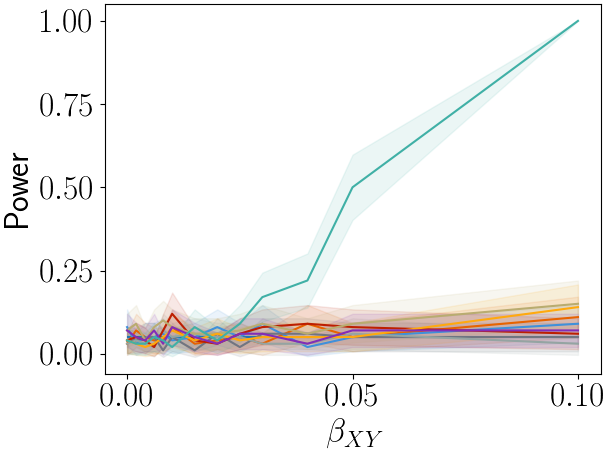}%
\includegraphics[width=0.32\linewidth]{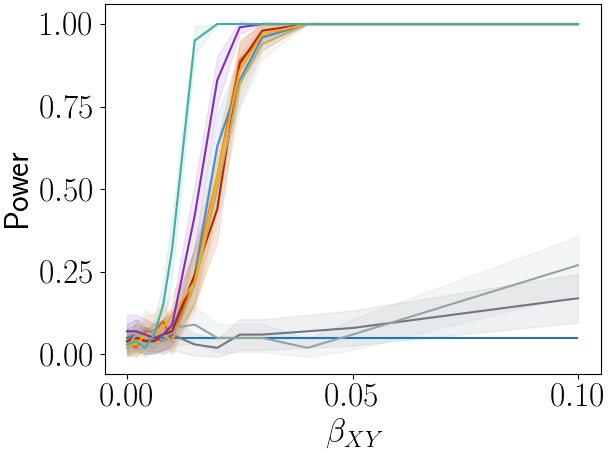}%
\includegraphics[width=0.32\linewidth]{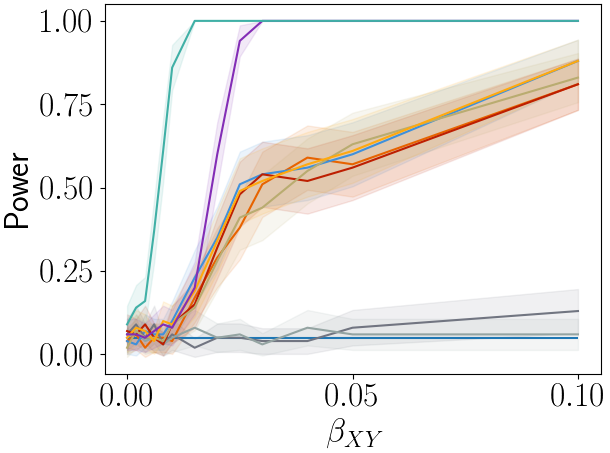}%

\end{subfigure}
\begin{subfigure}[H]{\linewidth}
\raisebox{1.5cm}{\rotatebox[origin=c]{90}{\scalebox{0.75}{$n=5000$}}}%
\includegraphics[width=0.32\linewidth]{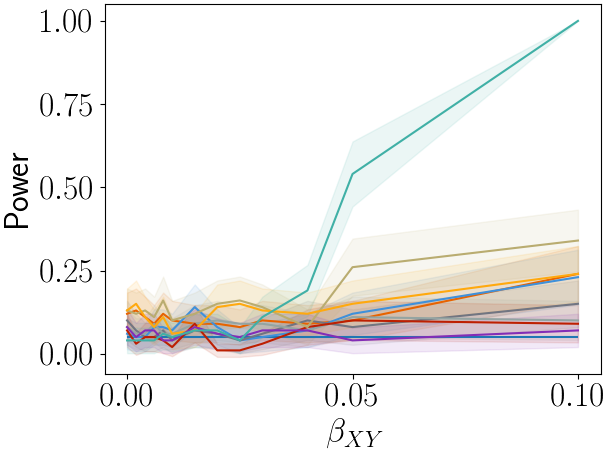}%
\includegraphics[width=0.32\linewidth]{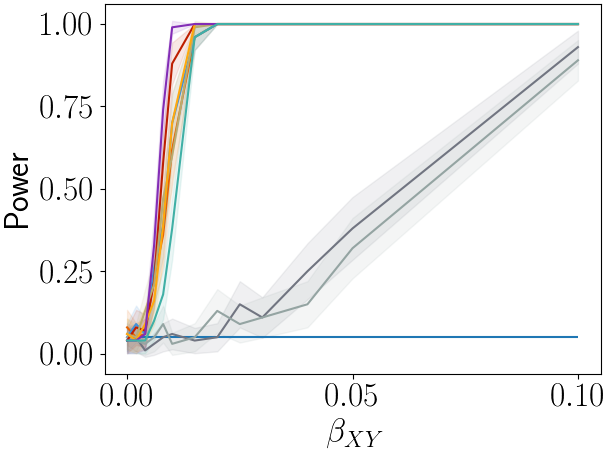}%
\includegraphics[width=0.32\linewidth]{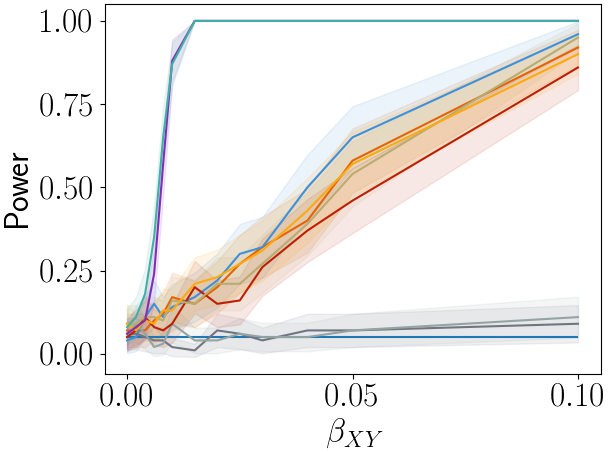}%

\end{subfigure}
\begin{subfigure}[H]{\linewidth}
\raisebox{1.5cm}{\rotatebox[origin=c]{90}{\scalebox{0.75}{$n=10000$}}}%
\includegraphics[width=0.32\linewidth]{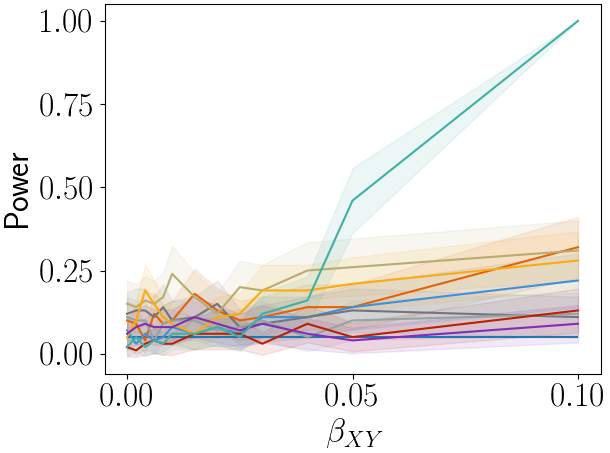}%
\includegraphics[width=0.32\linewidth]{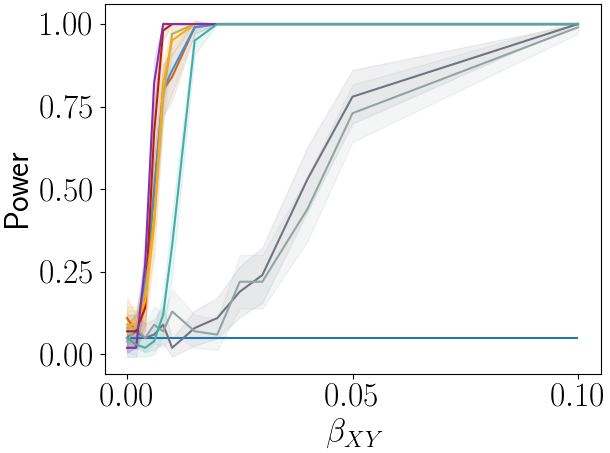}%
\includegraphics[width=0.32\linewidth]{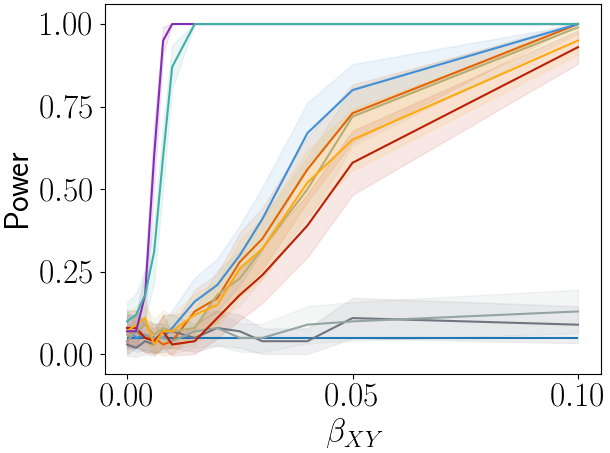}%
\end{subfigure}

\caption{bd-HSIC(NCE-$p^*$, mixing): {\color{colorA} \sampleline{line width=2pt}}  $\quad$ bd-HSIC(TRE-$p^*$, mixing): {\color{colorB} \sampleline{line width=2pt}} $\quad$ bd-HSIC(random uniform): {\color{colorC} \sampleline{line width=2pt}} $\quad$ bd-HSIC(RuLSIF): {\color{colorD} \sampleline{line width=2pt}} $\quad$ bd-HSIC(true weights): {\color{colorE} \sampleline{line width=2pt}} $\quad$ bd-HSIC(NCE-$p^*$ prod): {\color{colorG} \sampleline{line width=2pt}} $\quad$ bd-HSIC(TRE-$p^*$ product): {\color{colorH} \sampleline{line width=2pt}} $\quad$ bd-HSIC(RuLSIF product): {\color{colorI} \sampleline{line width=2pt}} $\quad$
    bd-CME: {\color{colorK} \sampleline{line width=2pt}}\newline Mixed treatment results. We find that RuLSIF has an incorrect size under the null. TRE-$p^*$ prod seems to have the best power while being calibrated under the null.}
\label{mixed_res}
\end{figure}

In our experiments, we compare against RuLSIF and randomly sampled uniform weights. We also compare between estimating the density ratio of the binary treatments separately (denoted with suffix ``prod") and all treatments simultaneously (no suffix). The ``mixed" suffix is a reference to the \textit{dimension-wise mixing} proposed in \cite{rhodes2020telescoping}, which is applied when using TRE-$p^*$. We present the results in \Cref{mixed_res}. We note that bd-CME has an inflated type 1 error rate for $d_Z=50$.

\subsection{Pitfalls}
\label{pitfalls}

\subsubsection{Choice of kernels, a cautionary tale}
\begin{figure}[H]
    \centering
    \begin{subfigure}[t]{\linewidth}
\begin{subfigure}[H]{0.04\linewidth}
\hfill
\end{subfigure}
\begin{subfigure}[H]{0.33\linewidth}
\centering
\rotatebox{0}{\scalebox{0.75}{$n=1000$}}
\end{subfigure}%
\begin{subfigure}[H]{0.33\linewidth}
\centering
\rotatebox{0}{\scalebox{0.75}{$n=5000$}}
\end{subfigure}%
\begin{subfigure}[H]{0.33\linewidth}
\centering
\rotatebox{0}{\scalebox{0.75}{$n=10000$}}
\end{subfigure}
\end{subfigure}
\begin{subfigure}{0.33\linewidth}
\includegraphics[width=\linewidth]{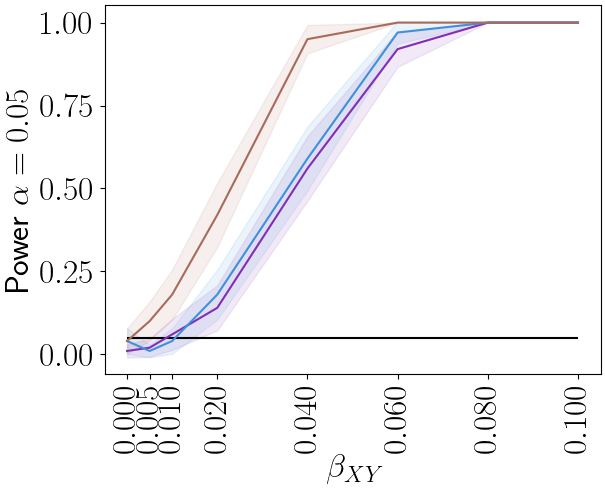}
\centering
\end{subfigure}%
\begin{subfigure}{0.33\linewidth}
\includegraphics[width=\linewidth]{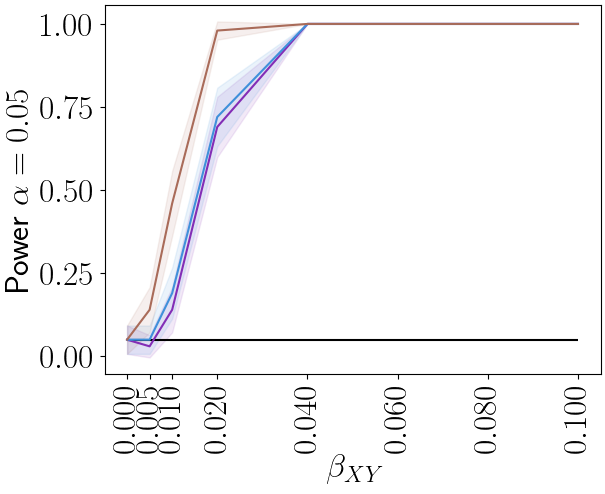}
\centering
\end{subfigure}%
\begin{subfigure}{0.33\linewidth}
\includegraphics[width=\linewidth]{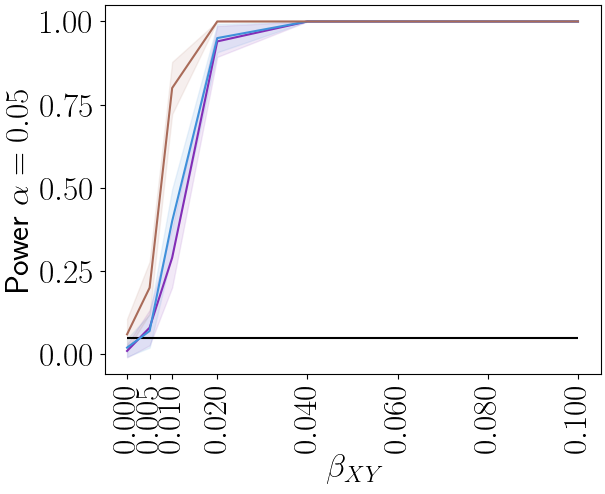}
\centering
\end{subfigure}%

    \caption{bd-HSIC(NCE-$p^*$): {\color{colorA} \sampleline{line width=2pt}} $\quad$bd-HSIC(true weights): {\color{colorE} \sampleline{line width=2pt}} $\quad$PDS: {\color{colorF} \sampleline{line width=2pt}}\newline Binary treatment for $n=1000,5000,10000$ using an linear kernel in bd-HSIC. Compared to \Cref{plot_1_bin}, bd-HSIC now has similar power to PDS.}
    \label{plot_1linear_bin}
\end{figure}

To improve the power of bd-HSIC, we could instead use a linear kernel. \Cref{plot_1linear_bin} illustrates that bd-HSIC has comparable power to PDS, when using a linear kernel for testing the do-null when one considers the univariate binary treatment case.

However, when applying the linear kernel to univariate data for a continuous treatment the test becomes uncalibrated. In fact, the linear kernel makes bd-HSIC much more sensitive to confounding, exhibited in \Cref{linear_break_rip}. 

\subsubsection{When does bd-HSIC break?}
\begin{figure}[!htb]
\centering
\begin{subfigure}{0.3\linewidth}
\includegraphics[width=\linewidth]{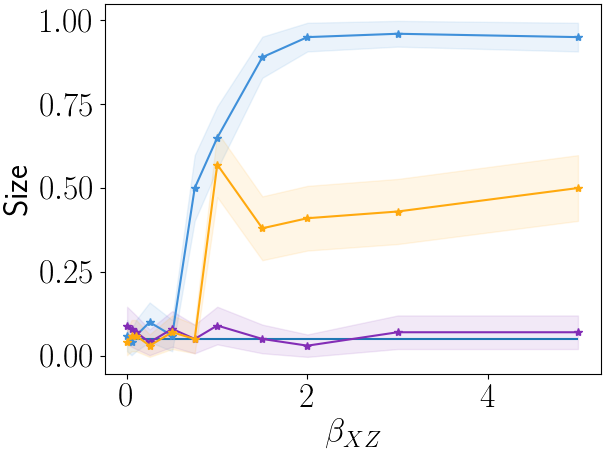}
\caption{Applying a linear kernel to univariate continuous treatment data. The linear kernel may cause the test to have incorrect size.}
\label{linear_break_rip}
\end{subfigure}\hfill
\begin{subfigure}{0.3\linewidth}
\includegraphics[width=\linewidth]{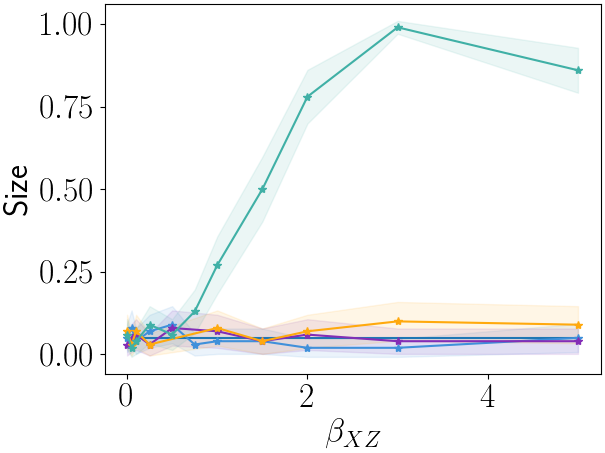}
\caption{Applying a RBF kernel to univariate continuous treatment data.\newline $\quad$ \newline $\quad$}
\end{subfigure}\hfill
\begin{subfigure}{0.3\linewidth}
\centering
\includegraphics[width=\linewidth]{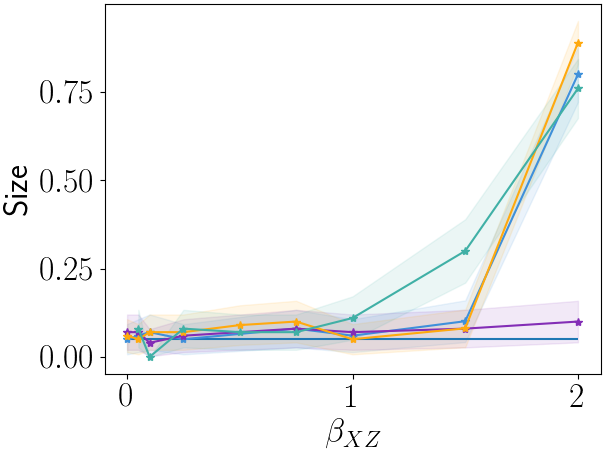}
\caption{bd-HSIC with estimated density ratios exhibits incorrect size for a certain amount of confounding. True weights are consistent.}
\label{bd_hsic_break}
\end{subfigure}
\caption{bd-HSIC(NCE-$p^*$): {\color{colorA} \sampleline{line width=2pt}}  $\quad$bd-HSIC(TRE-$p^*$): {\color{colorB} \sampleline{line width=2pt}} $\quad$bd-HSIC(true weights): {\color{colorE} \sampleline{line width=2pt}} $\quad$ bd-CME: {\color{colorK} \sampleline{line width=2pt}}}
\end{figure}

We demonstrate a typical failure mode of bd-HSIC, when the value of $\beta_{XZ}$ is so strong that the density ratio estimation fails. We show that NCE-$p^*$ and TRE-$p^*$ have incorrect size when $\beta_{XZ}$ becomes large enough in \Cref{bd_hsic_break}. Here we generate data under the null for $\beta_{XZ}\in\{0.0,0.05,0.1,0.15,0.25,0.5,0.75,1.0,1.5,2.0\}$ for $d_X=d_Y=3, d_Z=15$.

\subsection{Experiments on real-world data}

We apply bd-HSIC to two real-world data and compare against PDS and bd-CME.

\subsubsection{Lalonde data set experiments}

\begin{figure}[t]
\centering
\begin{subfigure}{0.33\linewidth}
\raisebox{2cm}{\rotatebox[origin=c]{90}{\scalebox{0.7}{Dummy outcome}}}%
\includegraphics[width=\linewidth]{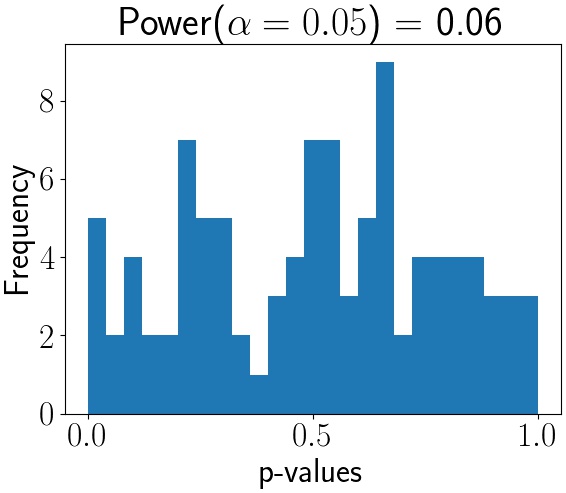}
\raisebox{2cm}{\rotatebox[origin=c]{90}{\scalebox{0.7}{Real earnings}}}%
\includegraphics[width=\linewidth]{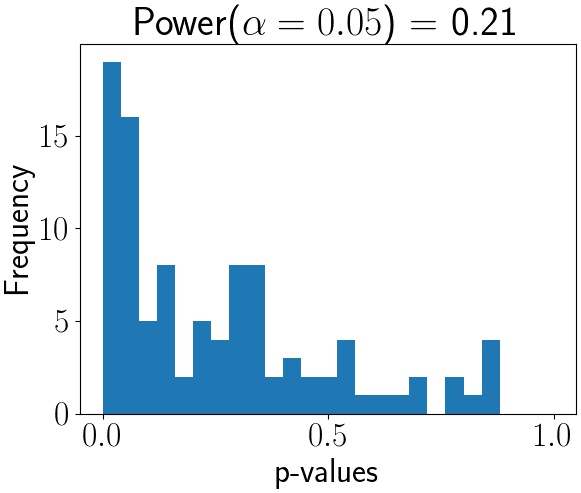}
\caption{\small{PDS }}
\centering
\end{subfigure}%
\begin{subfigure}{0.33\linewidth}
\centering
\includegraphics[width=\linewidth]{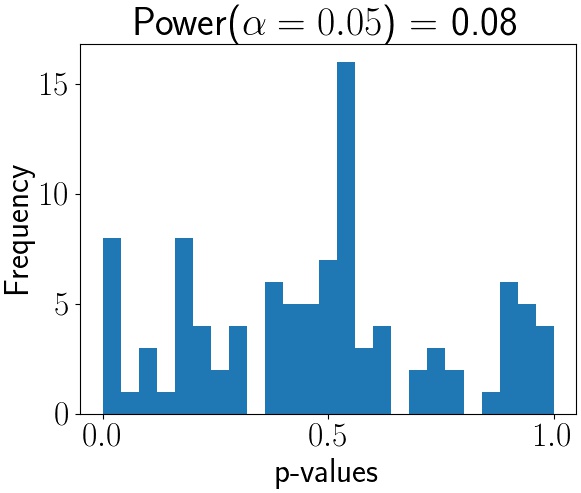}
\includegraphics[width=\linewidth]{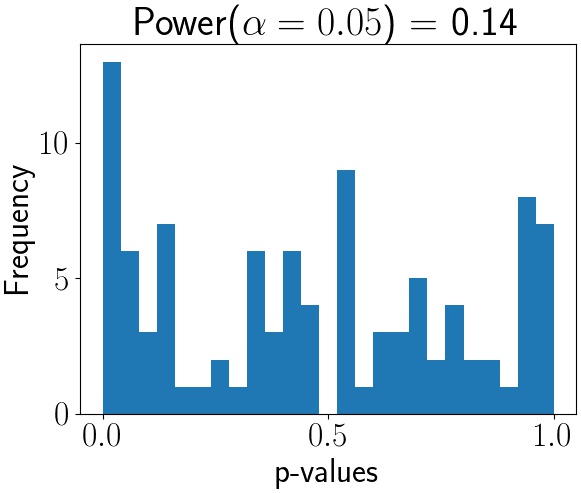}
\caption{bd-HSIC (TRE-$p^*$)}
\end{subfigure}
\begin{subfigure}{0.33\linewidth}
\centering
\includegraphics[width=\linewidth]{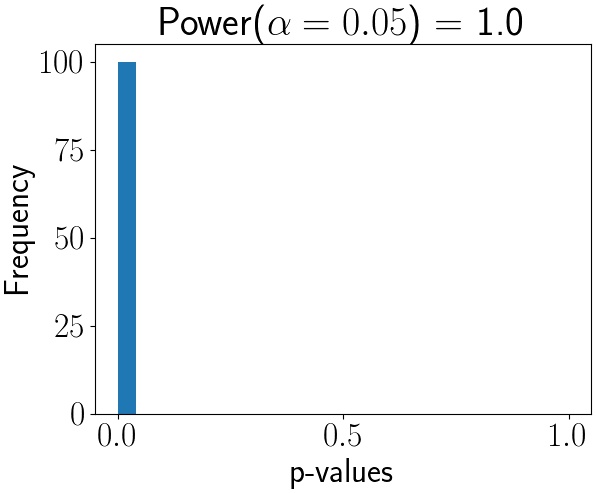}
\includegraphics[width=\linewidth]{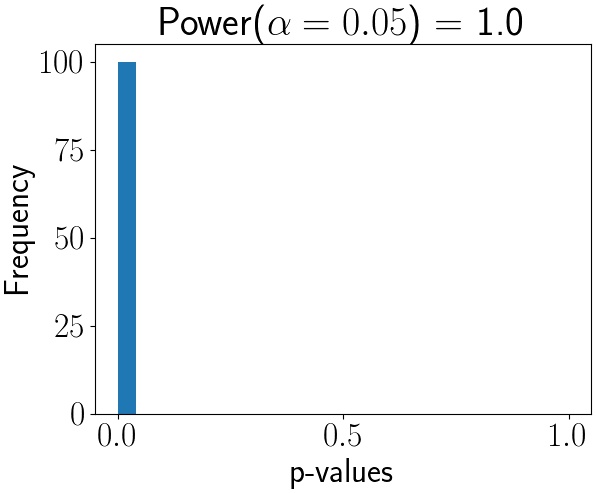}
\caption{bd-CME}
\end{subfigure}
\caption{PDS suggests that there exists a causal association between receiving training and real earnings. bd-CME appears to have incorrect size.}
\label{lalonde_comp}
\end{figure}

The Lalonde data set comes from a study that looked at the effectiveness of a job training program (the treatment) on the real earnings of an individual, a couple of years after completion of the program (the outcome). Each individual has several descriptive covariates such as age, academic background, which confound the relationship between the treatment and the outcome. We compare the power between PDS and bd-HSIC on the Lalonde data set in \Cref{lalonde_comp}. This is done by calculating the p-value on 100 bootstrap sampled subsets of the data set. We further generate a random independent dummy outcome to verify that our tests are calibrated. We note that both PDS and bd-HSIC have the correct type 1 control for dummy outcome, while their power is comparable for the real earnings outcome. 

\subsubsection{Twins data set experiments}
\begin{figure}[t]
\centering
\begin{subfigure}{0.33\linewidth}
\raisebox{2cm}{\rotatebox[origin=c]{90}{\scalebox{0.7}{Dummy outcome}}}%
\includegraphics[width=\linewidth]{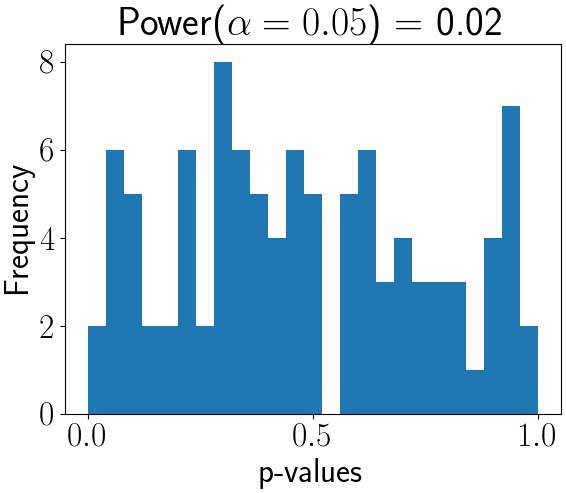}
\raisebox{2cm}{\rotatebox[origin=c]{90}{\scalebox{0.7}{Infant mortality}}}%
\includegraphics[width=\linewidth]{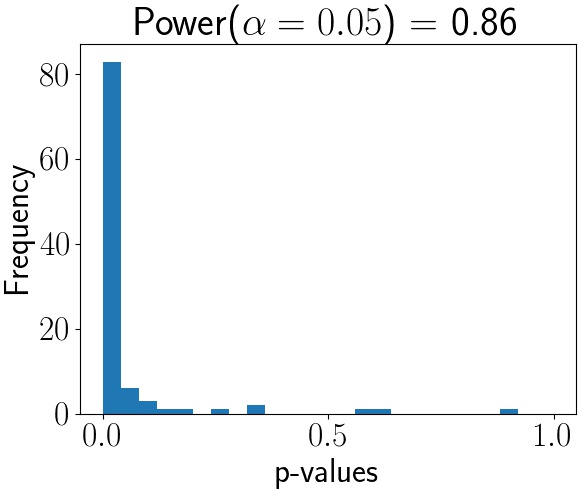}
\caption{\small{PDS }}
\centering
\end{subfigure}%
\begin{subfigure}{0.33\linewidth}
\centering
\includegraphics[width=\linewidth]{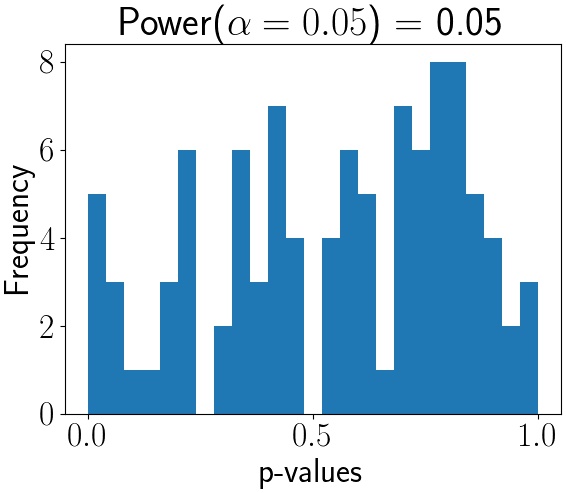}
\includegraphics[width=\linewidth]{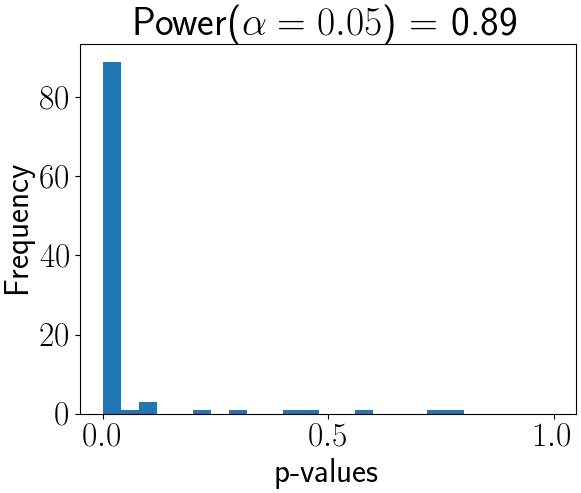}
\caption{bd-HSIC (TRE-$p^*$)}
\end{subfigure}
\begin{subfigure}{0.33\linewidth}
\centering
\includegraphics[width=\linewidth]{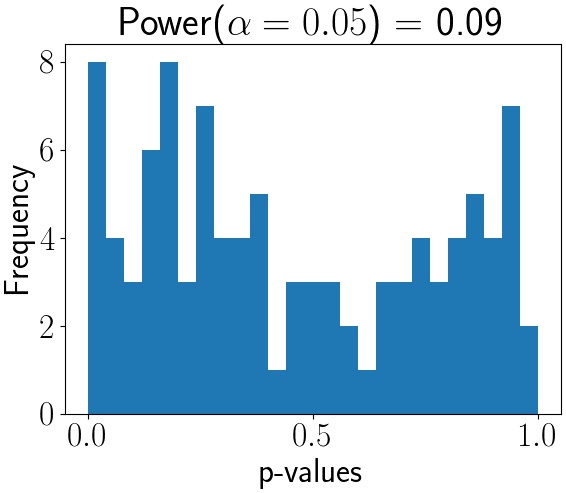}
\includegraphics[width=\linewidth]{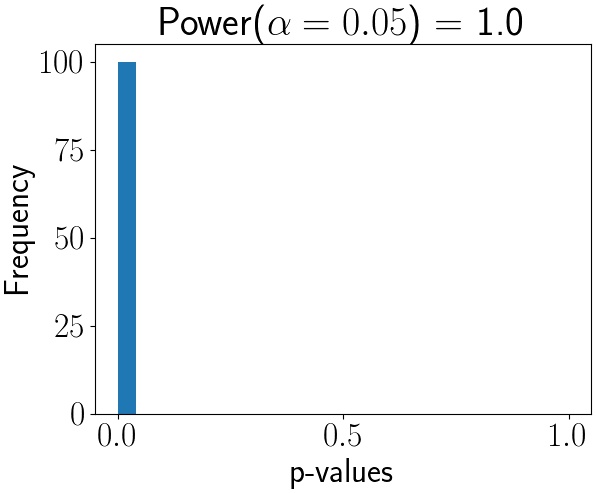}
\caption{bd-CME}
\end{subfigure}
\caption{All three methods suggest that there exists a causal association between weight and mortality.}
\label{twins_comp}
\end{figure}

The twins data set \citep{louizos2017causal} considers data of twin births in the US between 1989--1991. Here the treatment is being born the heavier twin and the outcome is mortality. Besides treatment and outcome, there are also descriptive confounders such as the smoking habits of the parents, education level of the parents, and medical risk factors of the children among many. For our experiments, we construct a slight variation of the experiment presented in \citet{louizos2017causal}, where we instead take the treatment to be $T=\left(\text{Weight}_{\text{heavier twin}}, \text{Weight}_{\text{lighter twin}},  \text{Weight}_{\text{heavier twin}}-\text{Weight}_{\text{lighter twin}}\right)$ and the outcome to be in the set  $\{-1, 0, 1\}$, where $-1$ indicates that the lighter twin died, 1 that the heavier twin died, and 0 that either, neither or both died; everything else is kept the same. Similar to the Lalonde data sets we calculate the p-value on 100 bootstrap sampled subsets of the data set and present results in \Cref{twins_comp}. Similar to the Lalonde data set, we generated a random independent dummy outcome to verify that our tests are calibrated. All three methods suggest there exists a causal association between infant weight and mortality.


\section{Conclusion}
\label{conclusion}
We present a novel non-parametric method termed backdoor-HSIC (bd-HSIC), which is an importance-weighted covariance-based statistic to test the causal null hypothesis, or \textit{do-null}. We first show that our proposed estimator for bd-HSIC is consistent. Experiments on a variety of synthetic data sets, including linear and non-linear dependencies, with different numbers of confounders and treatments, show that bd-HSIC is a flexible method with wider coverage of scenarios than parametric methods such as PDS. Finally, we compare bd-HSIC to PDS on two real-world data sets. Assuming a valid choice of confounders $Z$ satisfying the back door criterion, the test evaluates whether there is any effect on the distribution of the outcome.

A major benefit of bd-HSIC is that can serve as a powerful tool in causal inference as it complements parametric methods such as PDS. For example, PDS generally has better power when the underlying dependency is linear but fails if the dependency is symmetrically non-linear. Even Double Machine Learning, proposed in \cite{10.1111/ectj.12097} and a generalization of PDS, will not straightforwardly capture such non-linear causal dependencies without further assumptions. This is inherent to the design of Double Machine Learning, which is based on semi-parametric partially linear models. bd-HSIC is fully non-parametric and can be used as a general test for non-linear causal association, which is of broad interest to all statistically inclined sciences.  \newline \\ By combining with kernel conditional independence tests, this work could be extended to testing for null \emph{conditional} average treatment effects, as well as to testing more general \emph{nested} independence constraints \citep{richardson2023nested}. 

\clearpage

\acks{The authors would like to sincerely thank the reviewers and Silvia Chiappa for their feedback and suggestions that helped improve the paper. The authors are also thankful to Professor Tom Rainforth (University of Oxford) and Professor Ricardo Silva (University College London) for their suggestions. Robert Hu wrote the majority of the paper during his time as a PhD student at the Department of Statistics at the University of Oxford and was funded by Hennes \& Mauritz AB during his studies. 

} 

\appendix

\section{Remark on $p^*(y)$}
\label{pstar}

Why do we have to consider $H_{0}: p(y|\text{\it do}(x))=p^*(y)$ instead of $H_{0}: p(y|\text{\it do}(x))=p(y)$? We show in the proposition below that we need arbitrary distributions $p^*(y)$ to formulate the hypothesis in a distributional sense since we may have that the do-null hypothesis holds but $p(y\cmid\text{\it do}(x)) \neq p(y)$.
\begin{Proposition}
$p(y|\text{\it do}(x))=p(y) \quad\not\!\!\!\! \impliedby  Y \perp X \cmid \text{\it do}(X=x)$.

\begin{Proof}
We give an example as proof. Consider a backdoor setting. Let $X,Y,Z$ all be binary random variables, and assume that $X,Z$ are Bernoulli distributed with $p=0.5$. Then let's consider the following distribution:
\begin{equation}
    \begin{split}
        p(Y=1 \mid X=1,Z=1)&= p+\varepsilon\\
        p(Y=1 \mid X=0,Z=1)&= p \\
        p(Y=1 \mid X=1,Z=0)&= p - \varepsilon\\
        p(Y=1 \mid X=0,Z=0)&= p
    \end{split}
\end{equation}
Calculating $p(Y=1\mid do(X=x)) = \sum_Z p(Y=1\mid X,Z) p(Z) =  p$ for both $X=0,1$. Now let $p(X=1 \mid Z) = \frac{1}{4} + \frac{Z}{2}$. But then the marginal 
\begin{equation}
    \begin{split}
p(Y=1)& =\sum_{X,Z} p(Y=1 \mid X,Z) p(X\mid Z) p(Z)\\&=0.5(0.75(p+\varepsilon) + 0.25p  + 0.25(p-\varepsilon) + 0.75 p) = p + \frac{\varepsilon}{4}
    \end{split}
\end{equation}
Thus we may have that the do-null hypothesis holds but $p(y\cmid do(x))\neq p(y)$.

\end{Proof}
\end{Proposition}

\section{Consistency proofs}
\subsection{Proof of Theorem \ref{bigger_lemma}}
\label{lemma_proof}

\begin{Proof}
We first define the kernel mean embeddings used in our proposed estimator:
\begin{align*}
\begin{split}
         \mathbb{E}[W_{p^*}k(x,\cdot)l(y,\cdot)]= \mu_{xy}(\cdot) &= \int W_{p^*} k(x,\cdot) l(y,\cdot) \, d\mathbb{P}_{\mathcal{X}\otimes \mathcal{Y} \otimes \mathcal{X}}(x,y,z) \\&= \int \frac{p^*(x)}{p(x\cmid z)} k(x,\cdot) l(y,\cdot) \, d\mathbb{P}_{\mathcal{X}\otimes \mathcal{Y} \otimes \mathcal{X}}(x,y,z)
\end{split}
\end{align*}
\begin{equation*}
\mathbb{E}[k(X^{p^*},\cdot)] = \mu_{X^{p^*}}(\cdot) = \int k(X^{p^*},\cdot) \, d\mathbb{P}^*(x)
\end{equation*}
\begin{equation*}
    \mathbb{E}[W_{p^*}l(y,\cdot)] = \mu_{y}(\cdot) = \int \frac{p^*(x)}{p(x\cmid z)} l(y,\cdot) d\mathbb{P}_{\mathcal{X}\otimes \mathcal{Y} \otimes \mathcal{X}}(x,y,z).
\end{equation*}
Then we write the following for shorthand:
\begin{equation*}
    C_{p^*} = \mu_{xy}(\cdot) - \mu_{X^{p^*}}(\cdot)\otimes \mu_y(\cdot)
\end{equation*}
Thus
\begin{equation*}
    \mathbb{E}\left[\Vert C_{p^*} - \widehat{C}_{p^*}\Vert^2_{\mathrm{HS}} \right]
\end{equation*}
Which is
\begin{equation*}
\begin{split}
        &\mathbb{E}\Bigg[\Bigg\Vert\mu_{xy}(\cdot) - \mu_{X^{p^*}}(\cdot)\otimes\mu_y(\cdot) -\\& \bigg( \underbrace{\frac{1}{n}\sum_{i=1}^{n}\densratiodiscretei k(\cdot,x_{i})\otimes l(\cdot,y_{i})}_{\hat{\mu}_{xy}}-\underbrace{\left(\frac{1}{n}\sum_{j=1}^{n}k\left(\cdot,x_{j}^{p^*}\right)\right)}_{\hat{\mu}_{X^{p^*}}}\otimes \underbrace{\left(\frac{1}{n}\sum_{i=1}^{n}\densratiodiscretei l\left(\cdot,y_{i}\right)\right)}_{\hat{\mu}_y}\bigg) \Bigg\Vert^2 \Bigg].
\end{split}
\end{equation*}
Note that $w_i = \frac{p^*(x_i)}{p(x_i | z_i)}$, and for now we consider we have access to the true weights. Rearranging terms we get
\begin{equation*}
\begin{aligned}
    \mathbb{E}\left[\left\Vert \underbrace{\mu_{xy}(\cdot) - \hat{\mu}_{xy}(\cdot)}_{A} 
    + \underbrace{(\hat{\mu}_{X^{p^*}}(\cdot)\otimes\hat{\mu}_y(\cdot) - \mu_{X^{p^*}}(\cdot)\otimes\mu_y(\cdot))}_{B} 
    \right\Vert^2 \right] &= \\
    \mathbb{E}\left[ \langle A, A \rangle + 2 \langle A, B \rangle + \langle B, B \rangle \right].
\end{aligned}
\end{equation*}
The proof strategy is to obtain convergence rates for each term. First term $\langle A,A \rangle$:

\begin{equation*}
\begin{split}
        \mathbb{E}\left[\langle A,A\rangle \right] &= \mathbb{E}\left[\langle \mu_{xy}(\cdot),\mu_{xy}(\cdot)\rangle \right] -2\mathbb{E}\left[ \langle \mu_{xy}(\cdot),\hat{\mu}_{xy}(\cdot) \rangle \right] + \mathbb{E}\left[\langle \hat{\mu}_{xy}(\cdot),\hat{\mu}_{xy}(\cdot) \rangle\right] 
\end{split}
\end{equation*}
Let $x',y',z'\sim p$ be an independent copy of $x,y,z\sim p$. The first term is then
\begin{equation*}
    \begin{split}
        \mathbb{E}\left[\langle \mu_{xy}(\cdot),\mu_{x'y'}(\cdot)\rangle \right] = \mathbb{E}_{x,y,z,x',y',z'}\left[ \densratio \densratioprime k(x,x')l(y,y') \right],
    \end{split}
\end{equation*}
the second one:
\begin{align*}
        \mathbb{E}\left[\langle \mu_{xy}(\cdot),\hat{\mu}_{xy}(\cdot)\rangle \right]&=\mathbb{E}\left[\frac{1}{n}\sum_{i=1}^{n}\densratiodiscretei\mathbb{E}_{x,y,z}\left[\densratio k(x,x_{i}) l(y,y_{i})\right]  \right] \\
        &= \mathbb{E}_{x',y',z'}\left[\densratioprime\mathbb{E}_{x,y,z}[\densratio k(x,x') l(y,y')]  \right] \\
        &= \mathbb{E}_{x,y,z,x',y',z'}\left[ \densratio \densratioprime k(x,x')l(y,y') \right],
\end{align*}
and the final one:

\begin{equation*}
\begin{aligned}
    \mathbb{E}\left[\frac{1}{n^2} \sum_{i,j} w_i w_j k(x_i,x_j)l(y_i,y_j) \right] &= \\
    \underbrace{\mathbb{E}\left[\frac{1}{n^2} \sum_{i \neq j} w_i w_j k(x_i,x_j)l(y_i,y_j)\right]}_{(a)}
    &+ \underbrace{\mathbb{E}\left[\frac{1}{n^2} \sum_{i} w_i^2 k(x_i,x_i)l(y_i,y_i) \right]}_{(b)}.
\end{aligned}
\end{equation*}
Then $(a)$ is
\begin{equation*}
    (a) = \frac{(n-1)}{n} \mathbb{E}_{x,y,z,x',y',z'}\left[\densratio \densratioprime k(x,x')l(y,y')\right].
\end{equation*}
(b):
\begin{equation*}
\begin{split}
    (b)&=\frac{1}{n}\mathbb{E}\left[\left( \densratio \right)^2 k(x,x)l(y,y)\right] \\&\leq \frac{1}{n} \underbrace{ \sup_{x\in X}k(x,x) \sup_{y\in Y}k(y,y)}_{\text{Assumed to be bounded by } C} \underbrace{\mathbb{E}\left[{\left(\densratio\right)}^2\right]}_{\text{finite variance of density ratio, i.e. bounded by some constant $D$}} \\&=\frac{1}{n} C D
\end{split}
\end{equation*}
So for $\langle A,A \rangle$ the convergence rate is $\mathcal{O}(\frac{1}{n})$. $\langle A,B \rangle$:
\begin{equation*}
\begin{split}
       \langle A,B \rangle= \underbrace{\langle \mu_{xy}(\cdot),\hat{\mu}_{X^{p^*}}(\cdot)\otimes\hat{\mu}_{y}(\cdot)\rangle}_{(1)}+\underbrace{\langle \hat{\mu}_{xy}(\cdot),\mu_{X^{p^*}}(\cdot)\otimes\mu_{y}(\cdot)\rangle}_{(2)}\\-\underbrace{\langle\hat{\mu}_{xy}(\cdot),\hat{\mu}_{X^{p^*}}(\cdot)\otimes\hat{\mu}_{y}(\cdot)\rangle}_{(3)} -\underbrace{\langle\mu_{xy}(\cdot),\mu_{X^{p^*}}(\cdot)\otimes \mu_{y'}(\cdot)\rangle}_{(4)}  
\end{split}
\end{equation*}
Thus
\begin{equation*}
    \begin{split}
        (1)&= \mathbb{E}_{}\left[\frac{1}{n^2}\sum_{i,j} \densratiodiscretej \mathbb{E}_{x,y,z}\left[\densratio k(x,x_i^{p^*}) l(y,y_j)\right] \right]\\&=\mathbb{E}_{x',y',z'}\left[\densratioprime \mathbb{E}_{x,y,z}\left[\densratio k(x,x_i^{p^*}) l(y,y_j)\right] \right]
    \end{split}
\end{equation*}
\begin{equation*}
    \begin{split}
        (2)&= \mathbb{E}_{}\left[ \frac{1}{n}\sum_i^n \densratiodiscretei \mathbb{E}_{x_q}[k(x_i,X^{p^*})] \mathbb{E}_{x,y,z}\left[\densratio l(y,y_i)\right] \right] \\& = \mathbb{E}_{x',y',z'}\left[ \densratioprime \mathbb{E}_{x_q}[k(x',X^{p^*})] \mathbb{E}_{x,y,z}\left[\densratio l(y,y')\right] \right]
    \end{split}
\end{equation*}

\begin{equation*}
\begin{aligned}
    (3) &= \mathbb{E}\left[ \frac{1}{n^3} \sum_{i,k,j} \densratiodiscretei \densratiodiscretej k(x_i, X^{p^*}_k) l(y_i, y_j) \right] \\
    &= \mathbb{E}\left[ \frac{1}{n^3} \sum_{k, i \neq j} \densratiodiscretei \densratiodiscretej k(x_i, X^{p^*}_k) l(y_i, y_j) \right] \\
    &\quad + \mathbb{E}\left[ \frac{1}{n^3} \sum_{k, i = j} \left( \densratiodiscretei \right)^2 k(x_i, X^{p^*}_k) l(y_i, y_i) \right] \\
    &= \frac{n-1}{n} \mathbb{E}_{x', y', z'} \left[ \densratioprime \mathbb{E}_{x, y, z} \left[ \densratio \mathbb{E}_{X^{p^*}} \left[ k(x, X^{p^*}) \right] l(y, y') \right] \right] \\
    &\quad + \frac{1}{n} \mathbb{E}_{X^{p^*}, x, y, z} \left[ \left( \densratio \right)^2 k(x, X^{p^*}) l(y, y) \right].
\end{aligned}
\end{equation*}

\begin{equation*}
    \begin{split}
        (4)&= \mathbb{E}_{x,y,z}\left[ \densratio \mathbb{E}_{x_q}[k(x,X^{p^*})] \mathbb{E}_{x',y',z'}\left[\densratioprime l(y,y')\right] \right]
    \end{split}
\end{equation*}
We can use the same argument as in $\langle A,A \rangle$, consequently $\langle A,B \rangle \propto \mathcal{O}(\frac{1}{n})$ Let ${X^{p^*}}^{'},x',y',z'$ be independent copies of $X^{p^*},x,y,z$. Then
\begin{equation*}
\begin{split}
        \langle B,B \rangle= \underbrace{\langle\hat{\mu}_{X^{p^*}}(\cdot)\otimes\hat{\mu}_y(\cdot),\hat{\mu}_{X^{p^*}}(\cdot)\otimes\hat{\mu}_{y}(\cdot)\rangle}_{(1)}-\underbrace{2\langle\hat{\mu}_{X^{p^*}}(\cdot)\otimes\hat{\mu}_y(\cdot),\mu_{X^{p^*}}(\cdot)\otimes\mu_y(\cdot)\rangle}_{(2)} \\+ \underbrace{\langle\mu_{{X^{p^*}}^{'}}(\cdot)\otimes\mu_{y'}(\cdot),\mu_{X^{p^*}}(\cdot)\otimes\mu_y(\cdot)\rangle}_{(3)}
\end{split}
\end{equation*}
Where
\begin{equation*}
\begin{aligned}
    &(1) \\
    &= \mathbb{E}\left[ \frac{1}{n^2} \sum_{u,v} k(X^{p^*}_u, X^{p^*}_v) \frac{1}{n^2} \sum_{i,j} \densratiodiscretei \densratiodiscretej l(y_i, y_j) \right] \\
    &= \mathbb{E} \Bigg[ \frac{1}{n^2} \left( \sum_{u \neq v} k(X^{p^*}_u, X^{p^*}_v) + \sum_{u = v} k(X^{p^*}_u, X^{p^*}_u) \right) \\
    &\quad \times \frac{1}{n^2} \left( \sum_{i \neq j} \densratiodiscretei \densratiodiscretej l(y_i, y_j) + \sum_{i = j} \left( \densratiodiscretei \right)^2 l(y_i, y_i) \right) \Bigg] \\
    &= \underbrace{\mathbb{E}\left[ \frac{1}{n^4} \sum_{u \neq v} k(X^{p^*}_u, X^{p^*}_v) \sum_{i \neq j} \densratiodiscretei \densratiodiscretej l(y_i, y_j) \right]}_a \\
    &\quad + \underbrace{\mathbb{E}\left[ \frac{1}{n^4} \sum_{u = v} k(X^{p^*}_u, X^{p^*}_u) \sum_{i = j} \left( \densratiodiscretei \right)^2 l(y_i, y_i) \right]}_b \\
    &\quad + \underbrace{\mathbb{E}\left[ \frac{1}{n^4} \sum_{u \neq v} k(X^{p^*}_u, X^{p^*}_v) \sum_{i = j} \left( \densratiodiscretei \right)^2 l(y_i, y_i) \right]}_c \\
    &\quad + \underbrace{\mathbb{E}\left[ \frac{1}{n^4} \sum_{u = v} k(X^{p^*}_u, X^{p^*}_u) \sum_{i \neq j} \densratiodiscretei \densratiodiscretej l(y_i, y_j) \right]}_d.
\end{aligned}
\end{equation*}
Each term is then:

\begin{equation*}
\begin{aligned}
    a &= \mathbb{E}\left[\frac{1}{n^4} \sum_{u \neq v} k(X^{p^*}_u, X^{p^*}_v) \sum_{i \neq j} \densratiodiscretei \densratiodiscretej l(y_i, y_j) \right] \\
    &= \frac{(n-1)^2}{n^2} \mathbb{E}_{X^{p^*}, {X^{p^*}}'} \left[ k(X^{p^*}, {X^{p^*}}') \right] \mathbb{E}_{x, y, z, x', y', z'} \left[ \densratioprime \densratio l(y, y') \right] \\
    b &= \mathbb{E}\left[\frac{1}{n^4} \sum_{u = v} k(X^{p^*}_u, X^{p^*}_u) \sum_{i = j} \left( \densratiodiscretei \right)^2 l(y_i, y_i) \right] \\
    &= \frac{1}{n^2} \mathbb{E}_{X^{p^*}} \left[ k(X^{p^*}, X^{p^*}) \right] \mathbb{E}_{x, y, z} \left[ \left( \densratio \right)^2 l(y, y) \right] \\
    c &= \mathbb{E}\left[\frac{1}{n^4} \sum_{u \neq v} k(X^{p^*}_u, X^{p^*}_v) \sum_{i = j} \left( \densratiodiscretei \right)^2 l(y_i, y_i) \right] \\
    &= \frac{n-1}{n^2} \mathbb{E}_{X^{p^*}, {X^{p^*}}'} \left[ k(X^{p^*}, {X^{p^*}}') \right] \mathbb{E}_{x, y, z} \left[ \left( \densratio \right)^2 l(y, y) \right] \\
    d &= \mathbb{E}\left[\frac{1}{n^4} \sum_{u = v} k(X^{p^*}_u, X^{p^*}_u) \sum_{i \neq j} \densratiodiscretei \densratiodiscretej l(y_i, y_j) \right] \\
    &= \frac{n-1}{n^2} \mathbb{E}_{X^{p^*}} \left[ k(X^{p^*}, X^{p^*}) \right] \mathbb{E}_{x, y, z, x', y', z'} \left[ \densratioprime \densratio l(y, y') \right]
\end{aligned}
\end{equation*}

\begin{equation*}
    \begin{split}
        (2) &= \mathbb{E}\left[ \frac{1}{n}\sum_{i} \mathbb{E}_{X^{p^*}}[k(X^{p^*},X^{p^*}_i)]   \frac{1}{n}\sum_{j} \densratiodiscretej \mathbb{E}_{x,y,z}[\densratio l(y,y_j)]  \right]\\&= \mathbb{E}_{X^{p^*},{X^{p^*}}^{'}}\left[k(X^{p^*},{X^{p^*}}^{'})\right] \mathbb{E}_{x,y,z,x',y',z'}\left[\densratio\densratioprime l(y,y')  \right]
    \end{split}
\end{equation*}

\begin{equation*}
    \begin{split}
        (3) &= \mathbb{E}_{X^{p^*},{X^{p^*}}^{'}}\left[k(X^{p^*},{X^{p^*}}^{'})\right]  \mathbb{E}_{x,y,z,x',y',z'}\left[\densratio\densratioprime l(y,y')  \right]
    \end{split}
\end{equation*}

We note that the $ \mathbb{E}_{X^{p^*},{X^{p^*}}^{'}}\left[k(X^{p^*},{X^{p^*}}^{'})\right]  \mathbb{E}_{x,y,z,x',y',z'}\left[\densratio\densratioprime l(y,y')  \right]$-terms collapse for $(1),(2),(3)$. Thus:
\begin{equation*}
    \begin{split}
        &\langle B,B \rangle = \left(\frac{-2}{n}+\frac{1}{n^2}\right) \mathbb{E}_{X^{p^*},{X^{p^*}}^{'}}\left[k(X^{p^*},{X^{p^*}}^{'})\right]  \mathbb{E}_{x,y,z,x',y',z'}\left[\densratio\densratioprime l(y,y')  \right]\\& + \frac{1}{n^2}\mathbb{E}_{X^{p^*}}\left[ k(X^{p^*},{X^{p^*}})\right]\mathbb{E}_{x,y,z}\left[ \left(\densratio\right)^2 l(y,y) \right] \\&+ \frac{n-1}{n^2} \mathbb{E}_{X^{p^*},{X^{p^*}}^{'}}\left[ k(X^{p^*},{X^{p^*}}^{'})\right]\mathbb{E}_{x,y,z}\left[ \left(\densratio\right)^2 l(y,y) \right]\\&+\frac{n-1}{n^2} \mathbb{E}_{X^{p^*}}\left[ k(X^{p^*},{X^{p^*}})\right]\mathbb{E}_{x,y,z,x',y',z'}\left[\densratioprime \densratio l(y,y') \right]
    \end{split}
\end{equation*}
It suffices now to upper bound all the remaining expectations. First note that \newline $\mathbb{E}_{X^{p^*},{X^{p^*}}^{'}}\left[k(X^{p^*},{X^{p^*}}^{'})\right] \leq \sup_{X^{p^*},{X^{p^*}}^{'}}k(X^{p^*},{X^{p^*}}^{'}) \propto C_1$ and \newline  $ \mathbb{E}_{X^{p^*}}\left[ k(X^{p^*},{X^{p^*}})\right] \leq \sup_{X^{p^*}}k(X^{p^*},{X^{p^*}})\propto C_2$.\newline \\ Further $\mathbb{E}_{x,y,z,x',y',z'}\left[\densratioprime \densratio l(y,y') \right]\leq\mathbb{E}_{x',y',z'}\left[\densratioprime\right ] \mathbb{E}_{x,y,z}\left[\densratio\right] \sup_{y,y'}  l(y,y')=\sup_{y,y'}\propto C_3$. Finally $\mathbb{E}_{x,y,z}\left[ \left(\densratio\right)^2 l(y,y) \right] \leq C_4$ using the same arguments as before (finite variance of the density ratio). Then 
\begin{equation*}
    \begin{split}
        &\langle B,B \rangle = \left(\frac{-2}{n}+\frac{1}{n^2}\right) \mathbb{E}_{X^{p^*},{X^{p^*}}^{'}}\left[k(X^{p^*},{X^{p^*}}^{'})\right]  \mathbb{E}_{x,y,z,x',y',z'}\left[\densratio\densratioprime l(y,y')  \right]\\& + \frac{1}{n^2}\mathbb{E}_{X^{p^*}}\left[ k(X^{p^*},{X^{p^*}})\right]\mathbb{E}_{x,y,z}\left[ \left(\densratio\right)^2 l(y,y) \right] \\&+ \frac{n-1}{n^2} \mathbb{E}_{X^{p^*},{X^{p^*}}^{'}}\left[ k(X^{p^*},{X^{p^*}}^{'})\right]\mathbb{E}_{x,y,z}\left[ \left(\densratio\right)^2 l(y,y) \right]\\&+\frac{n-1}{n^2} \mathbb{E}_{X^{p^*}}\left[ k(X^{p^*},{X^{p^*}})\right]\mathbb{E}_{x,y,z,x',y',z'}\left[\densratioprime \densratio l(y,y') \right] \\& \leq \left(\frac{-2}{n}+\frac{1}{n^2}\right)C_1C_3 + \frac{1}{n^2}C_2C_4 + \frac{n-1}{n^2}C_1C_4 + \frac{n-1}{n^2}C_2C_3 \propto \mathcal{O}(\frac{1}{n})
    \end{split}
\end{equation*}
As $\widehat{C}_{p^*}$ is asymptotically unbiased in $L^2$ norm, it follows from Chebyshev's inequality that it is a consistent estimator. 
\end{Proof}
\subsection{Proof for Theorem \ref{big_theorem_consistency}}
\label{theorem_proof}
\begin{Proof}

Again consider:
\begin{equation*}
    \mathbb{E}\left[\Vert C_{p^*} - \widehat{C}_{p^*}\Vert^2_{\mathrm{HS}} \right]
\end{equation*}
However we take the estimator to be:
\begin{equation*}
    \widehat{C}_{p^*} = \frac{1}{n}\sum_{i=1}^{n}\hat{h}_n(x_i,z_i)k(\cdot,x_{i})\otimes l(\cdot,y_{i})-\left(\frac{1}{n}\sum_{j=1}^{n}k\left(\cdot,x_{j}^{p^*}\right)\right)\otimes\left(\frac{1}{n}\sum_{i=1}^{n}\hat{h}_n(x_i,z_i)l(\cdot,y_{i})\right).
\end{equation*}
Then we have 
\begin{equation*}
\begin{split}
        &\mathbb{E}\Bigg[\Bigg\Vert\mu_{xy}(\cdot) - \mu_{X^{p^*}}(\cdot)\otimes\mu_y(\cdot) - \\&\bigg( \underbrace{\frac{1}{n}\sum_{i=1}^{n}\hat{h}_n(x_i,z_i)k(\cdot,x_{i})\otimes l(\cdot,y_{i})}_{\hat{\mu}_{xy}}-\underbrace{\left(\frac{1}{n}\sum_{j=1}^{n}k\left(\cdot,x_{j}^{p^*}\right)\right)}_{\hat{\mu}_{X^{p^*}}}\otimes \underbrace{\left(\frac{1}{n}\sum_{i=1}^{n}\hat{h}_n(x_i,z_i)l(\cdot,y_{i})\right)}_{\hat{\mu}_y}\bigg) \Bigg\Vert^2 \Bigg]
\end{split}
\end{equation*}
We follow the same steps as in Theorem 1. 

\begin{equation*}
\begin{aligned}
    &\mathbb{E}\left[\left\Vert \underbrace{\mu_{xy}(\cdot) - \hat{\mu}_{xy}(\cdot)}_{A} 
    + \underbrace{\left(\hat{\mu}_{X^{p^*}}(\cdot) \otimes \hat{\mu}_y(\cdot) 
    - \mu_{X^{p^*}}(\cdot) \otimes \mu_y(\cdot)\right)}_{B} 
    \right\Vert^2 \right] \\
    &= \mathbb{E}\left[ \langle A, A \rangle 
    + 2 \langle A, B \rangle 
    + \langle B, B \rangle \right]
\end{aligned}
\end{equation*}
First term $\langle A,A \rangle$:

\begin{equation*}
\begin{split}
        \mathbb{E}\left[\langle A,A \rangle\right] &= \mathbb{E}\left[\langle \mu_{xy}(\cdot),\mu_{xy}(\cdot)\rangle \right] -2\mathbb{E}\left[ \langle \mu_{xy}(\cdot),\hat{\mu}_{xy}(\cdot) \rangle \right] + \mathbb{E}\left[\langle \hat{\mu}_{xy}(\cdot),\hat{\mu}_{xy}(\cdot) \rangle\right] 
\end{split}
\end{equation*}
Let $(x',y',z')\sim p$ be an independent copy of $(x,y,z)\sim p$.  Then the first term above is:
\begin{equation*}
    \begin{split}
        \mathbb{E}\left[\langle \mu_{xy}(\cdot),\mu_{x'y'}(\cdot)\rangle \right] = \mathbb{E}_{x,y,z,x',y',z'}\left[ \densratio \densratioprime k(x,x')l(y,y') \right];
    \end{split}
\end{equation*}
the second one is:
\begin{equation*}
    \begin{split}
        \mathbb{E}\left[\langle \mu_{xy}(\cdot),\hat{\mu}_{xy}(\cdot)\rangle \right]&=\mathbb{E}\left[\frac{1}{n}\sum_{i=1}^{n}\estdensi\mathbb{E}_{x,y,z}\left[\densratio k(x,x_{i}) l(y,y_{i})\right]  \right] \\& = \mathbb{E}_{x,y,z,x',y',z'}\left[ \estdens \densratioprime k(x,x')l(y,y') \right] \\&=
     (1+\mathcal{O}(\frac{1}{n^{\alpha}}))\mathbb{E}_{x,y,z,x',y',z'}\left[ \densratio \densratioprime k(x,x')l(y,y') \right];
    \end{split}
\end{equation*}
and the last term is:
\begin{equation*}
\begin{split}
    \lefteqn{\mathbb{E}\left[\frac{1}{n^2} \sum_{i,j} \estdensi\estdensj k(x_i,x_j)l(y_i,y_j) \right]}\\ 
    &=  \underbrace{\mathbb{E}\left[\frac{1}{n^2} \sum_{i\neq j} \estdensi\estdensj k(x_i,x_j)l(y_i,y_j)\right]}_{(a)}+\underbrace{\mathbb{E}\left[\frac{1}{n^2} \sum_{i} \estdensi^2 k(x_i,x_i)l(y_i,y_i) \right]}_{(b)}.  
\end{split}
\end{equation*}
We bound $(a)$:
\begin{equation*}
\begin{split}
        &(a) = \frac{(n-1)}{n} \mathbb{E}_{x,y,z,x',y',z'}\left[\estdens \estdensprime k(x,x')l(y,y')\right] \\&=\frac{(n-1)}{n} \mathbb{E}_{x,y,z,x',y',z'}\left[\densratio \densratioprime k(x,x')l(y,y')\right]\left(1+2\mathcal{O}(\frac{1}{n^{\alpha}}) +\mathcal{O}\left(\frac{1}{n^{2\alpha}}\right)\right)
\end{split}
\end{equation*}
then $(b)$
\begin{equation*}
\begin{split}
    (b)&=\frac{1}{n}\mathbb{E}\left[\left( \estdens \right)^2 k(x,x)l(y,y)\right] \\& \leq \frac{1}{n} \underbrace{ \sup_{x\in X}k(x,x) \sup_{y\in Y}k(y,y)}_{\text{Assumed to be bounded by } C} \underbrace{\mathbb{E}\left[{\left(\estdens\right)}^2\right]}_{\text{finite variance of density ratio, i.e.~bounded by some $D$}} \\&=\frac{CD}{n} \left(1+2\mathcal{O}(\frac{1}{n^{\alpha}}) +\mathcal{O}\left(\frac{1}{n^{2\alpha}}\right)\right)
\end{split}
\end{equation*}
So for $\langle A,A \rangle$ the convergence rate is $\mathcal{O}\left(\frac{1}{n^{\min(1,\alpha)}}\right)$, since that is the slowest decaying term. For the $\langle A,B \rangle$ part we have:
\begin{equation*}
\begin{split}
        \langle A,B \rangle= \underbrace{\langle \mu_{xy}(\cdot),\hat{\mu}_{X^{p^*}}(\cdot)\otimes\hat{\mu}_{y}(\cdot)\rangle}_{(1)}+\underbrace{\langle \hat{\mu}_{xy}(\cdot),\mu_{X^{p^*}}(\cdot)\otimes\mu_{y}(\cdot)\rangle}_{(2)}\\ -\underbrace{\langle\hat{\mu}_{xy}(\cdot),\hat{\mu}_{X^{p^*}}(\cdot)\otimes\hat{\mu}_{y}(\cdot)\rangle}_{(3)} -\underbrace{\langle\mu_{xy}(\cdot),\mu_{X^{p^*}}(\cdot)\otimes \mu_{y'}(\cdot)\rangle}_{(4)} 
\end{split}
\end{equation*}
Each part can then be written as:
\begin{equation*}
    \begin{split}
        (1)&= \mathbb{E}_{}\left[\frac{1}{n^2}\sum_{i,j} \estdensj \mathbb{E}_{x,y,z}\left[\densratio k(x,x_i^{p^*}) l(y,y_j)\right] \right]\\&=\mathbb{E}_{x',y',z'}\left[\estdensprime \mathbb{E}_{x,y,z}\left[\densratio k(x,x_i^{p^*}) l(y,y_j)\right] \right]\\&=\mathbb{E}_{x',y',z'}\left[\densratioprime \mathbb{E}_{x,y,z}\left[\densratio k(x,x_i^{p^*}) l(y,y_j)\right] \right]\left(1+\estasympslow\right)
    \end{split}
\end{equation*}
\begin{equation*}
    \begin{split}
        (2)&= \mathbb{E}_{}\left[ \frac{1}{n}\sum_i^n \estdensi \mathbb{E}_{x_q}[k(x_i,X^{p^*})] \mathbb{E}_{x,y,z}\left[\densratio l(y,y_i)\right] \right] \\& = \mathbb{E}_{x',y',z'}\left[ \estdensprime \mathbb{E}_{x_q}[k(x',X^{p^*})] \mathbb{E}_{x,y,z}\left[\densratio l(y,y')\right] \right]\\&= \mathbb{E}_{x',y',z'}\left[ \densratioprime \mathbb{E}_{x_q}[k(x',X^{p^*})] \mathbb{E}_{x,y,z}\left[\densratio l(y,y')\right]\right]\left(1+\estasympslow\right)
    \end{split}
\end{equation*}

\begin{equation*}
\begin{aligned}
    (3) &= \mathbb{E}\left[ \frac{1}{n^3} \sum_{i,k,j} \estdensi \estdensj k(x_i, X^{p^*}_k) l(y_i, y_j) \right] \\
    &= \mathbb{E}\left[ \frac{1}{n^3} \sum_{k,i \neq j} \estdensi \estdensj k(x_i, X^{p^*}_k) l(y_i, y_j) \right] \\
    &\quad + \mathbb{E}\left[\frac{1}{n^3} \sum_{k,i = j} \left( \estdensi \right)^2 k(x_i, X^{p^*}_k) l(y_i, y_i) \right] \\
    &= \left(1 + 2 \estasympslow + \estasymp\right) \frac{n-1}{n} 
    \mathbb{E}_{x', y', z'} \left[ \densratioprime \right. \\
    &\quad \left. \times \mathbb{E}_{x, y, z} \left[ \densratio 
    \mathbb{E}_{X^{p^*}} \left[ k(x, X^{p^*}) \right] l(y, y') \right] \right] \\
    &\quad + \left(1 + 2 \mathcal{O}\left(\frac{1}{n^{\alpha}}\right) + \mathcal{O}\left(\frac{1}{n^{2\alpha}}\right)\right) \frac{1}{n} 
    \mathbb{E}_{X^{p^*}, x, y, z} \left[ \left( \densratio \right)^2 k(x, X^{p^*}) l(y, y) \right].
\end{aligned}
\end{equation*}

\begin{equation*}
    \begin{split}
        (4)&= \mathbb{E}_{x,y,z}\left[ \densratio \mathbb{E}_{x_q}[k(x,X^{p^*})] \mathbb{E}_{x',y',z'}\left[\densratioprime l(y,y')\right] \right]
    \end{split}
\end{equation*}
We can use the same arguments as in $\langle A,A \rangle$, consequently $\langle A,B \rangle =  \mathcal{O}\left(\frac{1}{n^{\min(1,\alpha)}}\right)$. The final term $\langle B,B \rangle$ is
\begin{equation*}
\begin{split}
       \langle B,B \rangle= \underbrace{\langle\hat{\mu}_{X^{p^*}}(\cdot)\otimes\hat{\mu}_y(\cdot),\hat{\mu}_{X^{p^*}}(\cdot)\otimes\hat{\mu}_{y}(\cdot)\rangle}_{(1)}-\underbrace{2\langle\hat{\mu}_{X^{p^*}}(\cdot)\otimes\hat{\mu}_y(\cdot),\mu_{X^{p^*}}(\cdot)\otimes\mu_y(\cdot)\rangle}_{(2)}\\ + \underbrace{\langle\mu_{{X^{p^*}}^{'}}(\cdot)\otimes\mu_{y'}(\cdot),\mu_{X^{p^*}}(\cdot)\otimes\mu_y(\cdot)\rangle}_{(3)}, 
\end{split}
\end{equation*}
and we have
\begin{equation*}
\begin{split}
    &(1) = \\
    &\mathbb{E}\left[ \frac{1}{n^2} \sum_{u,v} k(X^{p^*}_{u}, X^{p^*}_{v}) \frac{1}{n^2} \sum_{i,j} \estdensi \estdensj l(y_i, y_{j}) \right] \\
    &= \mathbb{E}\Bigg[ \frac{1}{n^2} \left( \sum_{u \neq v} k(X^{p^*}_{u}, X^{p^*}_{v}) + \sum_{u = v} k(X^{p^*}_{u}, X^{p^*}_{u}) \right) \\
    &\quad \frac{1}{n^2} \left( \sum_{i \neq j} \estdensi \estdensj l(y_i, y_{j}) + \sum_{i = j} \estdensi^2 l(y_i, y_{i}) \right) \Bigg] \\
    &= \underbrace{\mathbb{E}\left[ \frac{1}{n^4} \sum_{u \neq v} k(X^{p^*}_{u}, X^{p^*}_{v}) \sum_{i \neq j} \estdensi \estdensj l(y_i, y_{j}) \right]}_{(a)} \\
    &\quad + \underbrace{\mathbb{E}\left[ \frac{1}{n^4} \sum_{u = v} k(X^{p^*}_{u}, X^{p^*}_{u}) \sum_{i = j} \estdensi^2 l(y_i, y_{i}) \right]}_{(b)} \\
    &\quad + \underbrace{\mathbb{E}\left[ \frac{1}{n^4} \sum_{u \neq v} k(X^{p^*}_{u}, X^{p^*}_{v}) \sum_{i = j} \estdensi^2 l(y_i, y_{i}) \right]}_{(c)} \\
    &\quad + \underbrace{\mathbb{E}\left[ \frac{1}{n^4} \sum_{u = v} k(X^{p^*}_{u}, X^{p^*}_{u}) \sum_{i \neq j} \estdensi \estdensj l(y_i, y_{j}) \right]}_{(d)}
\end{split}
\end{equation*}
Each of these terms is:
\begin{equation*}
\begin{aligned}
    (a) &= \mathbb{E}\left[\frac{1}{n^4} \sum_{u\neq v} k(X^{p^*}_{u}, X^{p^*}_{v}) \sum_{i\neq j} \estdensi \estdensj l(y_i, y_{j}) \right] \\
    &= \frac{(n-1)^2}{n^2} \mathbb{E}_{X^{p^*}, {X^{p^*}}^{'}} \left[ k(X^{p^*}, {X^{p^*}}^{'}) \right] \mathbb{E}_{x, y, z, x', y', z'} \left[ \densratioprime \densratio l(y, y') \right] \\
    &\quad \times \left(1 + 2 \estasympslow + \estasymp\right) \\
    (b) &= \mathbb{E}\left[\frac{1}{n^4} \sum_{u=v} k(X^{p^*}_{u}, X^{p^*}_{u}) \sum_{i=j} \estdensi^2 l(y_i, y_{i}) \right] \\
    &= \frac{1}{n^2} \mathbb{E}_{X^{p^*}} \left[ k(X^{p^*}, {X^{p^*}}) \right] \mathbb{E}_{x, y, z} \left[ \left( \densratio \right)^2 l(y, y) \right] \\
    &\quad \times \left(1 + 2 \mathcal{O}\left(\frac{1}{n^{\alpha}}\right) + \mathcal{O}\left(\frac{1}{n^{2\alpha}}\right)\right) \\
    (c) &= \mathbb{E}\left[\frac{1}{n^4} \sum_{u\neq v} k(X^{p^*}_{u}, X^{p^*}_{v}) \sum_{i=j} \estdensi^2 l(y_i, y_{i}) \right] \\
    &= \frac{n-1}{n^2} \mathbb{E}_{X^{p^*}, {X^{p^*}}^{'}} \left[ k(X^{p^*}, {X^{p^*}}^{'}) \right] \mathbb{E}_{x, y, z} \left[ \left( \densratio \right)^2 l(y, y) \right] \\
    &\quad \times \left(1 + 2 \mathcal{O}\left(\frac{1}{n^{\alpha}}\right) + \mathcal{O}\left(\frac{1}{n^{2\alpha}}\right)\right) \\
    (d) &= \mathbb{E}\left[\frac{1}{n^4} \sum_{u=v} k(X^{p^*}_{u}, X^{p^*}_{u}) \sum_{i\neq j} \estdensi \estdensj l(y_i, y_{j}) \right] \\
    &= \frac{n-1}{n^2} \mathbb{E}_{X^{p^*}} \left[ k(X^{p^*}, {X^{p^*}}) \right] \mathbb{E}_{x, y, z, x', y', z'} \left[ \densratioprime \densratio l(y, y') \right] \\
    &\quad \times \left(1 + 2 \estasympslow + \estasymp\right).
\end{aligned}
\end{equation*}
Returning to the expression for $\langle B,B\rangle$:
\begin{equation*}
    \begin{split}
        (2) &= \mathbb{E}\left[ \frac{1}{n}\sum_{i} \mathbb{E}_{X^{p^*}}[k(X^{p^*},X^{p^*}_i)]   \frac{1}{n}\sum_{j} \estdensj \mathbb{E}_{x,y,z}[\densratio l(y,y_j)]  \right]\\& = \mathbb{E}_{X^{p^*},{X^{p^*}}^{'}}\left[k(X^{p^*},{X^{p^*}}^{'})\right] \mathbb{E}_{x,y,z,x',y',z'}\left[\densratio\densratioprime l(y,y')  \right]\left(1+\estasympslow\right),
    \end{split}
\end{equation*}
and
\begin{equation*}
    \begin{split}
        (3) &= \mathbb{E}_{X^{p^*},{X^{p^*}}^{'}}\left[k(X^{p^*},{X^{p^*}}^{'})\right]  \mathbb{E}_{x,y,z,x',y',z'}\left[\densratio\densratioprime l(y,y')  \right].
    \end{split}
\end{equation*}
We use the same arguments as in Theorem \ref{bigger_lemma}. We note that the the sums collapse similarly in (1) with some added negligible terms converging faster than $\mathcal{O}(\frac{1}{n})$. What remains is then the slowest converging term $$\mathbb{E}_{X^{p^*},{X^{p^*}}^{'}}\left[k(X^{p^*},{X^{p^*}}^{'})\right] \mathbb{E}_{x,y,z,x',y',z'}\left[\densratio\densratioprime l(y,y')  \right]\estasympslow = \estasympslow$$ compared to $\mathcal{O}(\frac{1}{n})$. Hence $\langle B,B \rangle = \mathcal{O}\left(n^{-\min(1,\alpha)}\right)$. As $\langle A,A \rangle,\langle A,B \rangle,\langle B,B \rangle$ all have the same convergence rate of their bias terms, we conclude that $    \mathbb{E}\left[\Vert C_{p^*} - \widehat{C}_{p^*}\Vert^2_{\mathrm{HS}} \right] = \mathcal{O}\left(\frac{1}{n^{\min(1,\alpha)}}\right)$.
\end{Proof}
\subsection{Proof of Proposition \ref{prop_consistency}}
\label{proof_consistency}
\begin{Proof}
Since $\Vert\widehat{C}_{p^*}\Vert_{\text{HS}}^2(\psi)\geq 0$, it suffices by Markov's inequality to show that 
$$
\lim_{n\to \infty} \mathbb{E}\left[\Vert\widehat{C}_{p^*}\Vert_{\text{HS}}^2(\psi)\right]=0.
$$

Let let $w_i=\frac{p^*(x_i)}{p(x_i \mid z_i)}$. Define $\mathbf{M}_{k}(n):=\{1, \ldots, n\}^{k}$ as the $k$-fold Cartesian product of the set $\{1, \ldots, n\}$. We then have 
\begin{equation}
\begin{split}
    \Vert\widehat{C}_{p^*}\Vert_{\text{HS}}^2(\psi) &= \underbrace{\frac{1}{n^{2}} \sum_{i,j \in \mathbf{M}_{2}(n)} k(x_i,x_j)l(y_{\psi(i)},y_{\psi(j)})w_{i} w_{j}}_{A_n}
    \\ &+\underbrace{\frac{1}{n^{4}} \sum_{i,j,k,l \in \mathbf{M}_{4}(n)} k(x_k,x_l)l(y_{\psi(i)},y_{\psi(j)})w_{i} w_{j}}_{B_n} \\ &-2\underbrace{\frac{1}{n^{3}} \sum_{i,j,k \in \mathbf{M}_{3}(n)}w_i k(x_i,x_j)w_k l(y_{\psi(i)},y_{\psi(k)})}_{C_n},
\end{split}
    \end{equation}
which is abbreviated as $A_n+B_n-2C_n$. It suffices to show that $\lim _{n} \mathbb{E} A_{n}=\lim _{n} \mathbb{E} B_{n}=\lim _{n} \mathbb{E} C_{n} = \zeta$. Where 

$$
\zeta = \mathbb{E}\left [k(X,X^{'}) \right]  \mathbb{E}\left [l(Y,Y')  \right]
$$
where $X',Z'$ is an identical independent copy of $X,Z$ respectively. We employ the same strategy as in \cite{https://doi.org/10.1002/sta4.364} and partition the summing over the indices as
\begin{equation}
\begin{aligned}
A_n= \frac{1}{n^2} \sum_{i,j\in U(2, \psi, n)} k(x_i,x_j)l(y_{\psi(i)},y_{\psi(j)})w_{i} w_{j}+\frac{1}{n^2} \sum_{i,j\in R(2, \psi, n)} k(x_i,x_j)l(y_{\psi(i)},y_{\psi(j)})w_{i} w_{j}
\end{aligned}
\end{equation}
with
$$
U(2, \psi, n):=\left\{\left(i, j\right) \in M_2(n): \left(i,j,\psi\left(i\right), \psi\left(j\right)\right) \text{are 4 distinct elements}  \right \}
$$
and
$$
R(2, \psi, n):=M_2(n) \backslash U(2, \psi, n) .
$$
Here 4 distinct elements simply means that $i,j,\psi(i),\psi(j)$ are all different, i.e. 1,2,3,4. The main observation here is that as $n\to \infty$, almost all terms in sums $A_n,B_n,C_n$ will have distinct terms. We then take 

\begin{equation}
\begin{aligned}
\mathbb{E}\left[A_n\right]=& \mathbb{E}\left[\mathbb{E}\left[A_n \mid \psi\right]\right] \\
=& \mathbb{E}\left[\frac{1}{n^2} \sum_{U(2, \psi, n)} \mathbb{E}\left[k(x_i,x_j)l(y_{\psi(i)},y_{\psi(j)})w_{i} w_{j} \mid \psi\right]\right] \\
&+\mathbb{E}\left[\frac{1}{n^2} \sum_{R(2, \psi, n)} \mathbb{E}\left[k(x_i,x_j)l(y_{\psi(i)},y_{\psi(j)})w_{i} w_{j} \mid \psi\right]\right] \\
=& \mathbb{E}\left[\frac{|U(2, \psi, n)|}{n^2}\right]  \mathbb{E}\left[k(X,X^{'}) \right]  \mathbb{E}\left[l(Y,Y')  \right] +\mathbb{E}\left[\frac{|R(2, \psi, n)|}{n^2}\right] \mathcal{O}(1) 
 \rightarrow \zeta .
\end{aligned}
\end{equation}
Since the permutation occurs within $p(x\mid z)$, there will always be a dependency between $w_i,w_j$ and $l(y_{\psi(i)},y_{\psi(j)})$ meaning that distinct indices don't factorize directly unlike the setting in \cite{https://doi.org/10.1002/sta4.364}. However, by calculating the expectation we have that

\begin{equation}
\begin{aligned}
    &\mathbb{E}\left[k(x_i,x_j)l(y_{\psi(i)},y_{\psi(j)})w_{i} w_{j} \mid \psi\right]\\ &= \mathbb{E}\left[k(X,X')l(Y,Y')W W' \right] \\
    &= \int k(x,x')l(y,y')\frac{p^*(x)}{p(x\mid z)}\frac{p^*(x')}{p(x'\mid z')}p(x,y,z)p(x',y',z')dxdydzdx'dy'dz' \\
    &=\int k(x,x')l(y,y') p^*(x)p^*(x') \underbrace{\left(\int p(y\mid x,z)p(z)dz\right) \left(\int p(y'\mid x',z')p(z')dz' \right)}_{\text{$y$ and $x$ are rendered independent by having distinct indices}} dxdydx'dy' \\
    &=\int k(x,x')l(y,y')p^*(x)p^*(x')p(y)p(y')dxdx'dydy'\\&=\mathbb{E}\left[k(X,X^{'}) \right]  \mathbb{E}\left[l(Y,Y')  \right] = \zeta.
\end{aligned}
\end{equation}
Now we can repeat the argument in \cite{https://doi.org/10.1002/sta4.364} with 
\begin{equation}
\begin{aligned}
\frac{\mathbb{E}|U(2, \psi, n)|}{n^2} &= \frac{n(n-1)}{n^2} \cdot \mathbb{P}\left(\left(i,j,\psi(i), \psi(j)\right) \text{ are 4 distinct elements }\right) \\
&= \frac{n(n-1)}{n^2} \frac{\left(\begin{array}{c}
n-2 \\
2
\end{array}\right)\left(\begin{array}{c}
n-4 \\
2
\end{array}\right) \cdots\left(\begin{array}{c}
n-2 d+2 \\
2
\end{array}\right)}{\left(\begin{array}{c}
n \\
2
\end{array}\right)\left(\begin{array}{c}
n \\
2
\end{array}\right) \cdots\left(\begin{array}{l}
n \\
2
\end{array}\right)} \\
& \longrightarrow 1 .
\end{aligned}
\end{equation}

Hence $\lim _{n \rightarrow \infty} \mathbb{E}\left(A_n\right)= \mathbb{E}\left[k(X,X') \right]  \mathbb{E}\left[l(Y,Y')  \right] = \zeta$. We can repeat the argument for $B_n$ and $C_n$ and hence the sum of $\zeta$'s collapse into 0. Throughout the proof, we have assumed that $\mathbb{E}\left[k(X,X')l(Y,Y')W W' \right]<C$, where $C$ is some constant. 
\end{Proof}

\section{Proof of Proposition \ref{prop:estimator}}
\label{proof_thm1}
We first establish some ``RKHS calculus" before we proceed with the calculations.

\textit{Mean}
Some rules following Riesz representation theorem and RKHS spaces.
\begin{enumerate}
    \item $\langle\mu_x,f \rangle_{\mathcal{F}}=\mathbf{E}_x[\langle\phi(x),f \rangle_{\mathcal{F}}]=\mathbf{E}_x[f(x)]$
    \item $\langle\mu_y,g \rangle_{\mathcal{G}}=\mathbf{E}_y[\langle\psi(y),g \rangle_{\mathcal{G}}]=\mathbf{E}_y[g(x)]$
    \item $\|\mu_x\|^2_{\mathcal{F}}=\mathbf{E}_{x,x'}[\langle \phi(x),\phi(x')\rangle_{\mathcal{F}}]=\mathbf{E}_{x,x'}[k(x,x')]=\frac{1}{n^2}\sum_{i,j}k(x_i,x_j)$
\end{enumerate}

\textit{Tensor operator $\otimes$}

We may employ any $f \in \mathcal{F}$ and $g \in \mathcal{G}$ to define a tensor product operator $f \otimes g: \mathcal{G} \rightarrow \mathcal{F}$ as follows:
$(f \otimes g) h:=f\langle g, h\rangle_{\mathcal{G}} \quad$ for all $h \in \mathcal{G}$

\begin{Lemma}
For any $f_{1}, f_{2} \in \mathcal{F}$ and $g_{1}, g_{2} \in \mathcal{G}$ the following equation holds:
$\left\langle f_{1} \otimes g_{1}, f_{2} \otimes g_{2}\right\rangle_{\mathrm{HS}}=\left\langle f_{1}, f_{2}\right\rangle_{F}\left\langle g_{1}, g_{2}\right\rangle_{G}$
\end{Lemma}
Using this lemma, one can simply show the norm of $f \otimes g$ equals $\|f \otimes g\|_{\mathrm{HS}}^{2}=\|f\|_{\mathcal{F}}^{2}\|g\|_{\mathcal{G}}^{2}$ \newline \\ Let $\hat{\mu}_{P_x}=\frac{1}{n}\sum_{i=1}^nk(\mathbf{x}_i,\cdot)$ and $\hat{\mu}_{P_y}=\frac{1}{n}\sum_{i=1}^nl(\mathbf{y}_i,\cdot)$. We are now ready to proceed with the derivation.

\begin{Proof}
For 
\[
\widehat{C}_{p^*}=\frac{1}{n}\sum_{i=1}^{n}\tilde{w}_{i}k(\cdot,x_{i})\otimes l(\cdot,y_{i})-\left(\frac{1}{m_{x}}\sum_{j=1}^{m_{x}}k\left(\cdot,x_{j}^{p^*}\right)\right)\otimes\left(\frac{1}{n}\sum_{i=1}^{n}\tilde{w}_{i}l\left(\cdot,y_{i}\right)\right).
\]
we have the following:

\[
\Vert\widehat{C}_{p^*}\Vert^2 = A + B -2C
\]
where 

\[
A = \frac{1}{n^2} \sum_{i=1}^{n}\sum_{j=1}^{n}\tilde{w}_{i}\tilde{w}_{j}\left\langle k(\cdot,x_{i})\otimes l(\cdot,y_{i}),k(\cdot,x_{j})\otimes l(\cdot,y_{j})\right\rangle =\tilde{w}^{\top}\left(K\circ L\right)\tilde{w}=tr\left(D_{\tilde{w}}KD_{\tilde{w}}L\right),
\]

\[
B =\frac{1}{m_x^2n^2} \left\Vert \sum_{i=1}^{n}k\left(\cdot,x_{i}^{p^*}\right)\right\Vert ^{2}\left\Vert \sum_{i=1}^{n}\tilde{w}_{i}l\left(\cdot,y_{i}\right)\right\Vert ^{2}=\frac{1}{m_x^2n^2} (\mathbf{K}^{p^*})_{++}(\mathbf{L}\circ \tilde{W})_{++} 
\]
\begin{align*}
C &= \frac{1}{n^2m_x} \left\langle \sum_{i=1}^{n} \tilde{w}_{i} k(\cdot,x_{i}) \otimes l(\cdot,y_{i}), \left(\sum_{j=1}^{m_x} k(\cdot,x_{j}^{p^*})\right) \otimes \left(\sum_{j=1}^{n} \tilde{w}_{j} l(\cdot,y_{j})\right) \right\rangle \\
&= \frac{1}{n^2m_x} \sum_{i=1}^{n} \tilde{w}_{i} \left(\sum_{j=1}^{m_x} k(x_{i},x_{j}^{p^*})\right) \left(\sum_{r=1}^{n} \tilde{w}_{r} l(y_{i},y_{r})\right) \\
&= \tilde{w}^{\top} \left(K^q 1_{m_x} \circ L \tilde{w}\right) \\
&= \operatorname{tr} \left(D_{\tilde{w}} K^q 1_{m_x} \tilde{w}^{\top} L\right),
\end{align*}

It should be noted that we can choose the number of samples $m_x$ for $x_i^{p^*}\sim p^*$ to use. For practical purposes we set $m_x = n$.
\end{Proof}

\section{Optimal $c_{p^*}$ Choice}
\subsection{Derivation of univariate $c_{p^*}$}
\label{c_{p^*}_1_proof}
Suppose that we wish to choose an 'optimal' $c_{p^*}$ value for rescaling the distribution. One criterion for optimality is to maximize the effective sample size
$$
\text{ESS}:=\frac{\left(\sum_{i} w_{i}\right)^{2}}{\sum_{i} w_{i}^{2}}
$$
where $w_{i}=p^*\left(x ; \phi^{*}\right) / p_{X | Z}(x ; \phi)$ is the weight used to resample observations. This is essentially equivalent to minimizing the variance of the individual weights:
$$
\arg \min _{\phi \cdot} \operatorname{Var} w\left(X_{i}, Z_{i} ; \phi^{*}\right) .
$$
Note that
$$
\begin{aligned}
\mathbb{E} w\left(X_{i}, Z_{i} ; \phi^{*}\right) &=\int \frac{p^*\left(x, z, \phi^{*}\right)}{p_{X | Z}(x, z, \phi)} p_{X | Z}(x, z, \phi) d x d z \\
&=\int p^*\left(x, z, \phi^{*}\right) d x d z \\
&=1
\end{aligned}
$$
so this is equivalent to minimizing the squared expectation of $w()$.
If we assume that everything is Gaussian and that $X, Z$ have standard normal marginal distributions with correlation $\rho$, this becomes equivalent to minimizing
$$
\frac{1}{\tau^{2}} \mathbb{E}\left[\frac{\phi\left(\frac{X}{\tau}\right)}{\phi\left(\frac{X-\rho Z}{\sqrt{1-\rho^{2}}}\right)}\right]^{2}
$$
with respect to $\tau$. 
We can rewrite this expression as:
$$
\begin{aligned}
f(\tau) &=\frac{1}{2 \pi \sqrt{1-\rho^{2} \tau^{2}}} \iint_{-\infty}^{\infty} \exp \left\{-\frac{1}{2}\left(\begin{array}{l}
x \\
z
\end{array}\right)^{T}\left(\begin{array}{cc}
\frac{2}{\tau^{2}}-\frac{1}{1-\rho^{2}} & \frac{\rho}{1-\rho^{2}} \\
\frac{\rho}{1-\rho^{2}} & \frac{1-2 \rho^{2}}{1-\rho^{2}}
\end{array}\right)\left(\begin{array}{l}
x \\
z
\end{array}\right)\right\} d x d z \\
&=\frac{1}{\sqrt{1-\rho^{2} \tau^{2}}}\left|\begin{array}{cc}
\frac{2}{\tau^{2}}-\frac{1}{1-\rho^{2}} & \frac{\rho}{1-\rho^{2}} \\
\frac{\rho}{1-\rho^{2}} & \frac{1-2 \rho^{2}}{1-\rho^{2}}
\end{array}\right|^{-1 / 2} \\
&=\frac{1}{\tau^{2}}\left|\begin{array}{cc}
\frac{2\left(1-\rho^{2}\right)}{\tau^{2}}-1 & \rho \\
\rho & 1-2 \rho^{2}
\end{array}\right|^{-1 / 2}
\end{aligned}
$$

Note that minimizing $f$ is the same as maximizing $1 / f^{2}$, so we need to maximize
$$
\begin{aligned}
1 / f(\tau)^{2} &=\tau^{4}\left|\begin{array}{cc}
\frac{2\left(1-\rho^{2}\right)}{\tau^{2}}-1 & \rho \\
\rho & 1-2 \rho^{2}
\end{array}\right| \\
&\left.=\tau^{4}\left(\left(2 \frac{1-\rho^{2}}{\tau^{2}}-1\right)\left(1-2 \rho^{2}\right)-\rho^{2}\right)\right) \\
&=\tau^{2} 2\left(1-\rho^{2}\right)\left(1-2 \rho^{2}\right)-\left(1-\rho^{2}\right) \tau^{4} .
\end{aligned}
$$
This is maximized at $\tau^{2}=1-2 \rho^{2}$, or $\tau=\sqrt{1-2 \rho^{2}}$.

\subsection{Derivation of multivariate $c_{p^*}$}
\label{c_{p^*}_2_proof}
Now suppose that $(X, Z) \sim N_{p+p^*}(0, \Sigma)$ where we take $\Sigma = 
\begin{bmatrix}
\Sigma_{x x} & \Sigma_{x z}\\
\Sigma_{z x} & \Sigma_{z z}
\end{bmatrix}
$ and assume that $\Sigma_{x x}=I_{p}$ and $\Sigma_{z z}=I_{p^*}$ (again, this can be achieved by rescaling). By the same reasoning as above, we want to minimize the squared expectation of the weights, which amounts to minimizing

\[
\begin{aligned}
    f(\tau) = &\frac{1}{|T|} \iint_{-\infty}^{\infty} \exp \left\{ -\frac{1}{2} \begin{pmatrix} x \\ z \end{pmatrix}^{T} \begin{pmatrix} A & B \\ C & D \end{pmatrix} \begin{pmatrix} x \\ z \end{pmatrix} \right\} dx\, dz \\
    &\propto \frac{1}{|T|} \left| \begin{matrix} A & B \\ C & D \end{matrix} \right|^{-1/2}
\end{aligned}
\]

where

\[
\begin{aligned}
    A &= 2T^{-1} - (I_p - \Sigma_{xz} \Sigma_{zx})^{-1}, \\
    B &= (I_p - \Sigma_{xz} \Sigma_{zx})^{-1} \Sigma_{xz}, \\
    C &= \Sigma_{zx} (I_p - \Sigma_{xz} \Sigma_{zx})^{-1}, \\
    D &= I_{p^*} - \Sigma_{zx} (I_p - \Sigma_{xz} \Sigma_{zx})^{-1} \Sigma_{xz}.
\end{aligned}
\]
or maximizing
$$
|T|^{2}\left|\begin{array}{cc}
2 T^{-1}-\left(I_{p}-\Sigma_{x z} \Sigma_{z x}\right)^{-1} & \left(I_{p}-\Sigma_{x z} \Sigma_{z x}\right)^{-1} \Sigma_{x z} \\
\sum_{z x}\left(I_{p}-\Sigma_{x z} \Sigma_{z x}\right)^{-1} & I_{p^*}-\Sigma_{z x}\left(I_{p}-\Sigma_{x z} \Sigma_{z x}\right)^{-1} \Sigma_{x z}
\end{array}\right|
$$
Note that in this case we will choose a whole matrix $T$, rather than just a scaling constant, but we could simplify to assume that $T=c_{p^*} I_{p}$ for some scalar $c_{p^*}$. \newline \\ We take \newline $A := 2 T^{-1}-\left(I_{p}-\Sigma_{x z} \Sigma_{z x}\right)^{-1}$, $B:=\left(I_{p}-\Sigma_{x z} \Sigma_{z x}\right)^{-1} \Sigma_{x z}$, $D:=I_{p^*}-\Sigma_{z x}\left(I_{p}-\Sigma_{x z} \Sigma_{z x}\right)^{-1} \Sigma_{x z}$. The block matrix is then 
$$
M:=
\left|\begin{array}{cc}
A & B \\
B^\top & D
\end{array}\right|
$$
Assuming $D$ is invertible, we can use the Schur complement
\begin{equation*}
\operatorname{det}(M)=\operatorname{det}(D) \operatorname{det}\left(A-B D^{-1} B^\top\right)
\end{equation*}
Then
\begin{equation*}
   \operatorname{det}( A-B D^{-1} B^\top )= \operatorname{det}(2 T^{-1}-\left(I_{p}-\Sigma_{x z} \Sigma_{z x}\right)^{-1}-BD^{-1}B^\top).
\end{equation*}
which can easily be optimized with gradient descent with respect to $T=c_{p^*}I_p$. For estimated covariance matrices $\hat\Sigma = 
\begin{bmatrix}
\hat\Sigma_{xx} & \hat\Sigma_{xz}\\
\hat\Sigma_{zx} & \hat\Sigma_{zz}
\end{bmatrix}
$. It suffices to plug them in directly in the estimator.

\section{bd-CME test derivation}
\label{bd-cme}
We observe $\left\{ (x_{i},y_{i},z_{i})\right\} _{i=1}^{n}\sim p$,
some probability density on the joint space $\mathcal{X}\times\mathcal{Y}\times\mathcal{Z}$.
Define
\[
p(y|do(x))\propto\tilde{p}(y|do(x))=\int p(y|x,z)p(z)dz.
\]
Note that $p(y|do(x))\neq p(y|x)$ in general since $X$ and $Z$ need
not be independent. We would like to test
\[
H_{0}:\;p(y|do(x))\;\text{does not depend on }x
\]
versus the general alternative. Let $k,l,m$ be kernels on $\mathcal{X},\mathcal{Y},\mathcal{Z}$,
respectively. We first fit the conditional mean embedding
\begin{eqnarray*}
\mu_{p(\cdot|x,z)} & = & \int l(\cdot,y)p(y|x,z)dy,
\end{eqnarray*}
by learning a regression function $k(\cdot,x)\otimes m(\cdot,z)\mapsto l(\cdot,y)$.
This is given by $\hat{\mu}_{p(\cdot|x,z)}=\sum_{i=1}^{n}a(x,z)_{i}l(\cdot,y_{i})$,
where 
\[
a(x,z)=\left(K_{{\bf xx}}\circ M_{{\bf zz}}+\epsilon_{n}I\right)^{-1}\left(K_{{\bf x}x}\circ M_{{\bf z}z}\right),
\]
$\left[K_{{\bf xx}}\right]_{ij}=k(x_{i},x_{j})$ is the $n\times n$
Gram matrix and $K_{{\bf x}x}\in\mathbb{R}^{n}$, with $\left[K_{{\bf x}x}\right]_{i}=k(x_{i},x),$
and similarly for kernel $m$. Now, to obtain an estimate of

\begin{eqnarray*}
\mu_{\tilde{p}(\cdot|do(x))} & = & \int l(\cdot,y)\int p(y|x,z)p(z)dzdy,
\end{eqnarray*}
we simply average over the empirical $\hat{p}(z)$ to obtain: 
\[
\hat{\mu}_{\tilde{p}(\cdot|do(x))}=w(x)^{\top}L_{{\bf y}\cdot}=\sum_{i=1}^{n}w(x)_{i}l(\cdot,y_{i})=\frac{1}{n}\sum_{i=1}^{n}\sum_{j=1}^{n}a(x,z_{j})_{i}l(\cdot,y_{i}),
\]
i.e.

\[
w(x)=\frac{1}{n}\left(K_{{\bf xx}}\circ M_{{\bf zz}}+\epsilon_{n}I\right)^{-1}\left(K_{{\bf x}x}\circ M_{{\bf zz}}{\bf 1}\right).
\]
Denote 
\[
W_{{\bf x}}=\left[w(x_{1})\cdots w(x_{n})\right]^{\top}=\frac{1}{n}\left(K_{{\bf xx}}\circ{\bf 1}{\bf 1}^{\top}M_{{\bf zz}}\right)\left(K_{{\bf xx}}\circ M_{{\bf zz}}+\epsilon_{n}I\right)^{-1}.
\]
We also denote $\bar{w}=\frac{1}{n}\sum_{i=1}^{n}w(x_{i})$. Then
$\hat{\mu}_{\tilde{p}}=\bar{w}^{\top}L_{{\bf y}\cdot}$ estimates
the embedding of $\tilde{p}(y)=\int p(y|x,z)p(x)p(z)dxdz$. Now, under
the null, $\tilde{p}=\tilde{p}(\cdot|do(x)),\forall x$. Thus, we define
the statistic as the sum of the RKHS distances of the corresponding
embeddings 
\begin{eqnarray*}
S & = & \sum_{i=1}^{n}\left\Vert \hat{\mu}_{\tilde{p}(\cdot|do(x)=x_{i})}-\hat{\mu}_{\tilde{p}}\right\Vert _{\mathcal{H}_{l}}^{2}\\
 & = & \sum_{i=1}^{n}\left\Vert \left(w(x_{i})-\bar{w}\right)^{\top}L_{{\bf y}\cdot}\right\Vert _{\mathcal{H}_{l}}^{2}\\
 & = & \sum_{i=1}^{n}\text{Tr}\left[\left(w(x_{i})-\bar{w}\right)^{\top}L_{{\bf y}{\bf y}}\left(w(x_{i})-\bar{w}\right)\right]\\
 & = & \text{Tr}\left[L_{{\bf y}{\bf y}}\sum_{i=1}^{n}\left(w(x_{i})-\bar{w}\right)\left(w(x_{i})-\bar{w}\right)^{\top}\right]\\
 & = & \text{Tr}\left[L_{{\bf y}{\bf y}}W_{{\bf x}}^{\top}HW_{{\bf x}}\right]. 
\end{eqnarray*}
The test statistic $S$ uses a mean embedding that bears many similarities to \cite{10.1093/biomet/asad042}. In this context, we consider an RBF kernel on $Y$, which takes the outcome to the RKHS.

\section{Simulation algorithms}

\subsection{Binary Treatment}
\label{bin_generation}

Fix $\beta_{XY}>0$, and $\tau^{2}>0$.
$H_{0}$ case:
\begin{eqnarray*}
Z_{i} & \sim & \mathcal{N}\left(0,1\right),\\
X_{i}|Z_{i} & \sim & \text{Bernoulli}\left(\frac{1}{1+e^{-Z_{i}}}\right),\\
Y_{i}|Z_{i} & \sim & \mathcal{N}\left(\beta_{XY} Z_{i},\tau^{2}\right).
\end{eqnarray*}
$H_{1}$ case:
\begin{eqnarray*}
Z_{i} & \sim & \mathcal{N}\left(0,1\right),\\
X_{i}|Z_{i} & \sim & \text{Bernoulli}\left(\frac{1}{1+e^{-Z_{i}}}\right),\\
Y_{i}|Z_{i} & \sim & \mathcal{N}\left(\beta_{XY}\left(2X_{i}-1\right)\left|Z_{i}\right|,\tau^{2}\right).
\end{eqnarray*}
Note that in the alternative case, $X_{i}$ directly modulates the sign of the mean of $Y_{i}$. However, $2X_{i}-1$ will be strongly positively correlated with the sign of $Z_{i}$ implying that there is only a slight change in the dependence structure of $\left(X_{i},Y_{i},Z_{i}\right)$. In addition, all the marginals are the same.

\subsection{Continuous treatment}
\label{cont_generation}
We simulate data from the do-null and the alternative using rejection sampling \citep{evans2024parameterizing}, described in Algorithm \ref{generate_data}. To construct a data set where there is marginal dependence (\Cref{marg_break}) but the do-null holds, we replace the normal marginal distributions for $X,Y,Z$ in Algorithm \ref{generate_data} with exponential distributions, i.e.~$Y \sim \text{Exp}(X\beta_{XY})$. To generate data where there is conditional dependence but the do-null holds, we simply use the last set of parameters in Appendix \ref{cond_param} with normal marginal distributions.

\begin{algorithm}[H]
\SetAlgoLined
\KwInput{Number of samples $n$, dependencies $\beta_{XY}$, $\beta_{XZ}$, $\beta_{YZ}$, variance parameters $\theta,\phi\in \mathbb{R}^+$, dimensions $d_x,d_y,d_z$}
Initialize data container $\mathcal{D} = \{ \}$\\
Set $\mathbold{\beta}_{XZ}=[\underbrace{\beta_{XZ}}_{1:3},\underbrace{0}_{4:d_Z}]$ \\

 \While{$\#$ of samples $<n$}{
Set $p_X = \mathcal{N}(\textbf{0},\theta\cdot \phi\cdot \textbf{I}_{d_x})$
Sample $\{x_i\}_{i=1}^{N} \sim p_X$\\
Sample $\{y_i,z_i\}_{i=1}^{N} \sim \mathcal{N}(\textbf{0},\Sigma_{d_y+d_z})$\\
Transform 
$\{y_i'\}_{i=1}^{N} = \text{CDF}_{\mathcal{N}(0,1)}(\{y_i\}_{i=1}^{N})$\\
Define $p_{Y|X} = \mathcal{N}(X\beta_{Y},1)$\\
Set $\{y_i\}_{i=1}^{N} = \text{ICDF}_{p_{Y\mid X}}(\{y_i'\}_{i=1}^{N})$\\
Set $\mu_{X|Z} = Z\cdot \mathbold{\beta}_{XZ}$\\
Define $p_{X|Z} = \mathcal{N}(\mu_{X| Z},\phi)$\\
Calculate $\omega_i = \frac{p_{X| Z}(x_i)}{p_X(x_i)}$\\
Run rejection sampling using $\omega_i$ and obtain $\mathcal{D}' = \{x_i,y_i,z_i\}_{i=1}^{N'}\sim p^*$\\
Append data $\mathcal{D} = \mathcal{D} \cup \mathcal{D}'$
}
\KwReturn{$\mathcal{D}$}
\caption{Generating continuous data for $H_0$ and $H_1$}
\label{generate_data}
\end{algorithm}

\subsubsection{Parameter explanation}

There are several parameters used in the data generation algorithm primarily used to control for the difficulty of the problem and the ground truth hypothesis.

\begin{enumerate}[noitemsep]
    \item $\beta_{XY}$: Controls the dependency between $X$ and $Y$. A $\beta_{XY}>0$ implies $H_1$ ground truth and $\beta_{XY}=0$ implies $H_0$ ground truth
    \item $\beta_{XZ}$: Controls the dependency between $X$ and $Z$. A high $\beta_{XZ}$ implies a stronger dependency on $Z$ for $X$, implying a harder problem.
    \item $\beta_{YZ}$: Controls the dependency between $Y$ and $Z$. Fixed at a high value to ensure $Y$ is being confounded by $Z$.
    \item $\theta, \phi$: Controls the variance of $p_X$ and $p_{X| Z}$. $\theta > \phi$ ensures a higher \textit{Effective Sample Size} for true weights.
    \item $d_x,d_y,d_z$: Dimensionality of $X,Y,Z$. Higher dimensions imply a harder problem.  
    \item $\Sigma_{d_y+d_z}$: The covariance matrix that links $Z$ and $Y$. It should be noted that this covariance matrix is a function of $X$.
\end{enumerate}

\subsection{Mixed treatment}
\label{mixed_generation}
\begin{algorithm}[H]
\SetAlgoLined
\KwInput{Number of samples $n$, dependencies $\beta_{XY}$, $\beta_{XZ}$, $\beta_{YZ}$, variance parameters $\theta,\phi\in \mathbb{R}^+$, dimensions $d_x,d_y,d_z$}
Initialize data container $\mathcal{D} = \{ \}$\\
Set $\mathbold{\beta}_{XZ}=[\underbrace{\beta_{XZ}}_{1:3},\underbrace{0}_{4:d_Z}]$ \\
\While{$\#$ of samples $<n$}{
Set $p_X = \mathcal{N}(\textbf{0},\theta\cdot \phi\cdot \textbf{I}_{d_x})$\\ 
Set $p_X^{\tbin} =\text{Bin}(p=0.5)$\\ 
Sample $\{x_i^{\tcont}\}_{i=1}^{N} \sim p_X$\\
Sample $\{x_i^{\tbin}\}_{i=1}^{N} \sim \text{Bin}(p=0.5)$\\
Concatenate $X = X_{\tcont} \cup X_{\tbin}$ \\
Sample $\{y_i,z_i\}_{i=1}^{N} \sim \mathcal{N}(\textbf{0},\Sigma_{d_y+d_z})$\\
Transform
$\{y_i'\}_{i=1}^{N} = \text{CDF}_{\mathcal{N}(0,1)}(\{y_i\}_{i=1}^{N})$\\
Define $p_{Y| X} = \mathcal{N}(X\beta_{Y},1)$\\
Set $\{y_i\}_{i=1}^{N} = \text{ICDF}_{p_{Y| X}}(\{y_i'\}_{i=1}^{N})$\\
Set $\mu_{X| Z} = Z\cdot \mathbold{\beta}_{XZ}$\\
Set $\nu_{X| Z} = \frac{1}{1+e^{-\mu_{X| Z}}}$\\
Define $p_{X| Z} = \mathcal{N}(\mu_{X| Z},\phi)$\\
Define $p_{X| Z}^{\tbin} = \text{Bin}(p=\nu_{X| Z})$\\
Calculate $\omega_i = \frac{p_{X| Z}(x_i^{\tcont})}{p_X(x_i^{\tcont})}\cdot \frac{p_{X| Z}^{\tbin}(x_i^{\tbin})}{p_X^{\tbin}(x_i^{\tbin})}$\\
Run rejection sampling using $\omega_i$ and obtain $\mathcal{D}' = \{x_i,y_i,z_i\}_{i=1}^{N'}\sim p^*$\\
Append data $\mathcal{D} = \mathcal{D} \cup \mathcal{D}'$
}
\KwReturn{$\mathcal{D}$}
\caption{Generating mixed data for $H_0$ and $H_1$}
\label{generate_data_mixed}
\end{algorithm}
\section{Parameters for data generation}
\label{params_data_gen}
We provide parameters used in the data generation procedure for each type of treatment. Exact details can be found in the code base.
\textit{Binary treatment}

\begin{enumerate}
    \item Dependency $\beta_{XY}$: $[0.0,0.02,0.04,0.06,0.08,0.1]$
    \item Variance $\tau$: $1.0$
\end{enumerate}

\textit{Continuous treatment}
\begin{enumerate}
    \item $\beta_{XY}$: $[0.0,0.001,0.002,0.003,0.004,0.005,0.008,0.012,0.016,0.02]$
    \item $\beta_{XZ}$: $d_Z=1: 0.75,d_Z=3,15,50: 0.25$
    \item $\beta_{YZ}$: $[0.5,0.0]$ 
    \item $\theta, \phi$: $d_Z=1: (2,2)$, $d_Z=3: (4,2)$, $d_Z=15: (8,2)$, $d_Z=50: (16,2)$
    \item $\Sigma_{d_y+d_z}$: See code base for details
\end{enumerate}

\textit{Mixed treatment}
\begin{enumerate}
    \item $\beta_{XY}$: $[0.0,0.002,0.004,0.006,0.008, 0.01, 0.015,0.02,0.025,0.03,0.04,0.05,0.1]$
    \item $\beta_{XZ}$: $0.05$
    \item $\beta_{YZ}$: $[0.5,0.0]$ 
    \item $\theta, \phi$: $d_Z=2: (2,2)$, $d_Z=15: (16,2)$, $d_Z=50: (16,2)$ 
    \item $\Sigma_{d_y+d_z}$: See code base for details
\end{enumerate}

\textit{$X \not\perp Y$ data}

\begin{enumerate}
    \item $\beta_{XY}$: $[0.0]$
    \item $\beta_{XZ}$: $1.0$
    \item $\beta_{YZ}$: $[0.5,0.0]$ 
    \item $\theta, \phi$: $d_Z=1: (0.1,1.5)$ 
    \item $\Sigma_{d_y+d_z}$: See code base for details
\end{enumerate}

\textit{$X \not\perp Y\mid Z$ data}
\label{cond_param}
\begin{enumerate}
    \item $\beta_{XY}$: $[0.0]$
    \item $\beta_{XZ}$: $0.0$.
    \item $\beta_{YZ}$: $[-0.5,4.0]$ 
    \item $\theta, \phi$: $d_Z=1: (1.0,2.0)$ 
    \item $\Sigma_{d_y+d_z}$: See code base for details
\end{enumerate}


\clearpage

\bibliography{ref}





\end{document}